\documentclass[twocolumn]{aastex62}
\pdfoutput=1 
\usepackage{amsmath,amstext,amssymb,mathtools,graphicx,color}
\usepackage[T1]{fontenc}
\usepackage{apjfonts}
\usepackage[figure,figure*]{hypcap}
\usepackage{chngpage}

\graphicspath{{./}{plots/}}

\shorttitle{Composition of kilonova}
\shortauthors{W. Even et al.}
\begin{document}


\title{Composition Effects on Kilonova Spectra and Light Curves:  I}


\correspondingauthor{W. Even}
\email{even@lanl.gov}


\author[0000-0002-5412-3618]{Wesley Even}
\affiliation{Center for Theoretical Astrophysics, Los Alamos National Laboratory, Los Alamos, NM, 87545, USA}
\affiliation{Joint Institute for Nuclear Astrophysics - Center for the Evolution of the Elements, USA}
\affiliation{Computer, Computational, and Statistical Sciences Division, Los Alamos National Laboratory, Los Alamos, NM, 87545, USA}
\affiliation{Department of Physical Science, Southern Utah University, Cedar City, UT, 84720, USA}

\author[0000-0003-4156-5342]{Oleg Korobkin}
\affiliation{Center for Theoretical Astrophysics, Los Alamos National Laboratory, Los Alamos, NM, 87545, USA}
\affiliation{Joint Institute for Nuclear Astrophysics - Center for the Evolution of the Elements, USA}
\affiliation{Computer, Computational, and Statistical Sciences Division, Los Alamos National Laboratory, Los Alamos, NM, 87545, USA}

\author[0000-0003-2624-0056]{Christopher~L. Fryer}
\affiliation{Center for Theoretical Astrophysics, Los Alamos National Laboratory, Los Alamos, NM, 87545, USA}
\affiliation{Joint Institute for Nuclear Astrophysics - Center for the Evolution of the Elements, USA}
\affiliation{Computer, Computational, and Statistical Sciences Division, Los Alamos National Laboratory, Los Alamos, NM, 87545, USA}
\affiliation{The University of Arizona, Tucson, AZ 85721, USA}
\affiliation{Department of Physics and Astronomy, The University of New Mexico, Albuquerque, NM 87131, USA}
\affiliation{The George Washington University, Washington, DC 20052, USA}

\author[0000-0003-1087-2964]{Christopher~J. Fontes}
\affiliation{Center for Theoretical Astrophysics, Los Alamos National Laboratory, Los Alamos, NM, 87545, USA}
\affiliation{X Computational Physics Division, Los Alamos National Laboratory, Los Alamos, NM, 87545, USA}

\author[0000-0003-3265-4079]{R.~T. Wollaeger}
\affiliation{Center for Theoretical Astrophysics, Los Alamos National Laboratory, Los Alamos, NM, 87545, USA}
\affiliation{Computer, Computational, and Statistical Sciences Division, Los Alamos National Laboratory, Los Alamos, NM, 87545, USA}

\author[0000-0001-6893-0608]{Aimee Hungerford}
\affiliation{Center for Theoretical Astrophysics, Los Alamos National Laboratory, Los Alamos, NM, 87545, USA}
\affiliation{Joint Institute for Nuclear Astrophysics - Center for the Evolution of the Elements, USA}
\affiliation{X Computational Physics Division, Los Alamos National Laboratory, Los Alamos, NM, 87545, USA}

\author[0000-0002-5936-3485]{Jonas Lippuner}
\affiliation{Center for Theoretical Astrophysics, Los Alamos National Laboratory, Los Alamos, NM, 87545, USA}
\affiliation{Joint Institute for Nuclear Astrophysics - Center for the Evolution of the Elements, USA}
\affiliation{Computer, Computational, and Statistical Sciences Division, Los Alamos National Laboratory, Los Alamos, NM, 87545, USA}

\author[0000-0001-6432-7860]{Jonah Miller}
\affiliation{Center for Theoretical Astrophysics, Los Alamos National Laboratory, Los Alamos, NM, 87545, USA}
\affiliation{Computer, Computational, and Statistical Sciences Division, Los Alamos National Laboratory, Los Alamos, NM, 87545, USA}
\affiliation{Center for Nonlinear Studies, Los Alamos National Laboratory, Los Alamos, NM 87545, USA}

\author[0000-0002-9950-9688]{Matthew~R. Mumpower}
\affiliation{Center for Theoretical Astrophysics, Los Alamos National Laboratory, Los Alamos, NM, 87545, USA}
\affiliation{Joint Institute for Nuclear Astrophysics - Center for the Evolution of the Elements, USA}
\affiliation{Theoretical Division, Los Alamos National Laboratory, Los Alamos, NM, 87545, USA}

\author[0000-0002-9950-9688]{G.~Wendell Misch}
\affiliation{Center for Theoretical Astrophysics, Los Alamos National Laboratory, Los Alamos, NM, 87545, USA}
\affiliation{Joint Institute for Nuclear Astrophysics - Center for the Evolution of the Elements, USA}
\affiliation{Theoretical Division, Los Alamos National Laboratory, Los Alamos, NM, 87545, USA}

\begin{abstract}

The merger of neutron star binaries is believed to eject a wide range of heavy elements into the universe.  By observing the emission from this ejecta, scientists can probe the ejecta properties (mass, velocity and composition distributions).  The emission (a.k.a. kilonova) is powered by the radioactive decay of the heavy isotopes produced in the merger and this emission is reprocessed by atomic opacities to optical and infra-red wavelengths.  Understanding the ejecta properties requires calculating the dependence of this emission on these opacities.  The strong lines in the optical and infra-red in lanthanide opacities have been shown to significantly alter the light-curves and spectra in these wavelength bands, arguing that the emission in these wavelengths can probe the composition of this ejecta.  Here we study variations in the kilonova emission by varying individual lanthanide (and the actinide uranium) concentrations in the ejecta.  The broad forest of lanthanide lines makes it difficult to determine the exact fraction of individual lanthanides.  Nd is an exception.  Its opacities above 1 micron are higher than other lanthanides and observations of kilonovae can potentially probe increased abundances of Nd.  Similarly, at early times when the ejecta is still hot (first day), the U opacity is strong in the 0.2-1 micron wavelength range and kilonova observations may also be able to constrain these abundances.

\end{abstract}

\keywords{}


\section{Introduction} \label{sec:intro}

Even before it was known that gamma-ray bursts (GRBs) were of extragalactic origin, astronomers proposed that the merger of either a binary neutron star (NS/NS) or a black hole--neutron star binary (BH/NS) could produce these energetic explosions~\citep{narayan92}.  Under the accretion disk paradigm, these mergers are expected to produce short-duration gamma-ray bursts.  By assuming mergers produce the bulk of short-duration GRBs~\citep{popham99}, population synthesis studies demonstrated that these bursts should occur beyond their star-formation regions, and in some cases, beyond their host galaxies~\citep{fryer99,bloom99}.  The observed spatial distribution matches these predictions~\citep{fong13}, and concurrent gravitational wave and gamma-ray observations demonstrate that these mergers produce not only gamma-rays~\citep{abbott17a,abbott17b} but also relativistic jets~\citep{mooley18a,mooley18b}.  All of this evidence has gradually produced a near universal acceptance that these compact mergers are behind most short-duration GRBs.

Compact mergers have also been suggested as an r-process site~\citep{lattimer74}, and increasingly detailed studies support this as a leading source of r-process elements~\citep[e.g.][]{freiburghaus99,korobkin12,bauswein13a,2016MNRAS.460.3255R,2017ARNPS..6701916T,metzger14}.  Models of the r-process-rich merger ejecta coupled with merger rates from the gravitational-wave detection of GW170817 demonstrate that compact mergers could well be the dominant r-process source in the universe~\citep[e.g.][]{abbott17c,cote18}.  However, calculating the ejected mass and its composition is not straightforward.  Although the initial ejecta (tidally or dynamically ejected) is very likely to be sufficiently neutron rich to produce a strong r-process signature, details of the nuclear physics can significantly impact the exact quantities of the isotopes produced~\citep{mumpower16, mumpower18, zhu18, vassh19}.  After the initial merger, a disk forms that drives further outflows~\cite{surman04, surman06, surman08}.  If the remnant compact core does not immediately collapse to a black hole, neutron star accretion can also drive outflows.  Neutron star accretion has been studied in detail in the context of supernova fallback, and the ejected mass can be as much as 25\% of the accreted material~\citep{fryer06,fryer09}.  The late-time ejecta composition (disk and neutron star ejecta) can be substantially altered by neutrinos, which reset the electron fraction as this material flows outward.  With higher electron fractions, the composition can range wildly, producing everything from iron peak elements to transuranic isotopes \citep{2013MNRAS.435..502F,2015MNRAS.448..541J,bovard17,2018PhRvD..98f3007F,miller19}.  At this time, modeling alone cannot place strong constraints on the precise elemental distributions of merger ejecta.

The optical and infrared observations of GW170817 provided a first glimpse at merger ejecta~\citep{2017Sci...358.1565E,2017PASJ...69..102T,2017ApJ...850L..37P,2017Natur.551...75S,2017Natur.551...80K,troja17,2017Natur.551...64A,2017ApJ...848L..18N,2017ApJ...848L..19C,2017ApJ...848L..17C,kasliwal17,abbott17c}.  The high optical luminosity in the first day suggested a large amount of low-neutron-fraction (high-$Y_e$) ejecta with a composition closer to iron peak elements.  The late-time (roughly 1 week) peak in the infrared suggested a neutron-rich component (presumably from the dynamical ejecta) that includes a broad lanthanide distribution with high opacities in the optical and near-IR bands.  On the surface, it appears that both red and blue ejecta components as outlined by \cite{metzger14} are present in GW170817.  In principle, features in the optical and infrared observations can be used to probe the ejecta composition, but different studies have predicted very different masses of low- and high-neutron-fraction ejecta~\citep[for a review, see][]{cote18}.  One difficulty in inferring the ejecta mass from the observed emission has been differences in the opacities used by the community.  For example, some groups use simple, constant opacities while others use opacities derived from detailed atomic physics calculations.  Even those calculations that employ detailed opacities typically resort to using a few well-studied elements as surrogates to represent an entire suite of lanthanide opacities.

Without a detailed understanding of the opacities and their effects on the emission, we cannot definitively infer the ejecta composition.  In this paper, we conduct the first study of the emission from kilonova using a complete set of lanthanide opacities~\citep{fontes19}.  We use two basic heavy-element profiles based on the solar r-process and simulated merger yields, testing both the case where these heavy elements comprise the bulk of the composition (mimicking dynamical ejecta) and a small fraction of the composition (closer to late-time ejecta).  We vary the ejecta composition to search for observational features in the kilonova spectra and light curves that might help observers discriminate between compositions.  Section~\ref{sec:caclulations} describes our calculations of the ejecta characteristics and details our simulation tools.  Section~\ref{sec:results} describes the light curves and spectra from our simulation grid, focusing on the physics behind the results that are most sensitive to the composition.  Our conclusions summarize the primary results and provide a link to the spectral and light-curve database produced in this study.

\section{Calculations}
\label{sec:caclulations}

For this sensitivity study, we use a simple 1-dimensional ejecta profile with a homologous outflow.
We consider two abundance patterns: one based on solar r-process residuals, and one computed from network calculations of NS merger dynamical ejecta.  In this section, we describe this initial profile and the transport model and its assumptions.

\subsection{Initial Conditions}

The late-time ejecta is either driven by disk outflows (neutrino- and viscosity-driven) or neutron star accretion.  A number of simulations have been developed to model these outflows~\citep[e.g.][]{2013MNRAS.435..502F,2015MNRAS.448..541J,bovard17,2018PhRvD..98f3007F,miller19}, each with simplifying assumptions.  At this time, no model produces the ejecta properties perfectly (e.g. none develop the highly-relativistic jets observed in GW170817), but models of the bulk of the mass ejected at late times suggest slower velocities than the dynamical ejecta.

The evolution of this homologous outflow is described by the following equations:
\begin{subequations}
  \label{eq1:pmods}
  \begin{gather}
    v(r,t) = \frac{r}{t} \;\;, \\
    \rho(r,t) = \rho_0\left(\frac{t}{t_0}\right)^{-3}
    \left(1 - \frac{r^2}{(v_{\max}t)^2}\right)^3 \;\;, \\
    T(r,t) = T_0\left(\frac{t}{t_0}\right)^{-1}
    \left(1 - \frac{r^2}{(v_{\max}t)^2}\right) \;\;,
  \end{gather}
\end{subequations}
where $t_0=10,000$ s and $v_{\max}=0.4c$, with $c$ the speed of light.
For an ejecta mass of $M_{\rm ej}=0.01$ M$_{\odot}$, the initial peak density is $\rho_0=2.24\times10^{-13}$ g/cm$^3$.  The initial peak temperature $T_0$ is set by the residual thermal energy from nucleosynthesis and the energy deposited from the decay of radioactive elements.  Following \cite{wollaeger18}, we set $T_0=5.7\times10^4$ K.

\subsection{Composition}

\begin{figure*}
  \centering
  \includegraphics[width=0.48\textwidth]{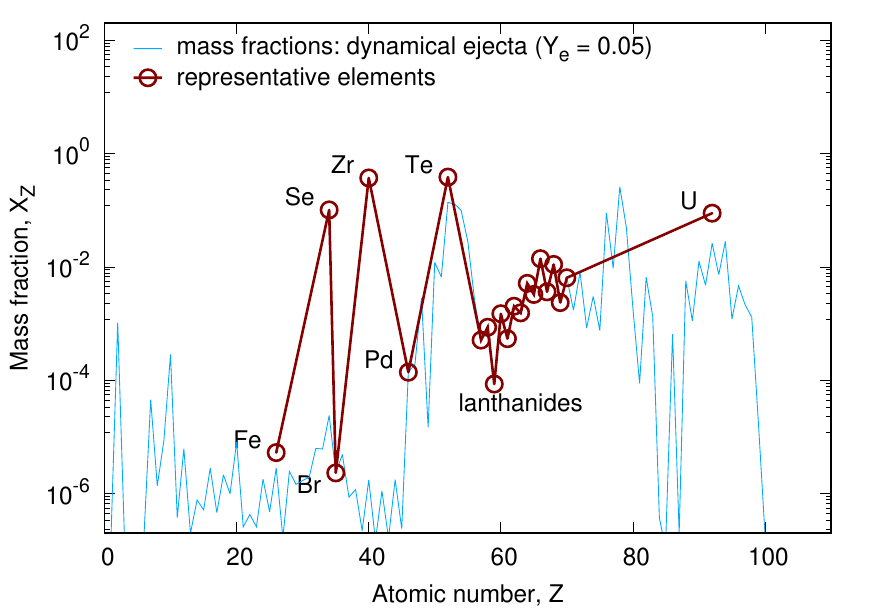}
  \includegraphics[width=0.48\textwidth]{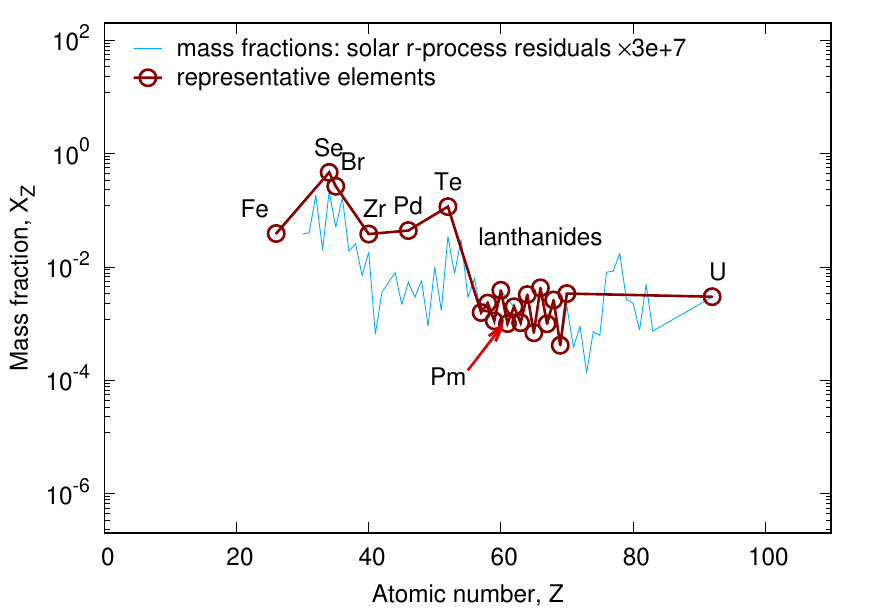}
  \caption{Abundance in mass fraction versus atomic number.
    Left panel: the main r-process abundance pattern, as expected to be produced in dynamical ejecta;
    Right panel: rescaled solar r-process residuals, with added Pm with the mass fraction of
    $X_{\rm Pm} = 10^{-3}$.
  }
  \label{fig:composition}
\end{figure*}

In these models, we study two different ejecta compositions.  The first is based on the solar r-process pattern in the sun, scaling the mass fraction up by $3\times10^7$ to produce an r-process dominated abundance pattern.  Because Pm has no stable isotopes, the solar system has only trace amounts of this element.  However, it will be present in the merger, so we also raise the Pm abundance in this model.  We term this abundance pattern our "solar r-process" ejecta.  Our other composition model consists of the yields from a dynamical ejecta reaction network calculation.  These compositions are shown in Figure~\ref{fig:composition}.

Starting from these base compositions, we vary the abundance of individual elements (focusing on lanthanides) to determine the dependence of the light curves and spectra on these species; the results are summarized in Tables \ref{tab:lanth} and \ref{tab:solaru}.  We also vary the total heavy r-process yields to generate models that are closer to what we expect from the late-time ejecta; table \ref{tab:solarla} summarizes this study.

\begin{table}
\scriptsize
\begin{tabular}{l|c|r|r}
\hline\hline
 & atomic & dynamcial ejecta (Y$_e$=0.05) & solar r-process residuals  \\
element & number & mass fraction & mass fraction \\
\hline
Fe & 26 & 5.32e-6 & 3.90e-2 \\
Se & 34 & 1.01e-1 & 4.66e-1 \\
Br & 35 & 2.32e-6 & 2.65e-1 \\
Zr & 40 & 3.72e-1 & 3.81e-2 \\
Pd & 46 & 1.39e-4 & 4.37e-2 \\
Te & 52 & 3.85e-1 & 1.16e-1 \\
La & 57 & 5.11e-4 & 1.57e-3 \\
Ce & 58 & 8.66e-4 & 2.30e-3 \\
Pr & 59 & 8.59e-5 & 1.12e-3 \\
Nd & 60 & 1.50e-3 & 3.86e-3 \\
Pm & 61 & 5.42e-4 & 9.99e-4 \\
Sm & 62 & 2.03e-3 & 1.98e-3 \\
Eu & 63 & 1.55e-3 & 1.04e-3 \\
Gd & 64 & 5.13e-3 & 3.26e-3 \\
Tb & 65 & 3.27e-3 & 6.88e-4 \\
Dy & 66 & 1.40e-2 & 4.26e-3 \\
Ho & 67 & 3.64e-3 & 9.96e-4 \\
Er & 68 & 1.11e-2 & 2.62e-3 \\
Tm & 69 & 2.34e-3 & 4.14e-4 \\
Yb & 70 & 6.44e-3 & 3.39e-3 \\
U  & 92 & 8.83e-2 & 2.99e-3 \\
\hline
\end{tabular}
\caption{
Effective mass fractions of representative elements used to compute opacities.  The third column is the main r-process abundance pattern expected to be produced in dynamical ejecta.  The fourth column is the rescaled (by a factor of $3\times10^7$) solar r-process residuals, amended with Pm given the mass fraction $X_{\rm Pm} = 10^{-3}$.  This data is displayed visually in Figure~\ref{fig:composition}.}

\label{tab:comp}
\end{table}

In calculating our opacities, we employ the full suite of lanthanide opacities from~\cite{fontes19}, but we do not have detailed opacities for the majority of elements lighter and heavier than the lanthanides.  For the heavier elements (e.g. actinides), we use uranium as a surrogate.  For elements ranging from the iron peak to the lanthanides, we use Fe, Se, Zr, Pd, and Te as surrogates.  For the purpose of computing opacity, the effective abundance of a surrogate element is the sum of the abundances of the elements it represents; these effective abundances are shown in Figure~\ref{fig:composition} and Table~\ref{tab:comp}.

\subsection{Light-Curve Calculations}

\begin{figure}[ht]
  \centering
  \includegraphics[width=\columnwidth]{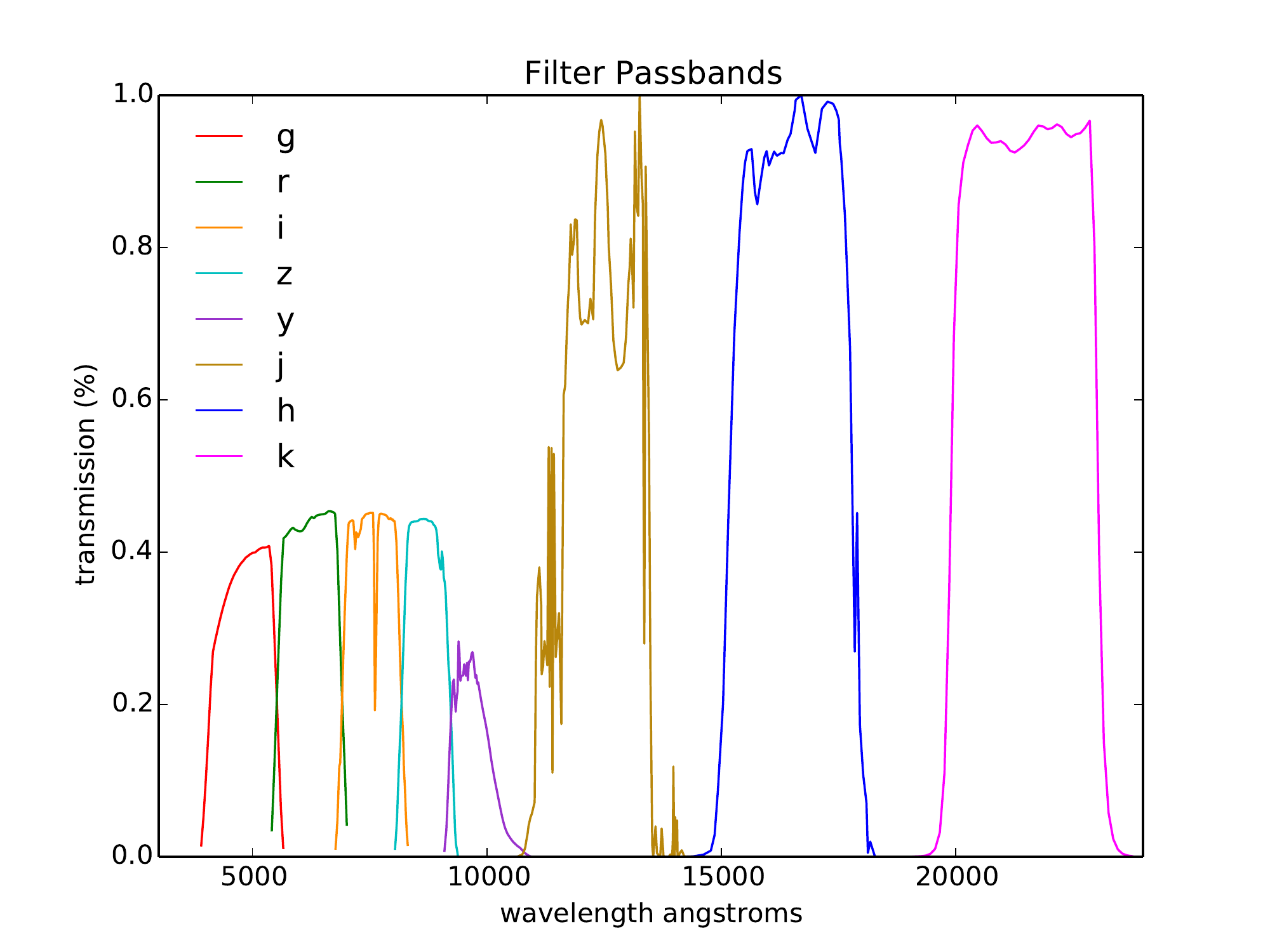}
  \caption{
    Frequency pass bands used in the analysis of this paper.
    The optical {\it grizy} and nIR {\it JHK} passbands are from the LSST and 2MASS filters, respectively.}
  \label{fig:filter}
\end{figure}
For radiative transfer, we use {\tt SuperNu}~\citep{wollaeger2013,wollaeger2014} with tabular opacities from the LANL suite of relativistic atomic physics codes~\citep{fontes2015b}.  {\tt SuperNu} is a multi-dimensional, multi-group Monte Carlo transport method.  It has already been used extensively on kilonova light curves in 1- and 2-dimensions to match both the emission from GW170817~\citep{troja17,tanvir17} and in an initial suite of sensitivity studies~\citep{kasliwal17,wollaeger18} using a reduced set of opacities from~\cite{fontes17}.  For the calculations in this paper, we limit our study to 1-dimensional models with 1024 energy groups.  In calculating colors, we use color band filters for the g, r, i z, y, J, H and K bands (Figure~\ref{fig:filter}).

\begin{figure}
  \centering
  \includegraphics[width=0.9\columnwidth]{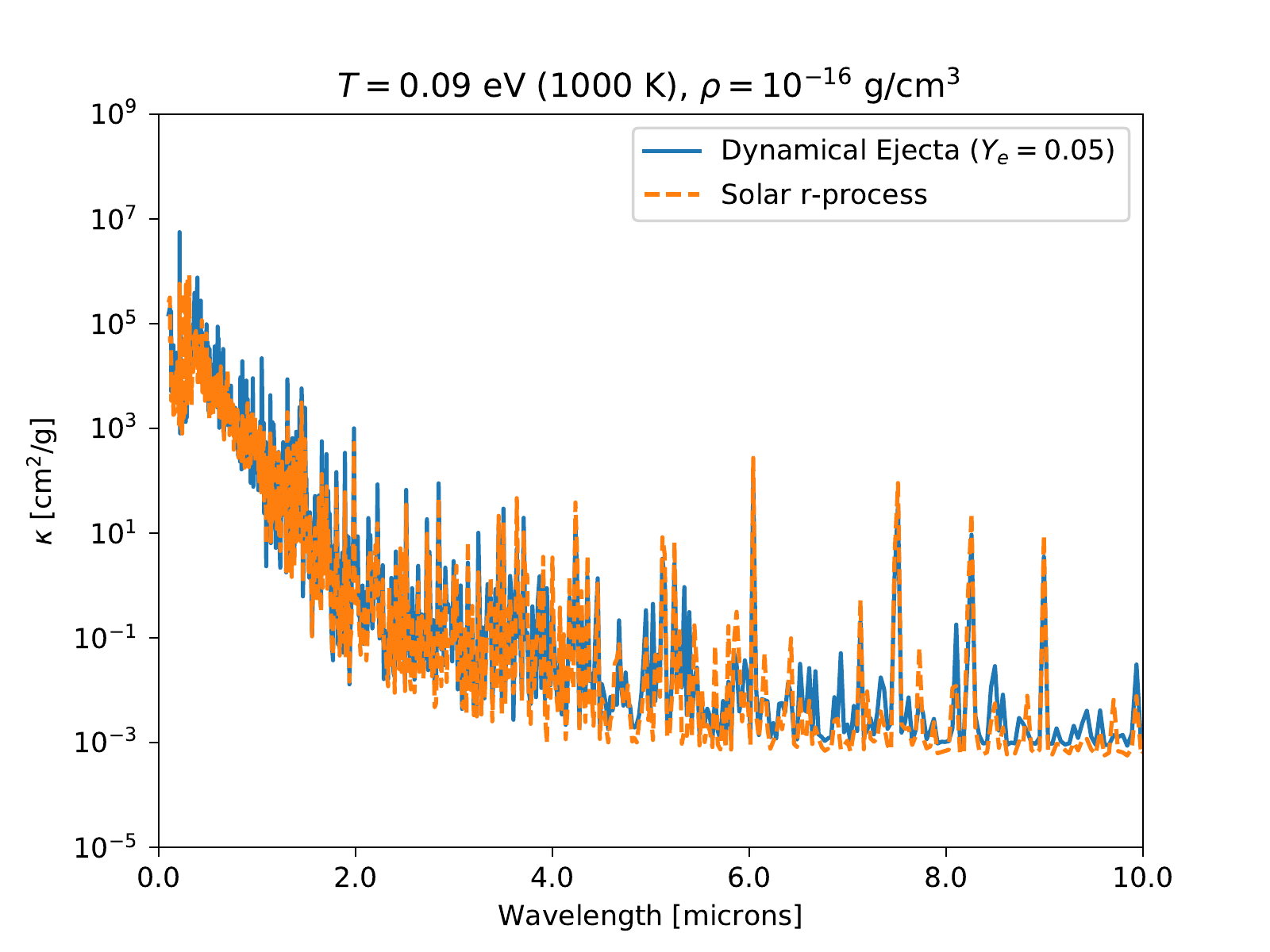}
  \caption{Opacities for our two base abundances:  solar and simulation dynamical ejecta at }
  \label{fig:opac}
\end{figure}
In this project, we used the preferred set of binned opacities from \cite{fontes19}.  Figure~\ref{fig:opac} shows the structure of the opacities binned at the resolution of our calculations.  It shows the opacities of our two base abundances at a temperature of 0.09~eV and a density of $10^{-16} {\rm g\,cm^{-3}}$.  The huge number of lines within the wavelength range plotted here (primarily due to the Lanthanides), makes it very difficult to pick out differences.  We have binned this forest of lines to the energy resolution in our simulations and, even with this coarse binning, it is difficult to distinguish individual elements via specific lines.  In this paper, we examine the dependence of the spectra and light curves on individual lanthanides in order to develop a means to probe the abundances in observed kilonovae.

\section{Spectra and Light Curves}
\label{sec:results}

Throughout this work, magnitudes and fluxes are computed for an event at a standard distance of 10 pc, so magnitudes are absolute magnitudes.

\subsection{Role of Lanthanides}
\label{sec:lanthanide}

\begin{figure*}[!htp]
  \begin{center}
    \includegraphics[width=\textwidth]{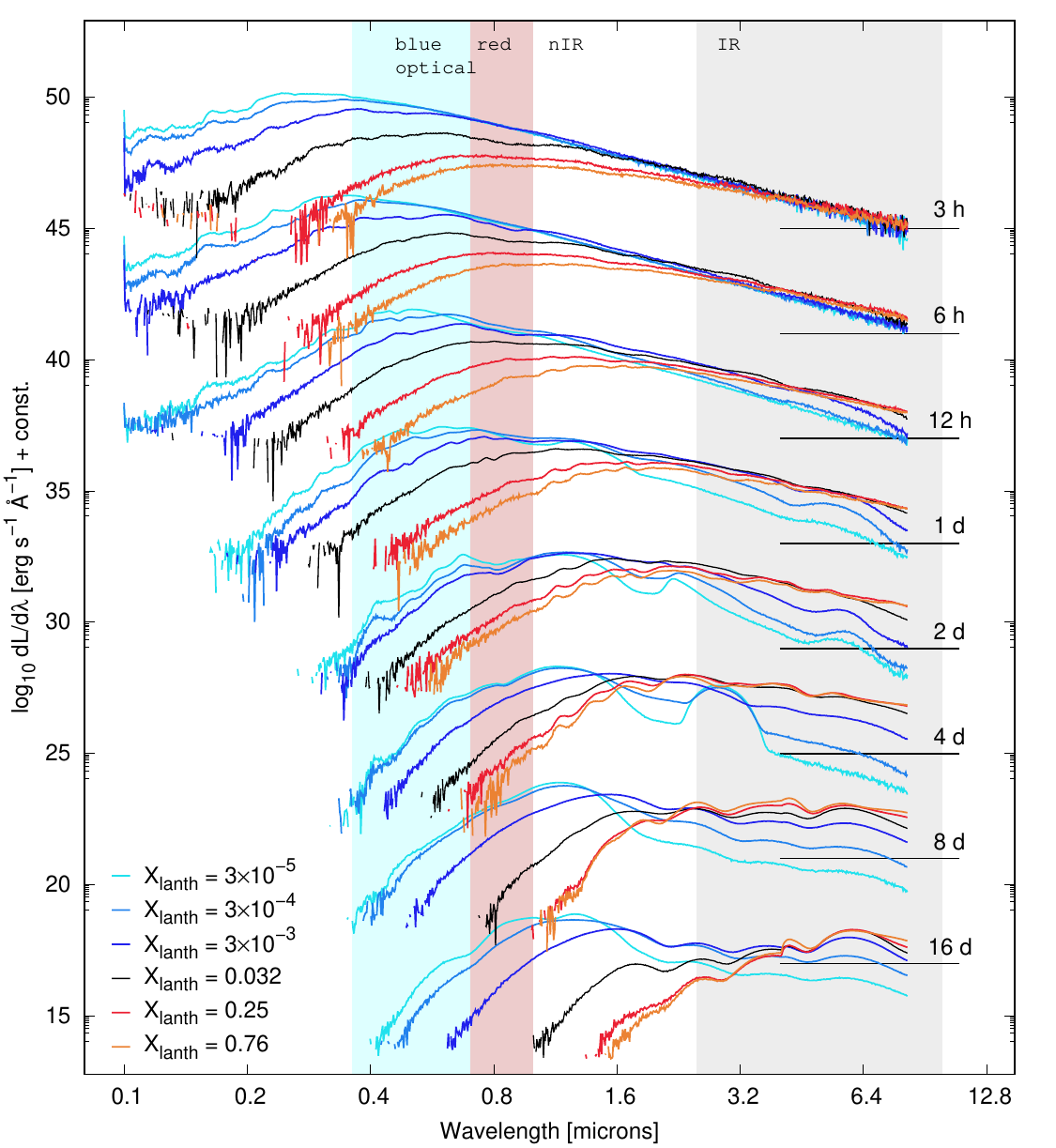}
  \end{center}
  \caption{Sensitivity of kilonova spectra to the mass fraction of lanthanides and uranium ($Z\ge57$).  The baseline model here represents the solar r-process residuals and contains total mass fraction ${X_{\rm lanth}=0.032}$ of lanthanides.  The constant shifting the y axis is set to zero for the spectra at 1\,d.  The black horizontal lines correspond to the same flux level for each model (the difference between each line denotes the constant shift).}
  \label{fig:lanthspec}
\end{figure*}

\begin{figure*}
  \begin{tabular}{ccc}
    \includegraphics[width=0.32\textwidth, trim=.2cm 5.0cm .05cm 4.0cm,clip]{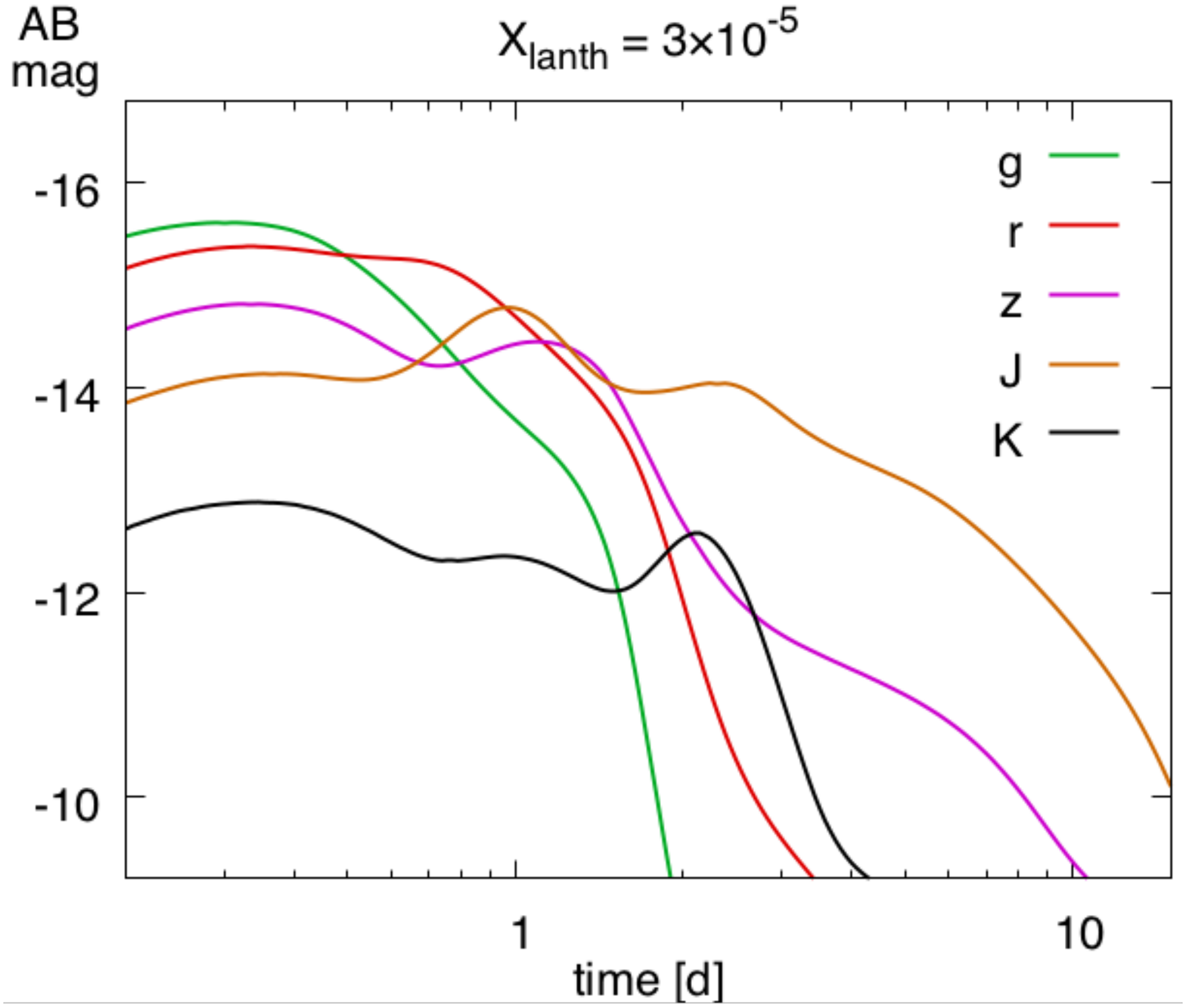} &
    \includegraphics[width=0.32\textwidth, trim=.2cm 5.0cm .05cm 4.0cm,clip]{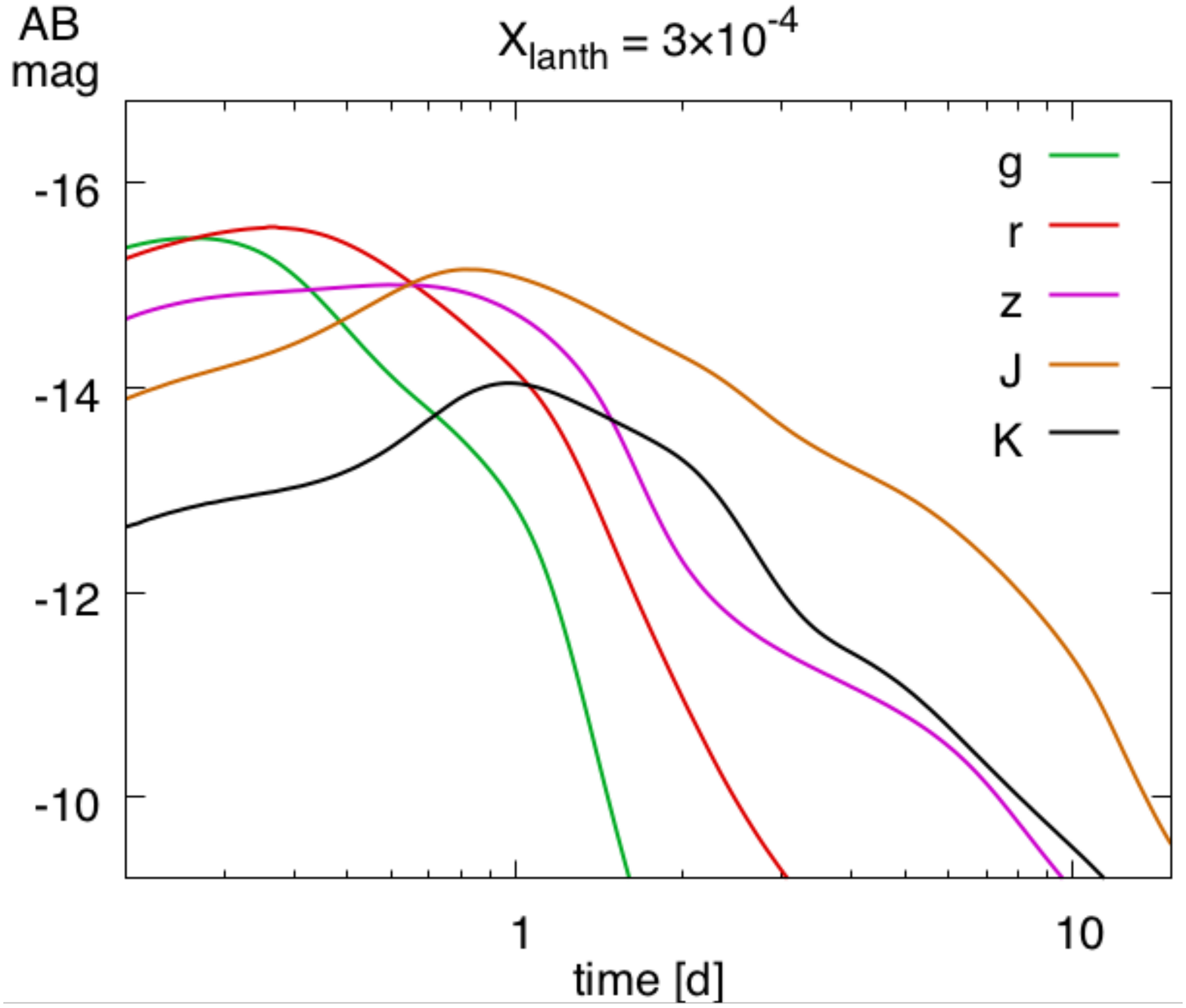} &
    \includegraphics[width=0.32\textwidth, trim=.2cm 5.0cm .05cm 4.0cm,clip]{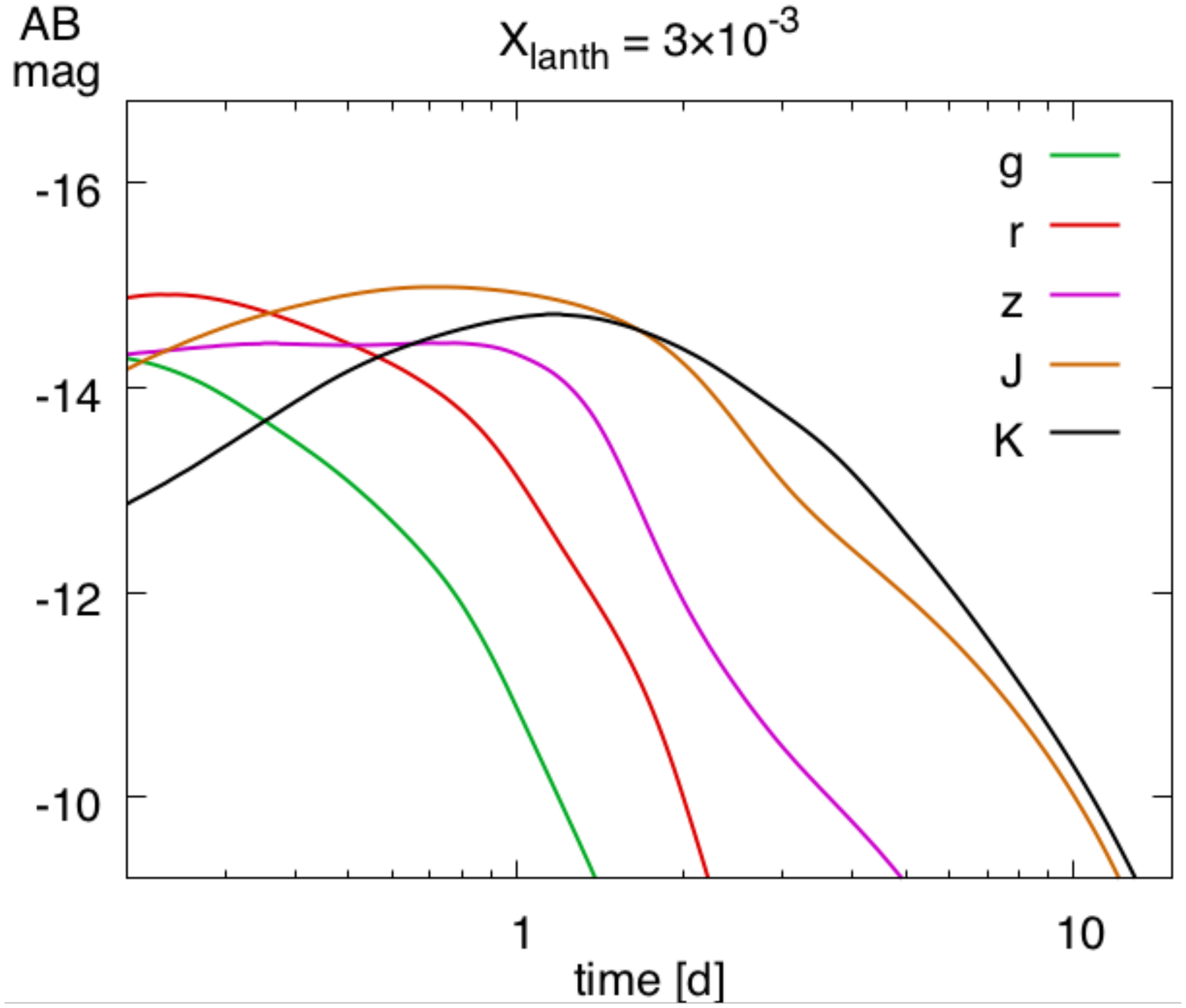}
  \\
    \includegraphics[width=0.32\textwidth, trim=.2cm 5.0cm .05cm 4.0cm,clip]{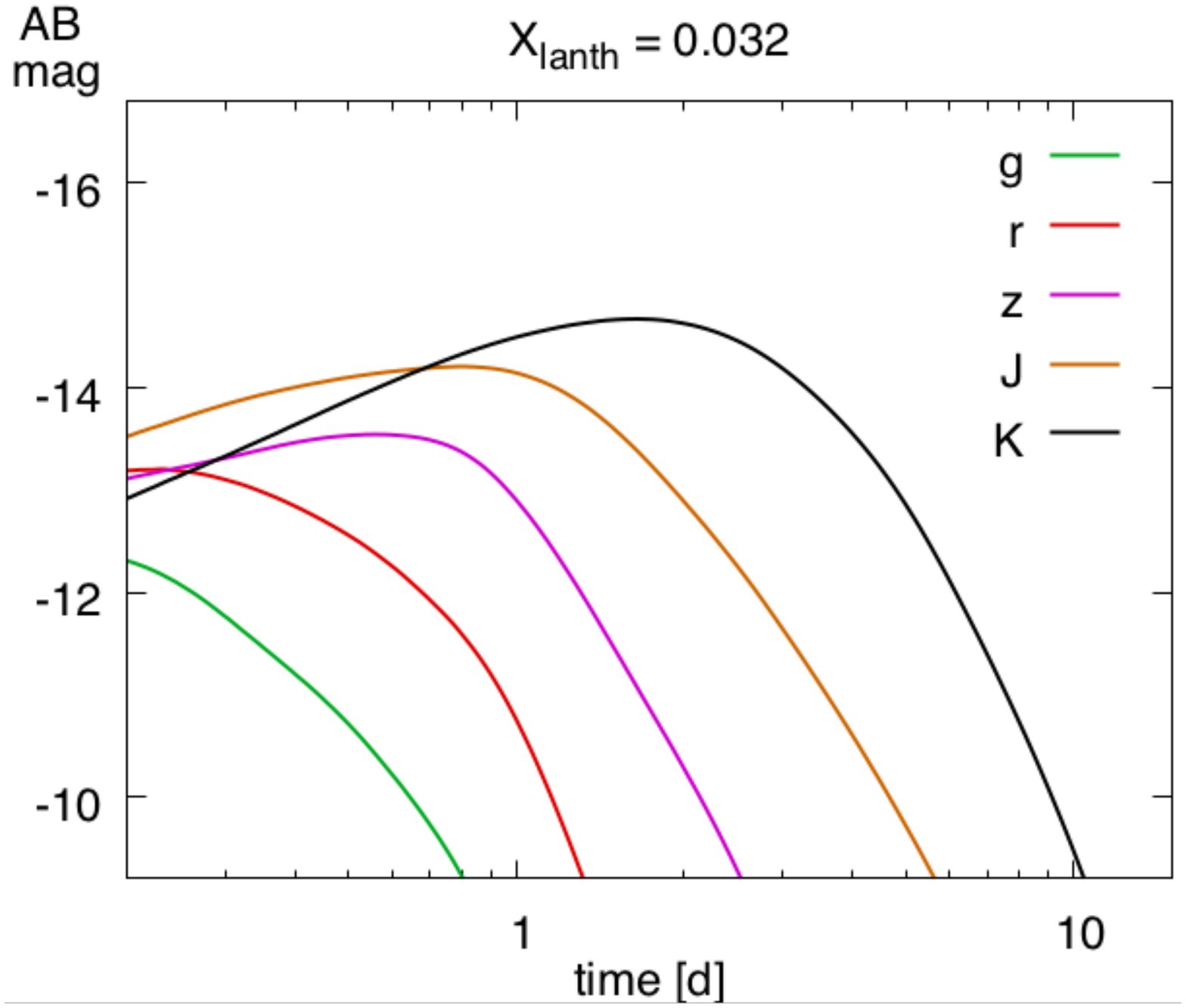} &
    \includegraphics[width=0.32\textwidth, trim=.2cm 5.0cm .05cm 4.0cm,clip]{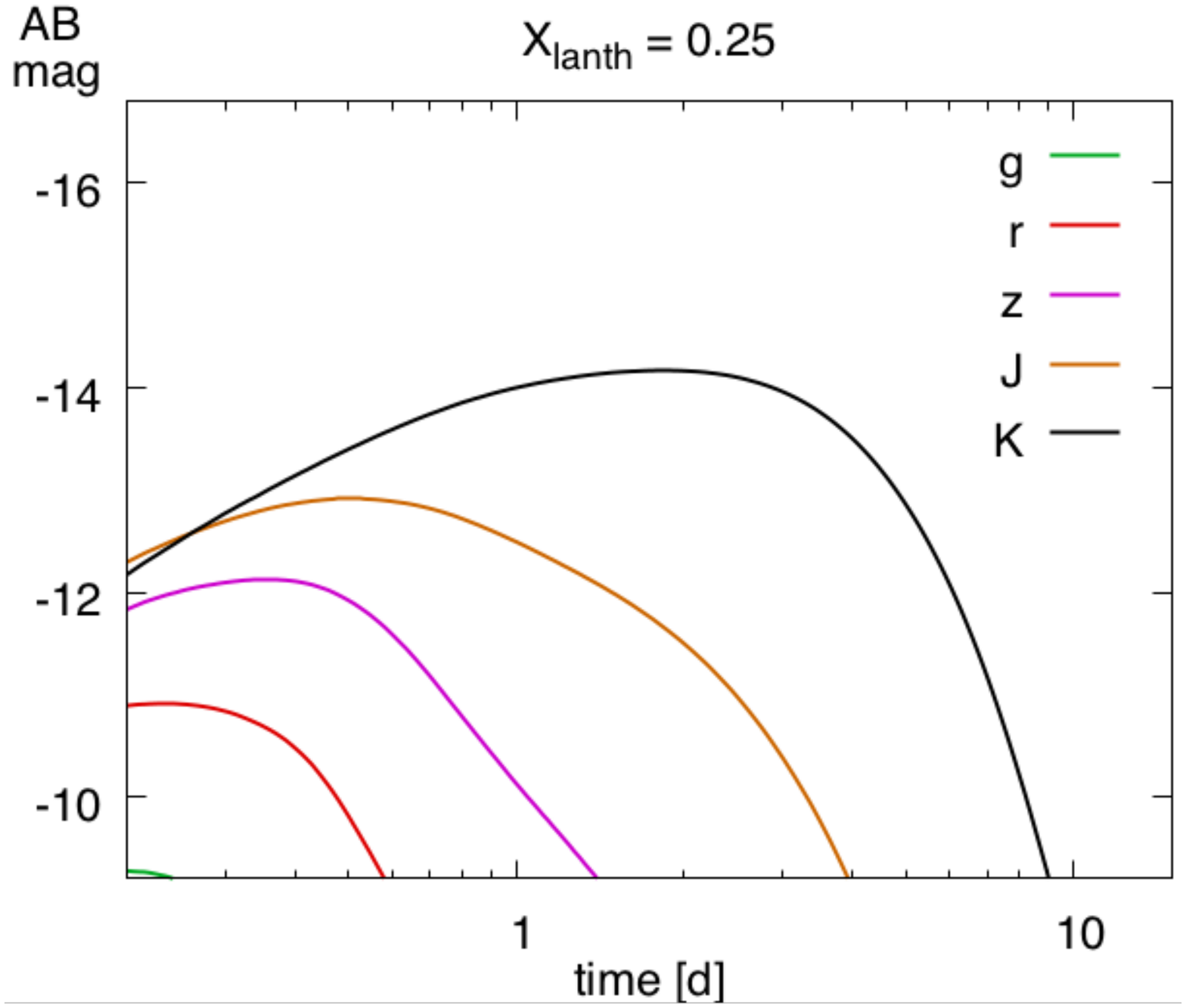} &
    \includegraphics[width=0.32\textwidth, trim=.2cm 5.0cm .05cm 4.0cm,clip]{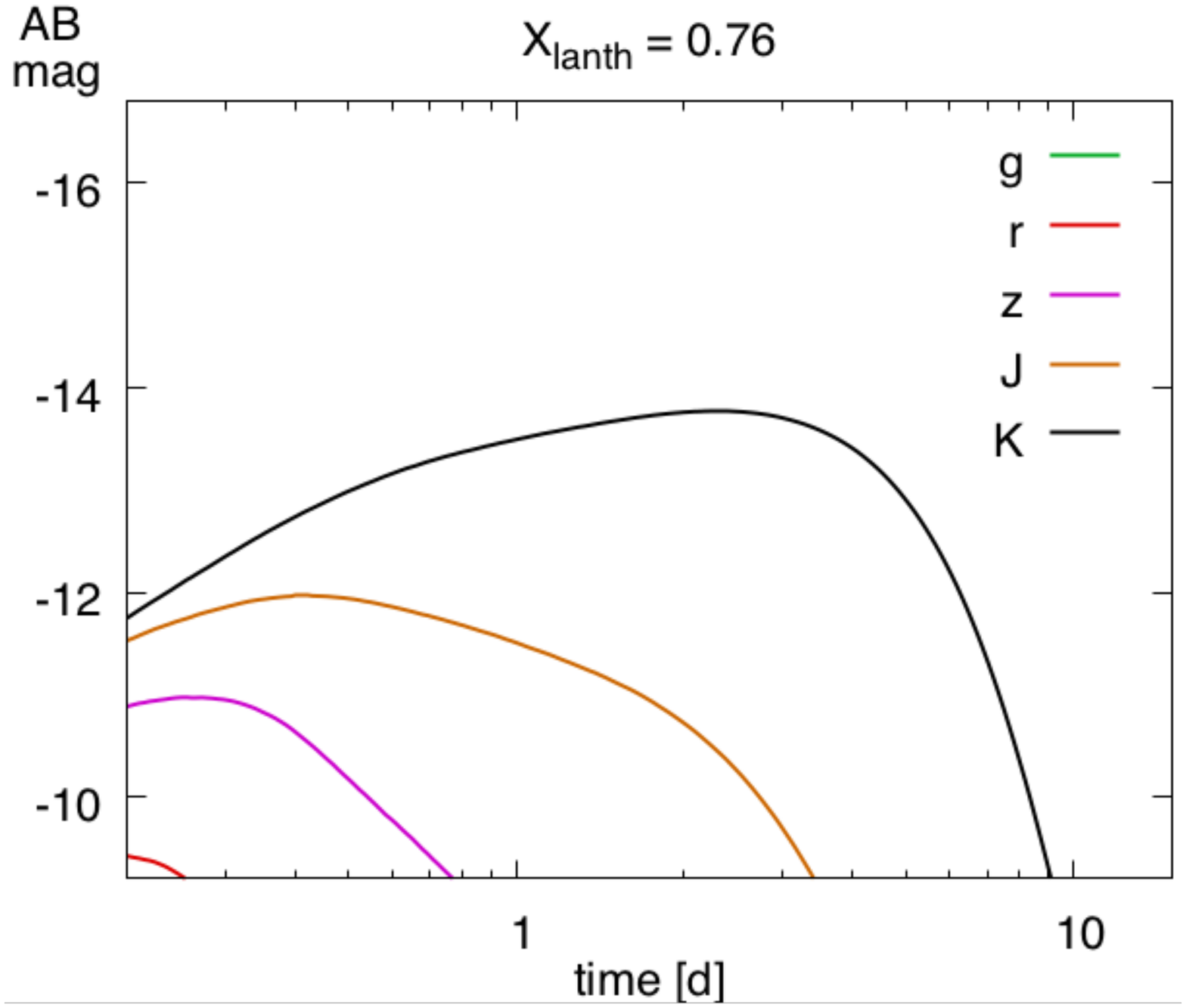}
  \\
  \end{tabular}
  \caption{Sensitivity of the broadband light curves to the mass fraction of lanthanides, relative to the solar r-process residuals, in the order of raising lanthanide fraction.}
  \label{fig:lanthmag}
\end{figure*}

Before we study the roles of individual lanthanides, we first discuss the impact of lanthanides as a group on kilonova emission spectra and observable light curves.  Figure~\ref{fig:lanthspec} shows the spectra at different times for six different lanthanide abundance fractions; we obtained these abundances from renormalized solar r-process residuals with the lanthanides rescaled by powers of ten, giving lanthanide fractions from $3\times10^{-5}$ to $0.75$.  Figure~\ref{fig:lanthmag} shows color band light curves resulting from these spectra.  As has been shown by several groups
\citep[e.g.][]{2017Natur.551...80K}, increased lanthanide fractions reduce the peak in the ultraviolet and optical bands while increasing the emission in the infrared.  The g band peaks strongly at early times if the lanthanide fraction is below $3\times10^{-4}$, but higher lanthanide fractions obscure this peak.  If the lanthanide fraction exceeds $3\times10^{-3}$, the peak emission is delayed, reddened, and dimmed.  The peak magnitude for high lanthanide fractions occurs in the K and redder bands, and it is two magnitudes dimmer than the g-band peak for low lanthanide fractions.  The infrared peak after 5\,d in GW170817 has been used to argue that lanthanides---and hence heavy r-process elements---were present in its ejecta.

\subsection{Solar r-Process Ejecta}

\begin{figure*}[!htp]
  \begin{center}
  \begin{tabular}{cccc}
    \hspace{-9mm}
    \vspace{-2mm}\includegraphics[width=0.3\textwidth, trim=.2cm 3.5cm .05cm 4.5cm,clip]{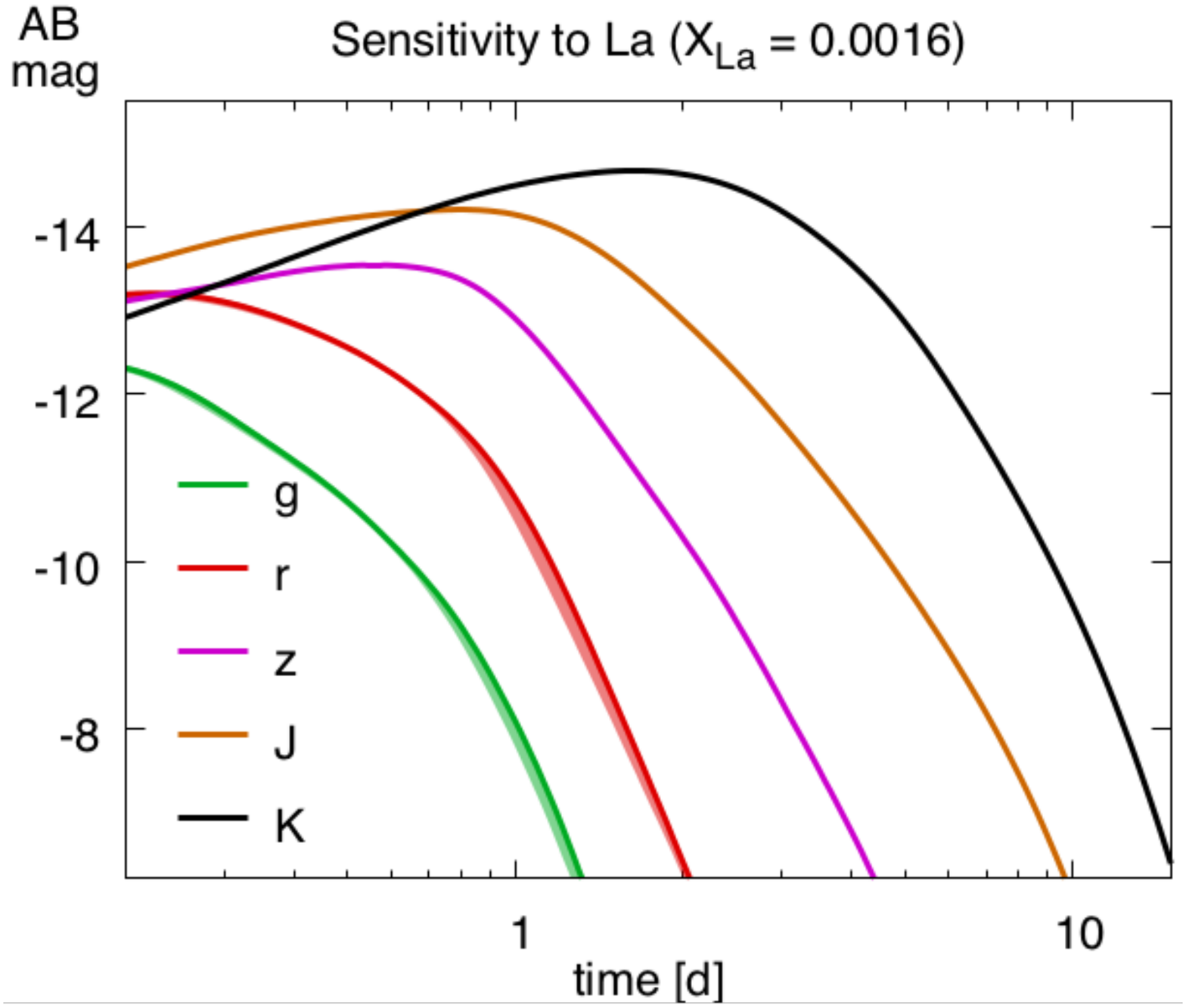} &
    \vspace{-2mm}\includegraphics[width=0.3\textwidth, trim=.2cm 3.5cm .05cm 4.5cm,clip]{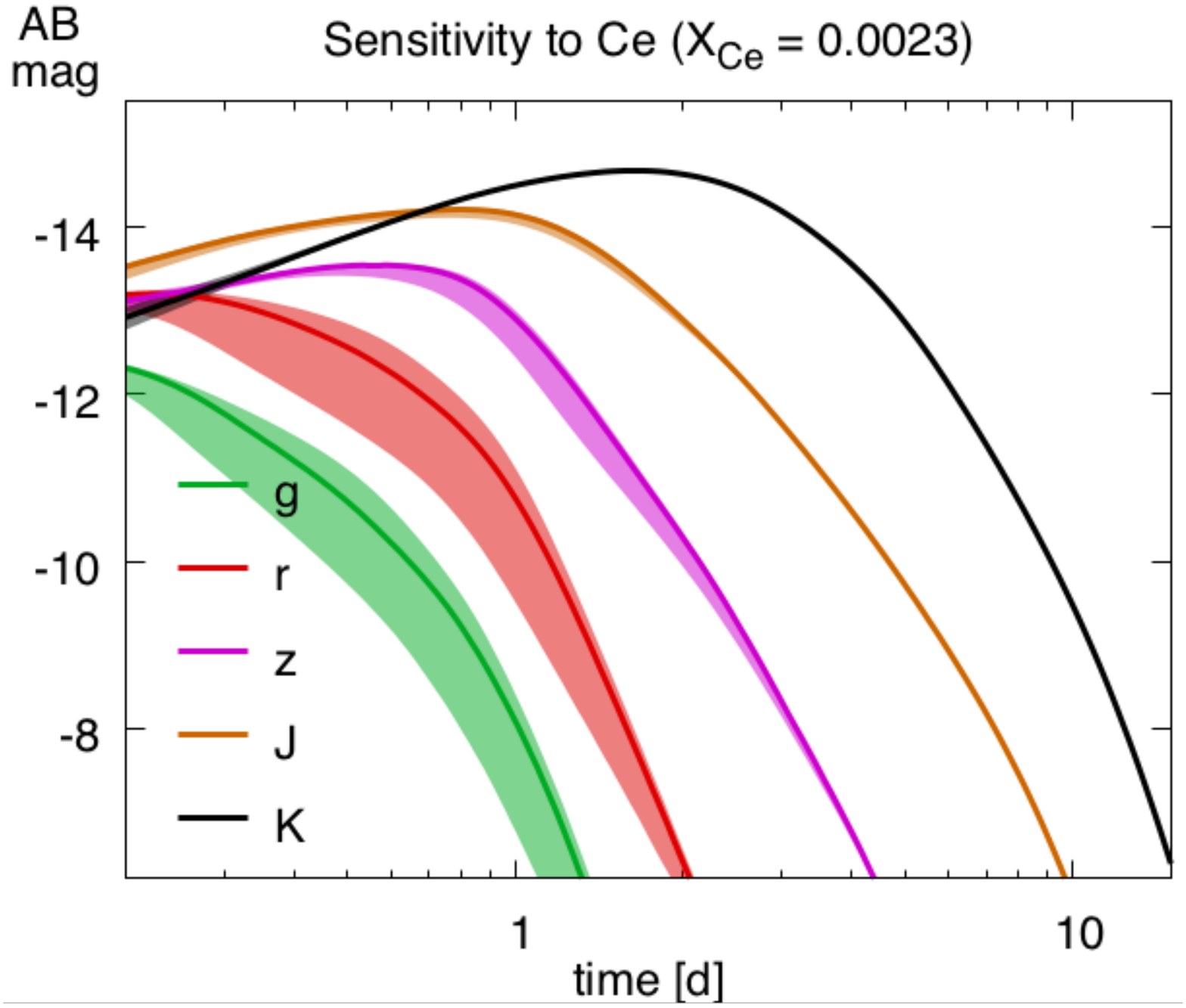} &
    \vspace{-2mm}\includegraphics[width=0.3\textwidth, trim=.2cm 3.5cm .05cm 4.5cm,clip]{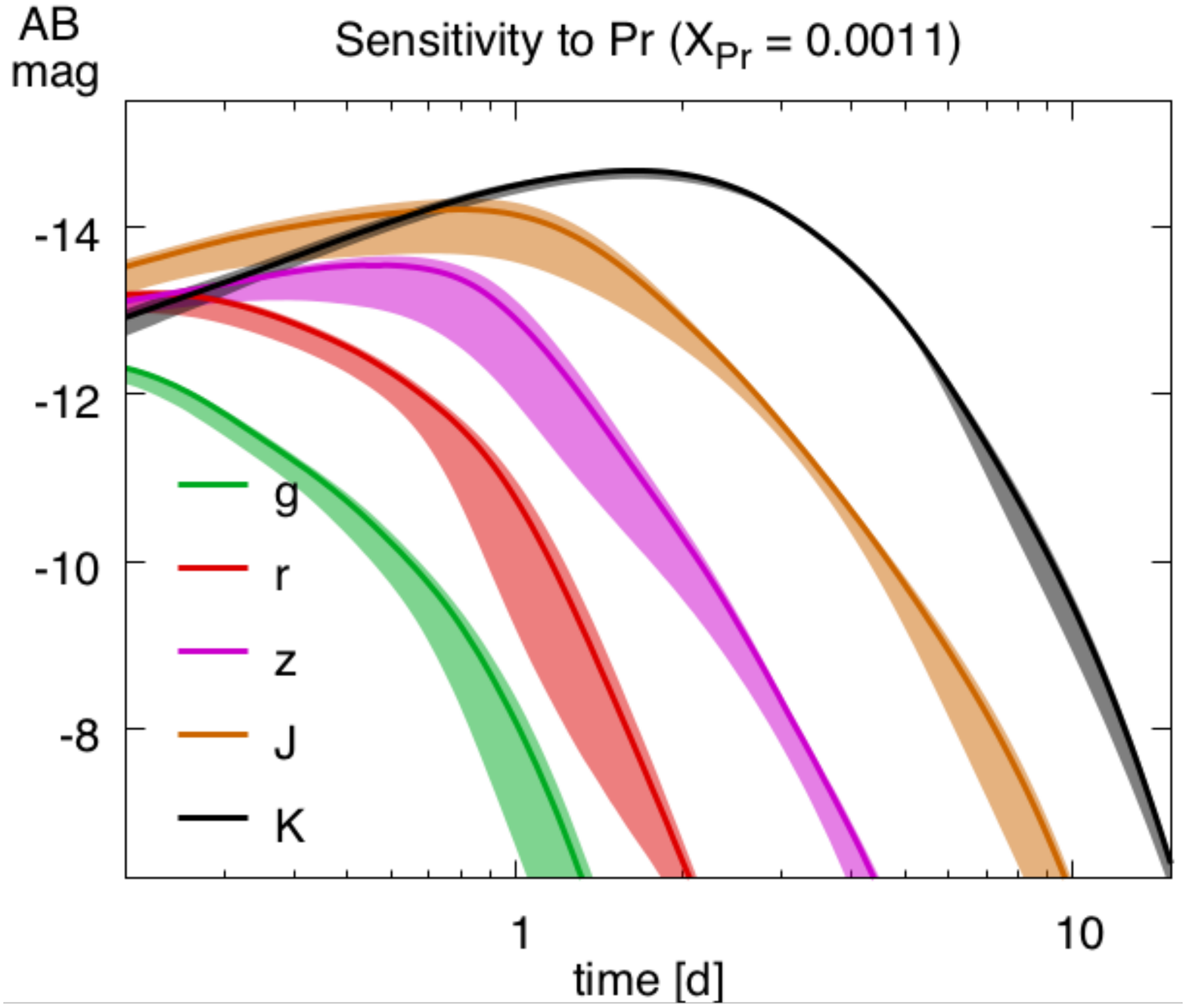}
    \\
    \hspace{-9mm}
    \vspace{-2mm}\includegraphics[width=0.3\textwidth, trim=.2cm 3.5cm .05cm 4.0cm,clip]{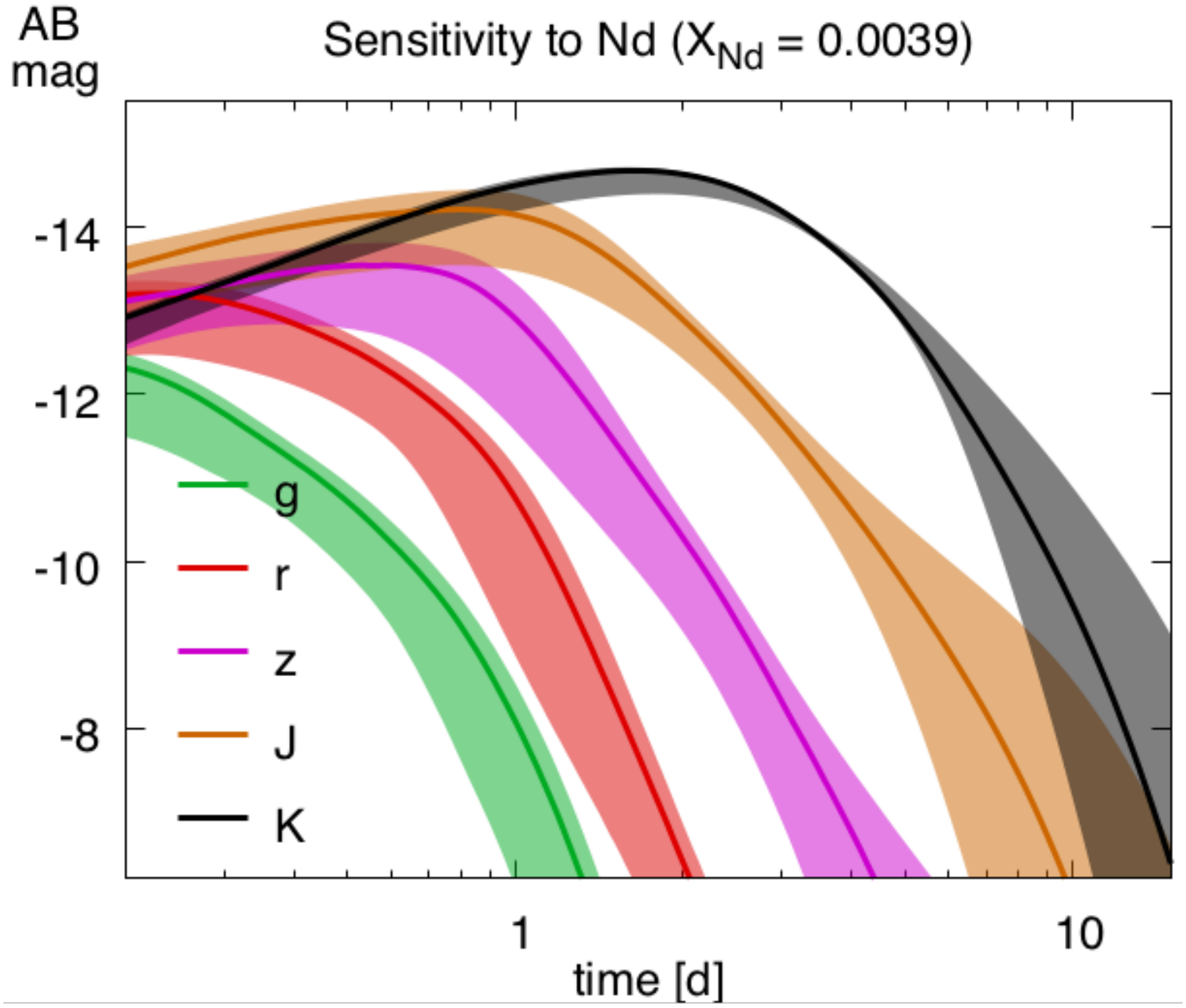} &
    \vspace{-2mm}\includegraphics[width=0.3\textwidth, trim=.2cm 3.5cm .05cm 4.0cm,clip]{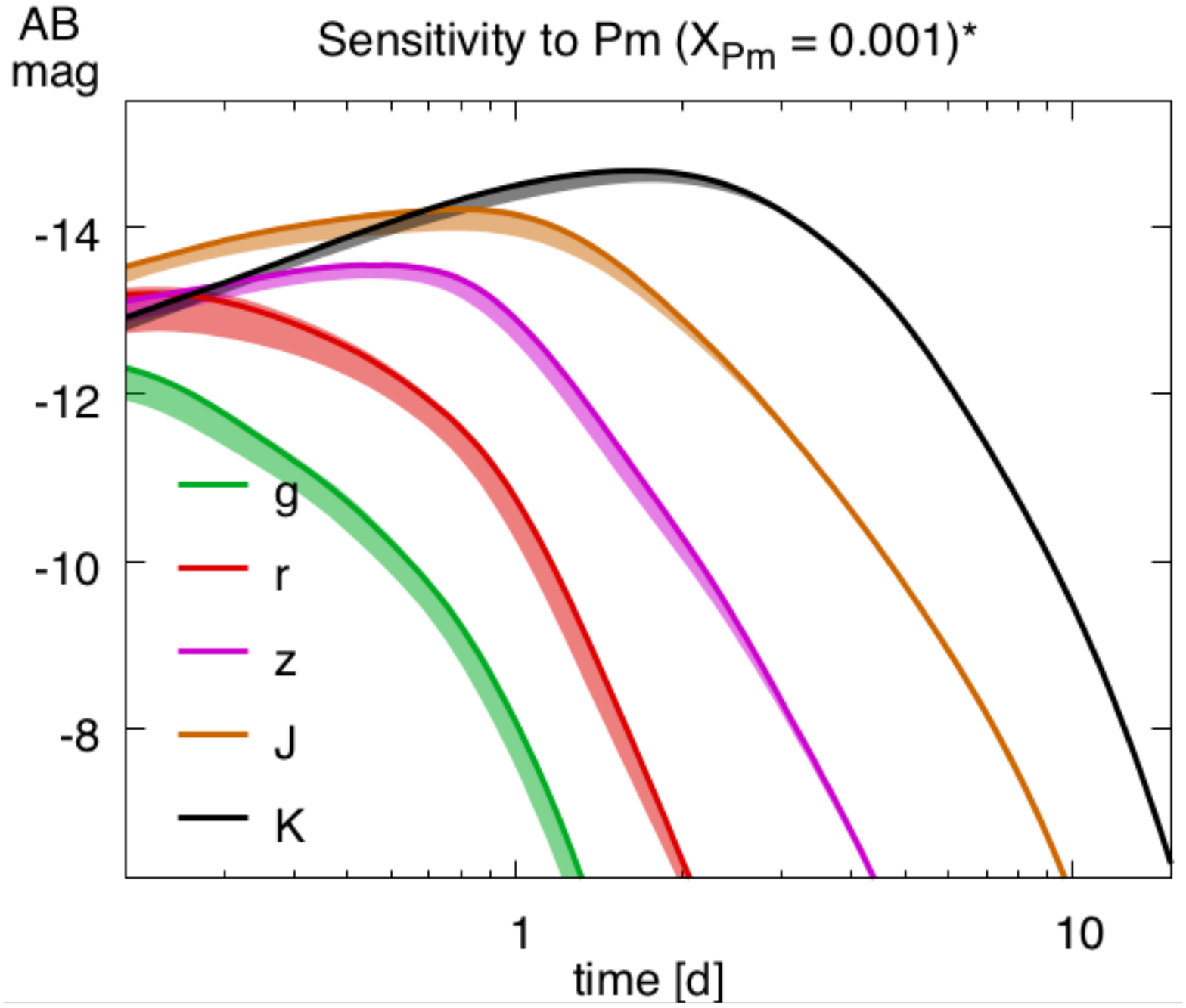} &
    \vspace{-2mm}\includegraphics[width=0.3\textwidth, trim=.2cm 3.5cm .05cm 4.0cm,clip]{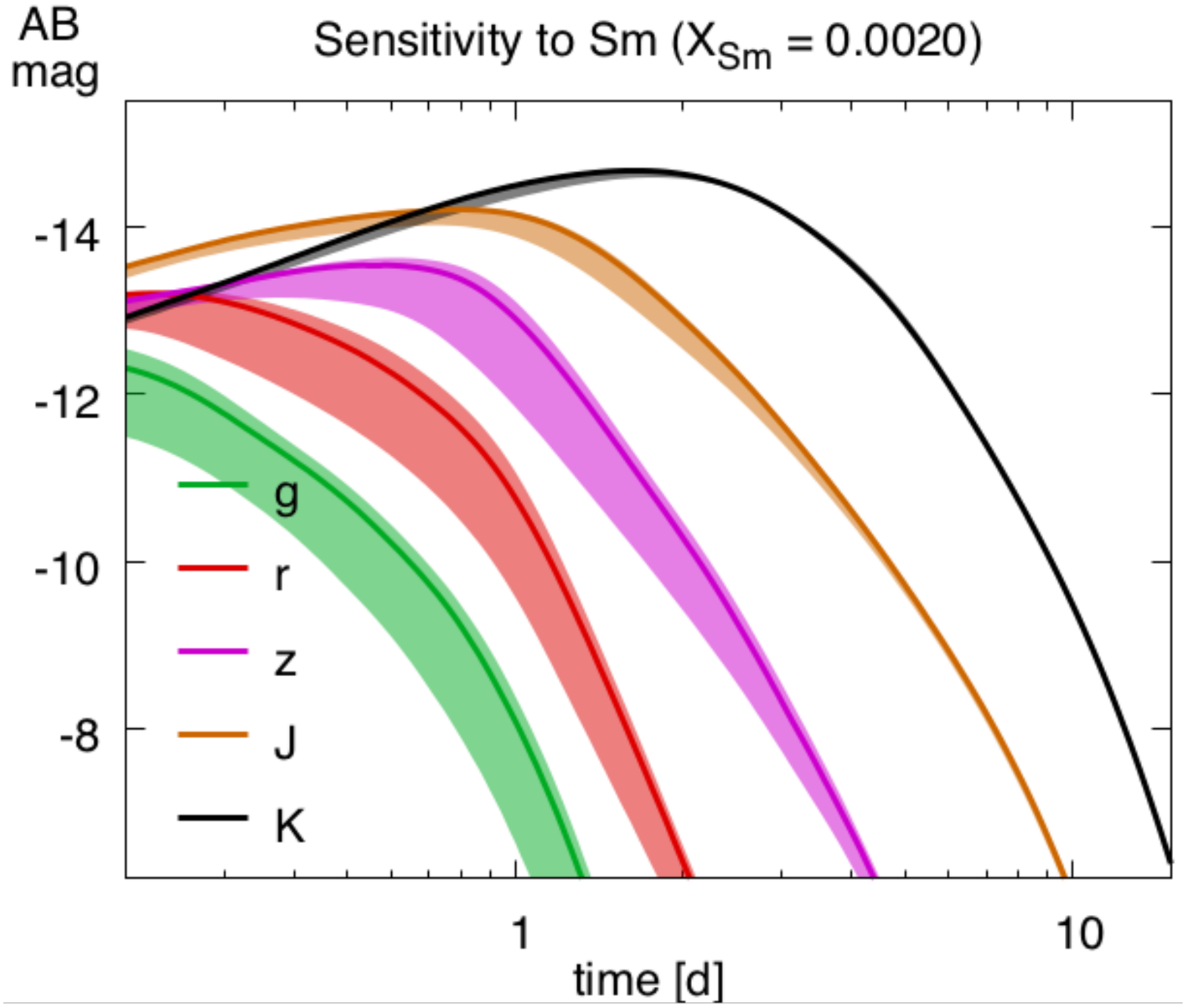}
    \\
    \hspace{-9mm}
    \vspace{-2mm}\includegraphics[width=0.3\textwidth, trim=.2cm 3.5cm .05cm 4.0cm,clip]{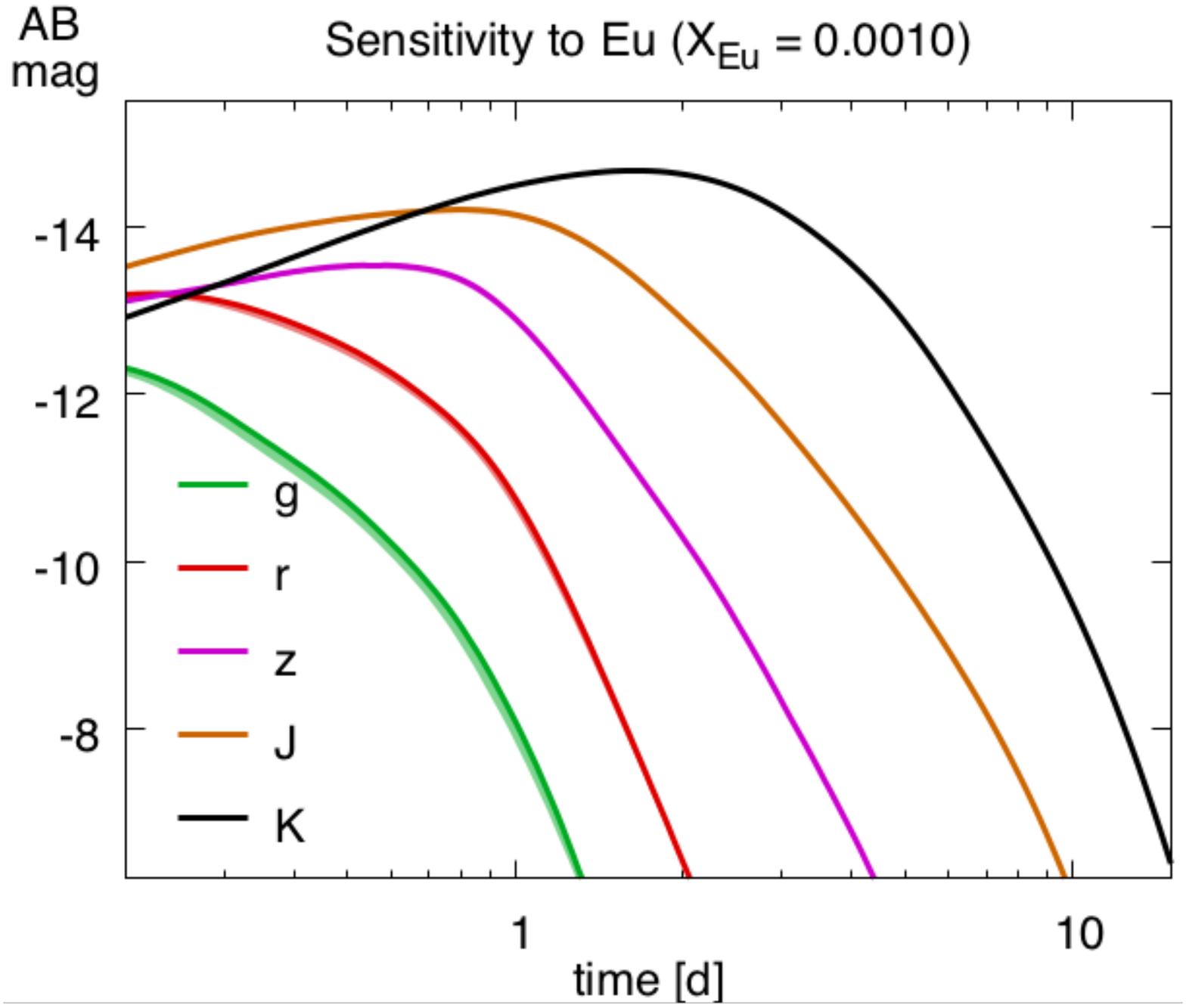} &
    \vspace{-2mm}\includegraphics[width=0.3\textwidth, trim=.2cm 3.5cm .05cm 4.0cm,clip]{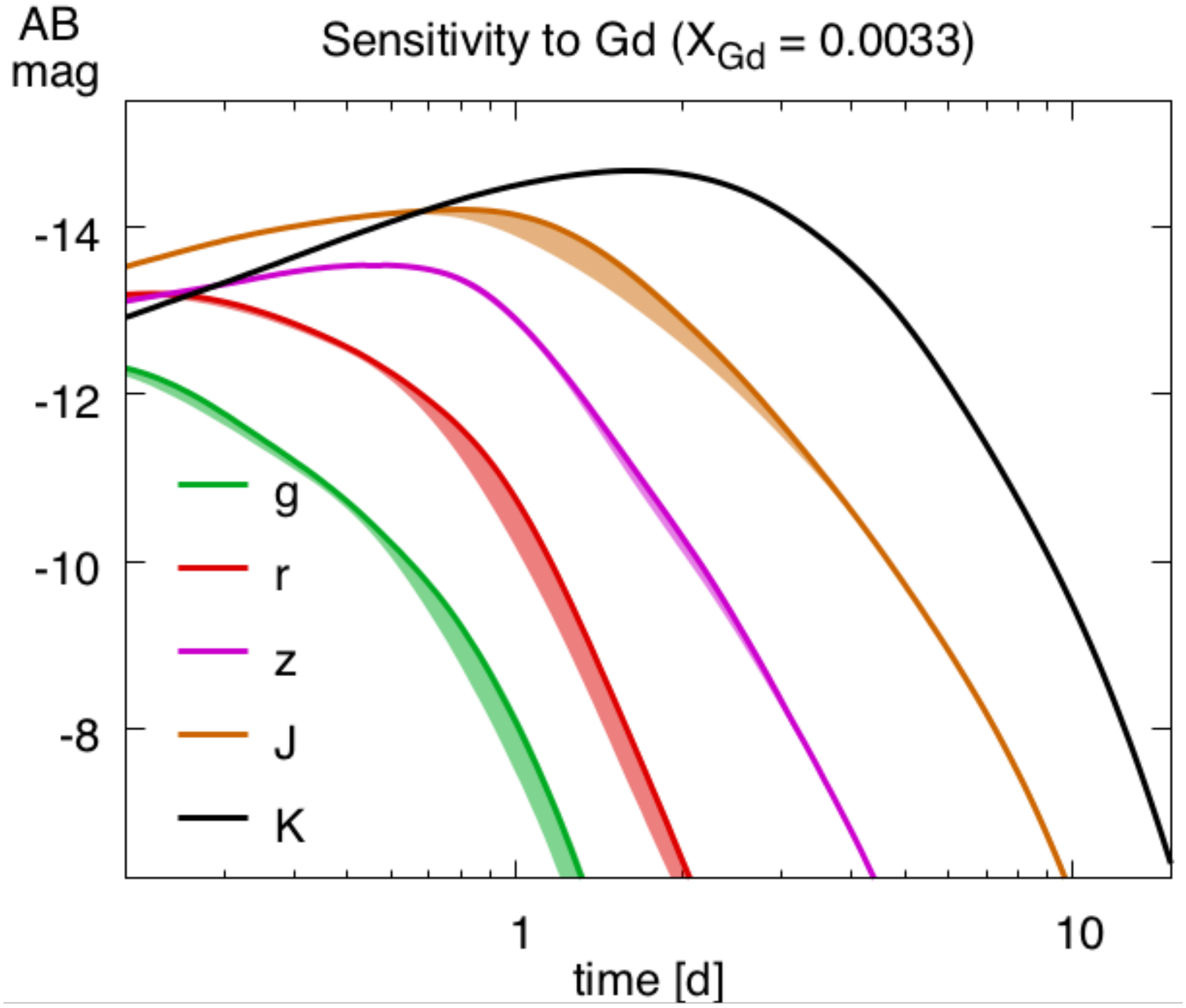} &
    \vspace{-2mm}\includegraphics[width=0.3\textwidth, trim=.2cm 3.5cm .05cm 4.0cm,clip]{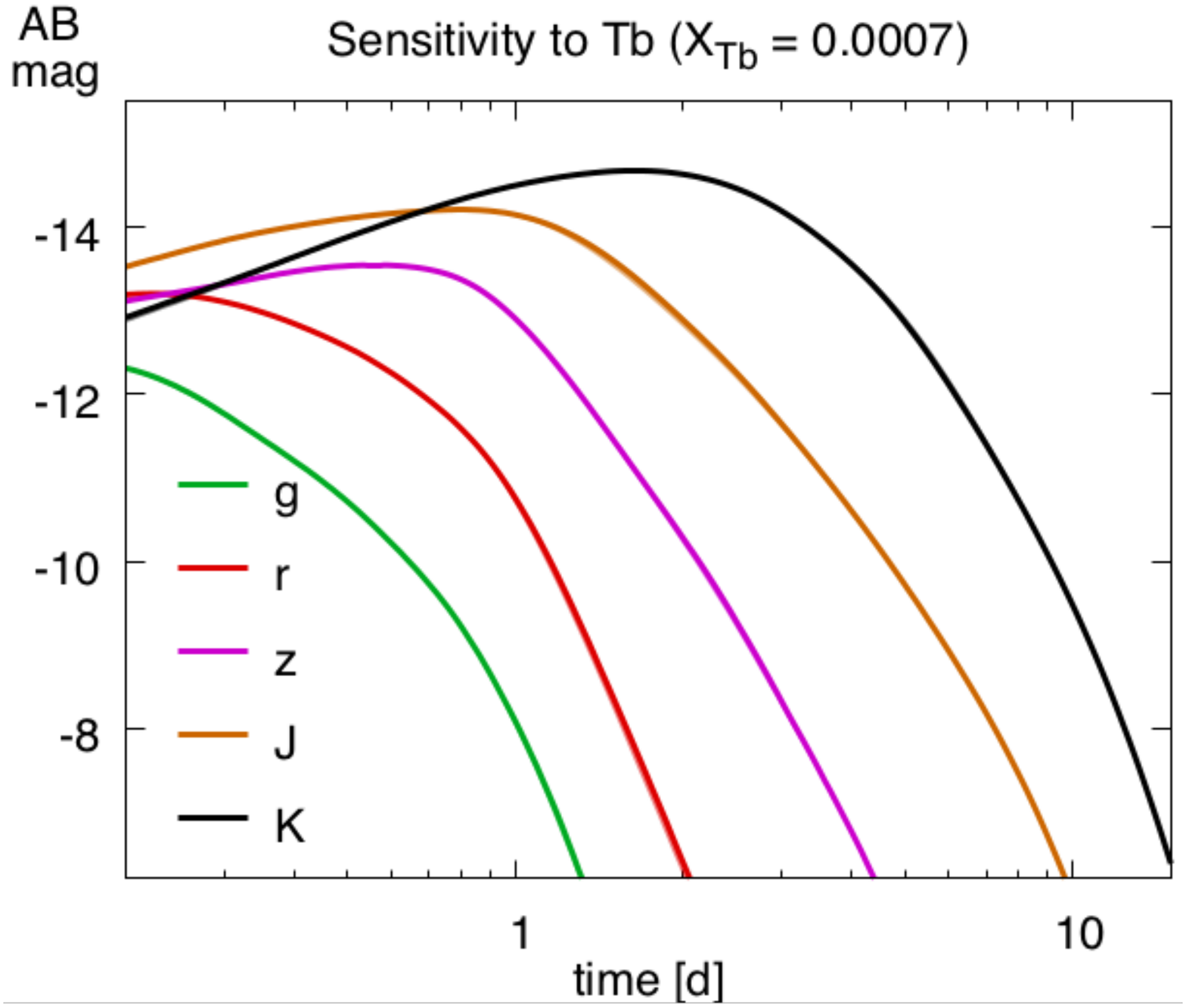}
    \\
    \hspace{-9mm}
    \vspace{-2mm}\includegraphics[width=0.3\textwidth, trim=.2cm 3.5cm .05cm 4.0cm,clip]{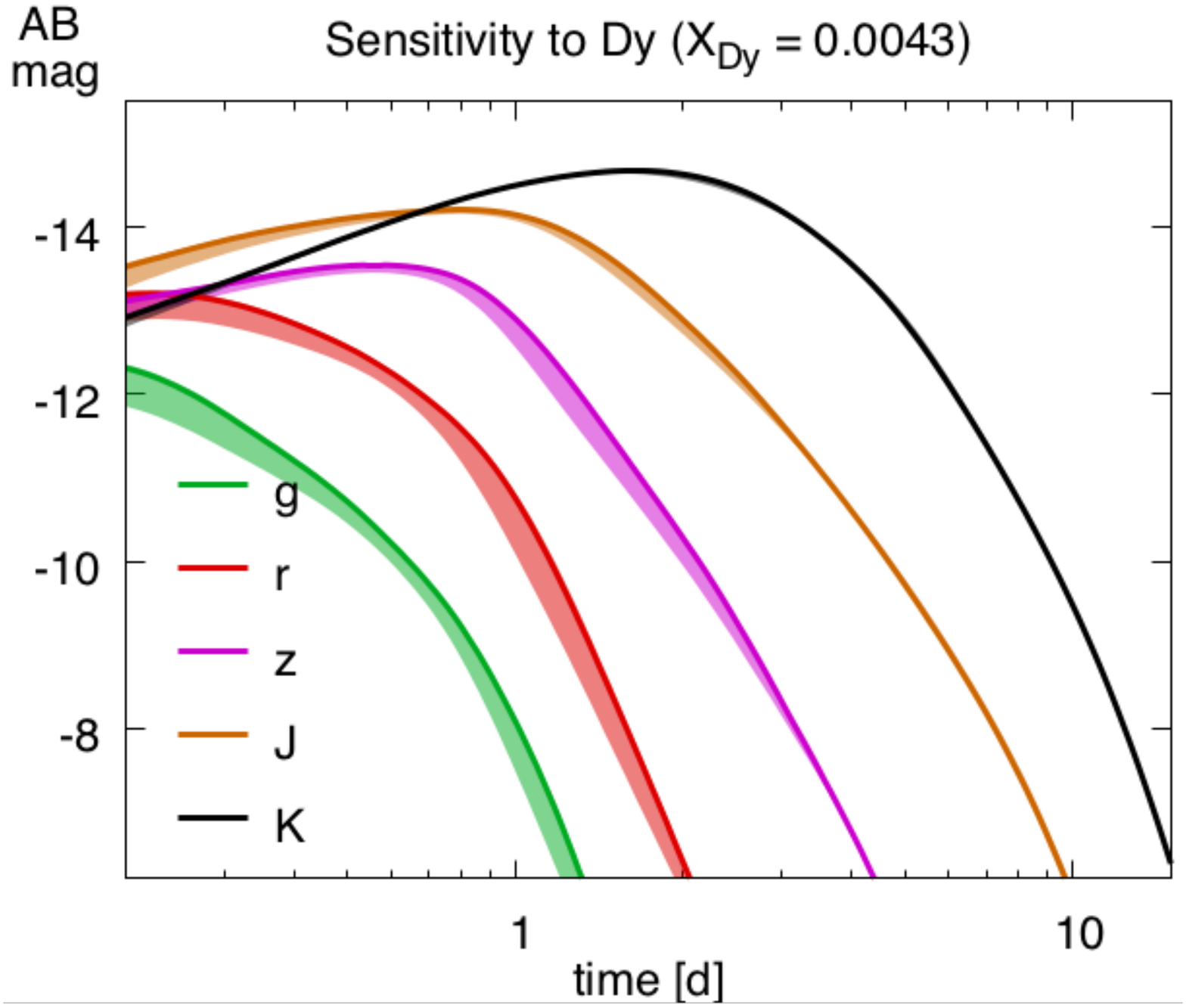} &
    \vspace{-2mm}\includegraphics[width=0.3\textwidth, trim=.2cm 3.5cm .05cm 4.0cm,clip]{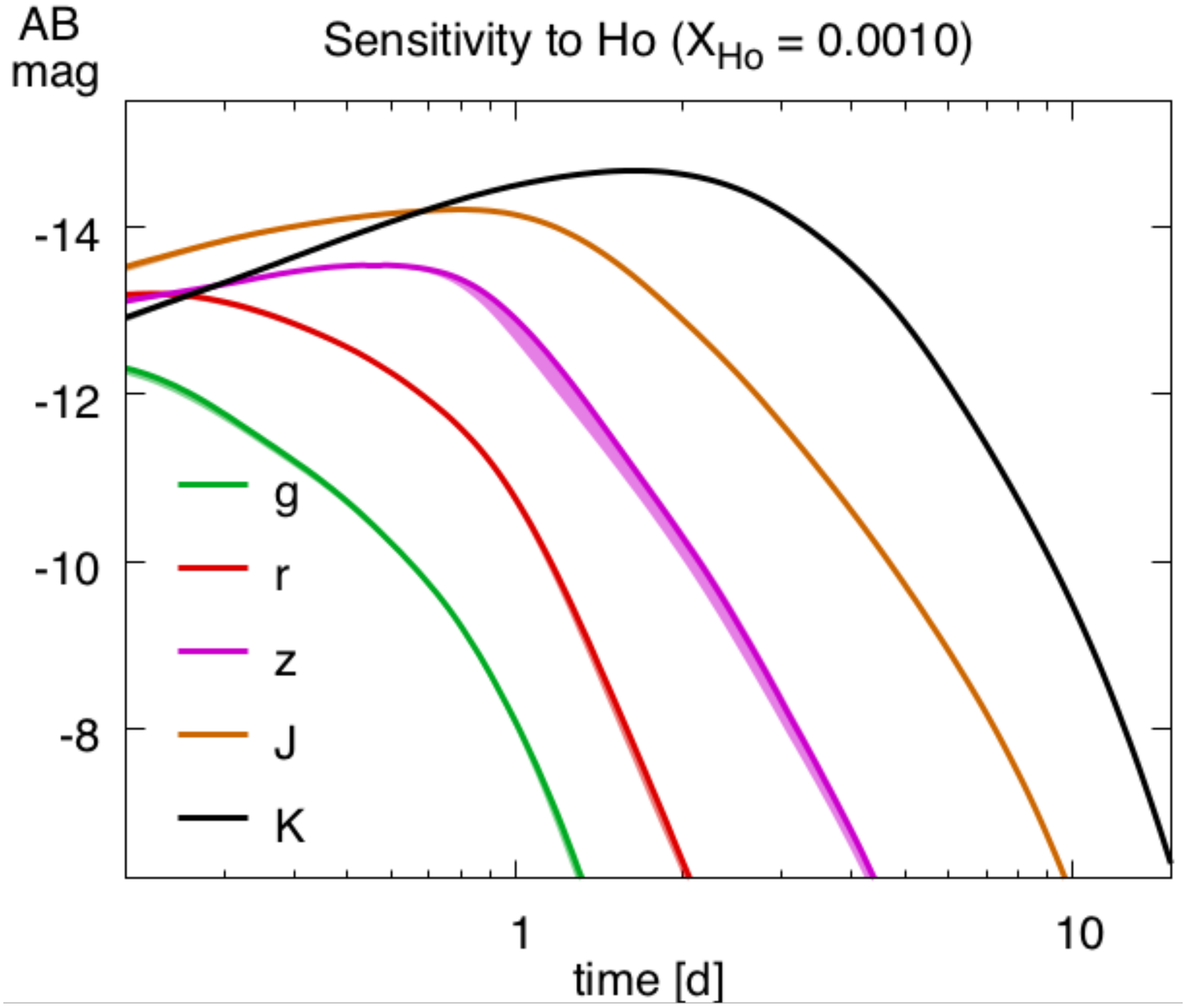} &
    \vspace{-2mm}\includegraphics[width=0.3\textwidth, trim=.2cm 3.5cm .05cm 4.0cm,clip]{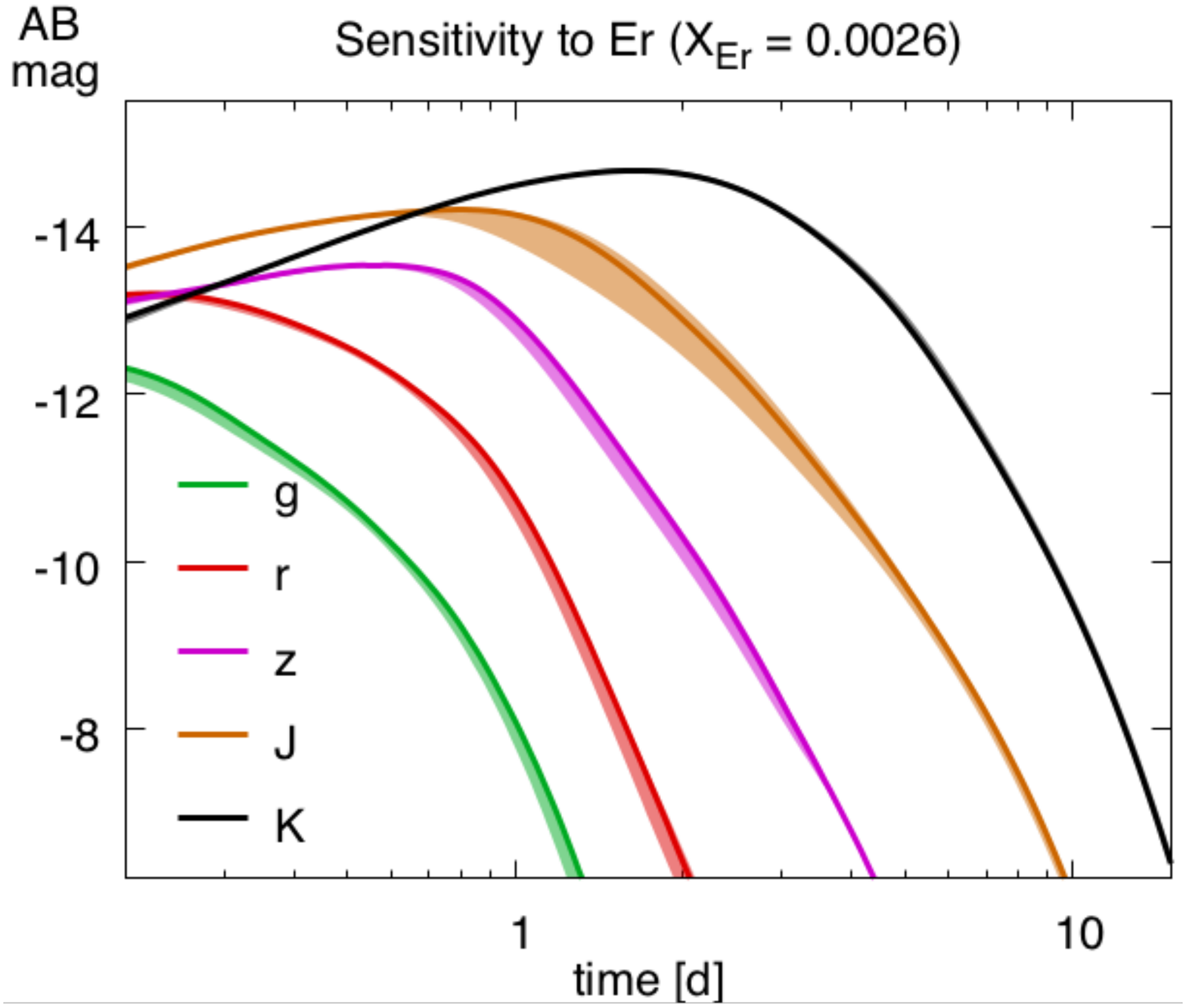}
    \\
    \hspace{-9mm}
    \vspace{-2mm}\includegraphics[width=0.3\textwidth, trim=.2cm 3.5cm .05cm 4.0cm,clip]{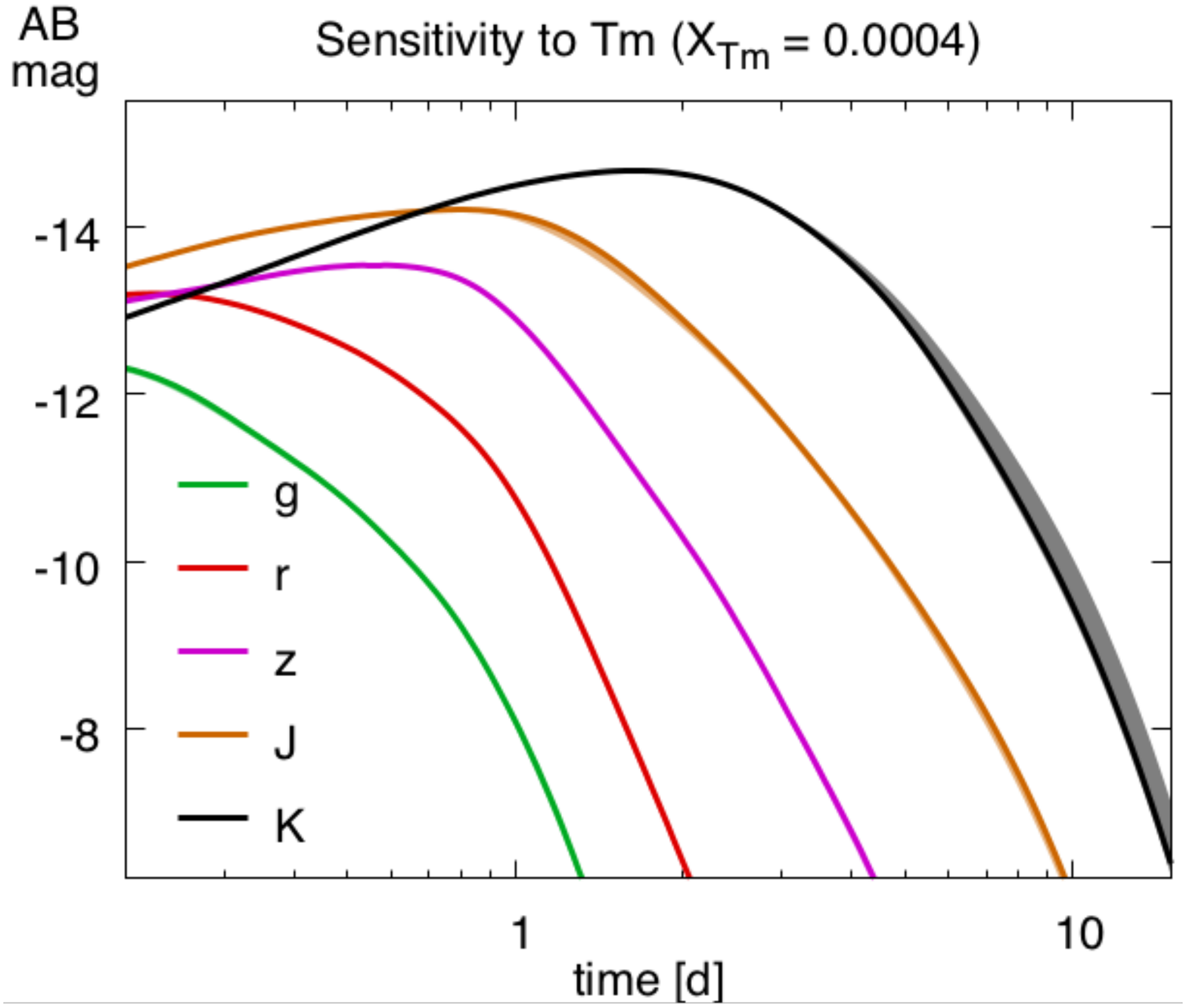} &
    \vspace{-2mm}\includegraphics[width=0.3\textwidth, trim=.2cm 3.5cm .05cm 4.0cm,clip]{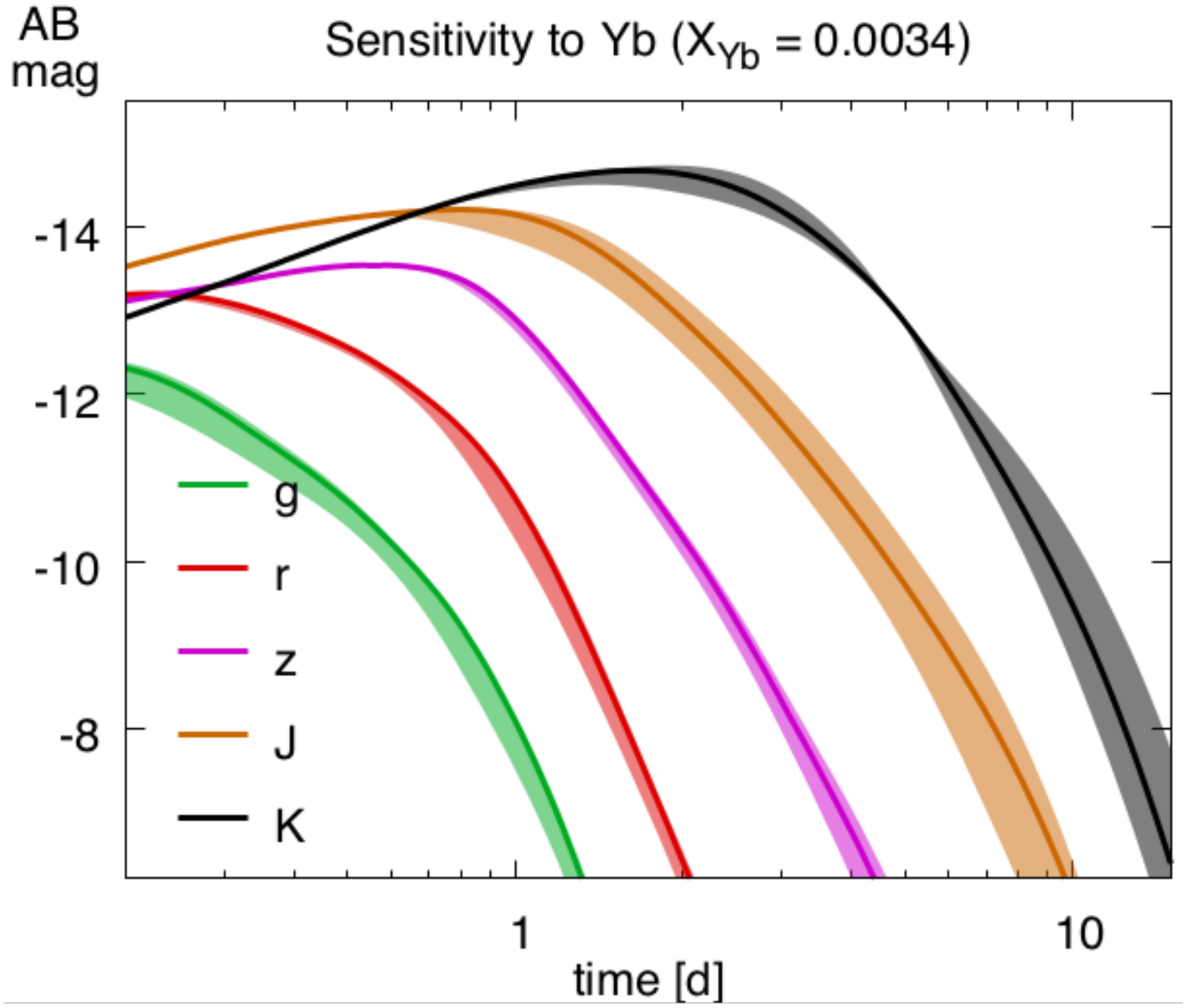} &
    \vspace{-2mm}\includegraphics[width=0.3\textwidth, trim=.2cm 3.5cm .05cm 4.0cm,clip]{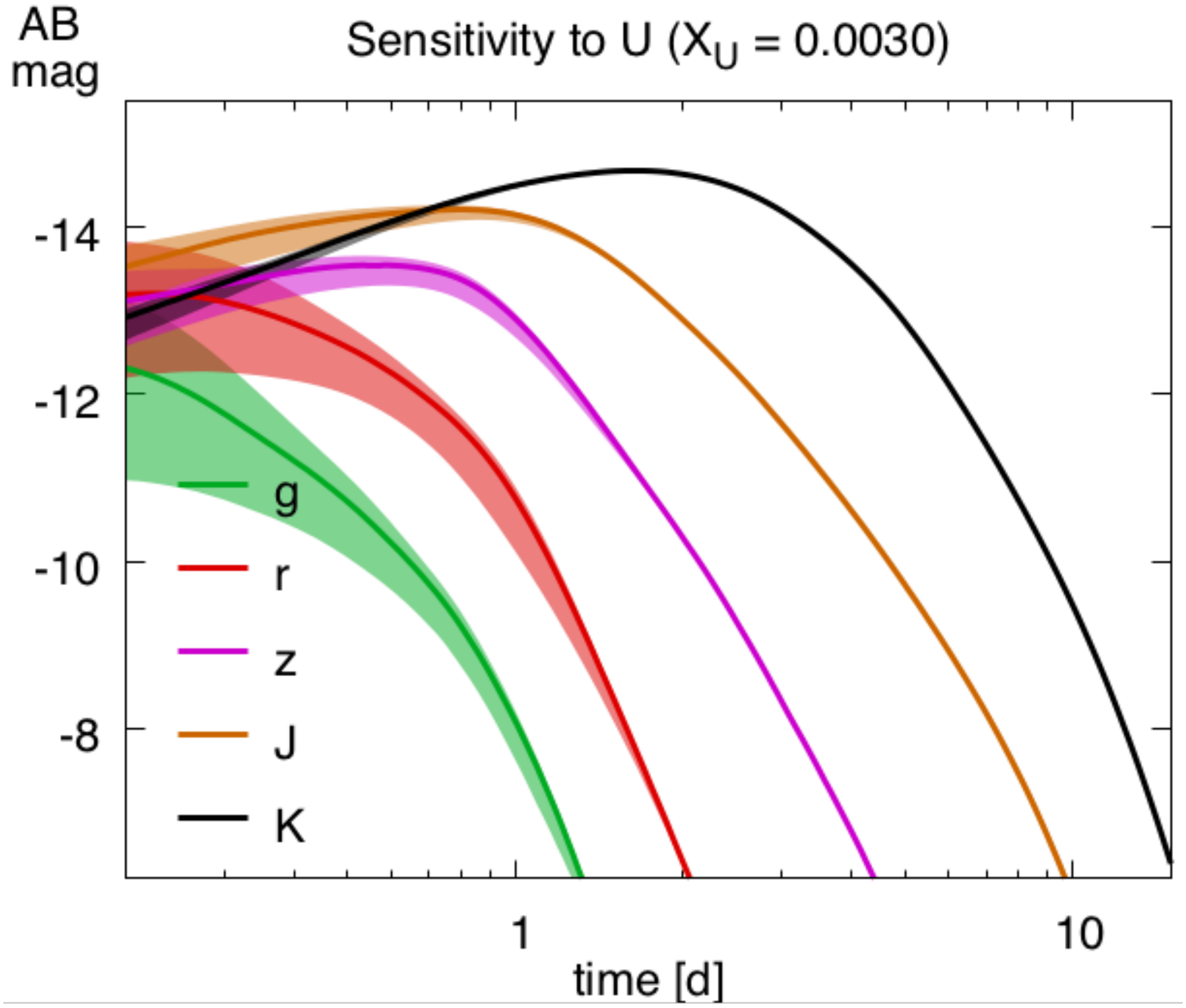}
  \end{tabular}
  \end{center}
  \caption{Impact of presence of individual lanthanides on the broadband magnitudes of a kilonova with solar abundances. The value in brackets indicates the mass fraction of a specific element. The range corresponds to the light curve sensitivity to increasing or decreasing the mass fraction by a factor of 10.}
    \label{fig:solarmag}
\end{figure*}

The focus of this paper is to study the roles of \emph{individual} lanthanide species in the kilonova spectra.  We varied individual lanthanide abundances in our r-process ejecta to search for features specific to each element.  Due to the forest of lines in each lanthanide, we do not expect strong line features, but it is possible that the spectra can be affected if one species is over- or under-produced in the ejecta.  Such spectral effects, in addition to supplying information about relative abundances, may provide constraints on nuclear physics properties in this nuclear mass region \citep{mumpower16,mumpower17,orford18,vilen18}.

Figure~\ref{fig:solarmag} shows the effect on kilonova light curves in the g, r, z, J, and K bands for each lanthanide species.  We vary the mass fraction of each element up and down by one order of magnitude from the standard value (which is listed at the top of each plot).  We include uranium to test the importance of actinides.  For the most part, kilonova light curves do not depend significantly on the abundance of individual elements, even as we raise and lower their mass fractions by an order of magnitude.  Nd is a notable exception.

\begin{figure*}
  \begin{tabular}{cc}
    \includegraphics[width=0.48\textwidth]{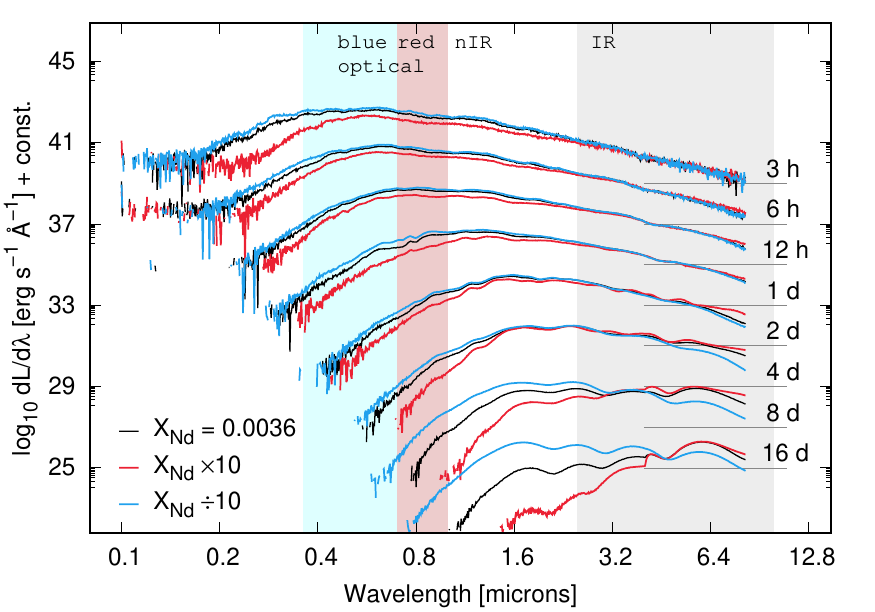} &
    \includegraphics[width=0.48\textwidth]{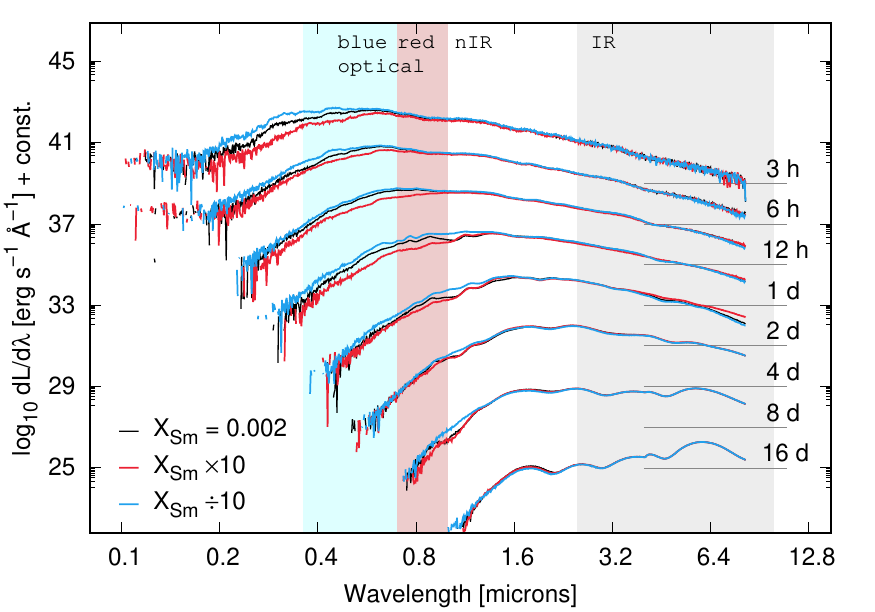}
  \\
    \includegraphics[width=0.48\textwidth]{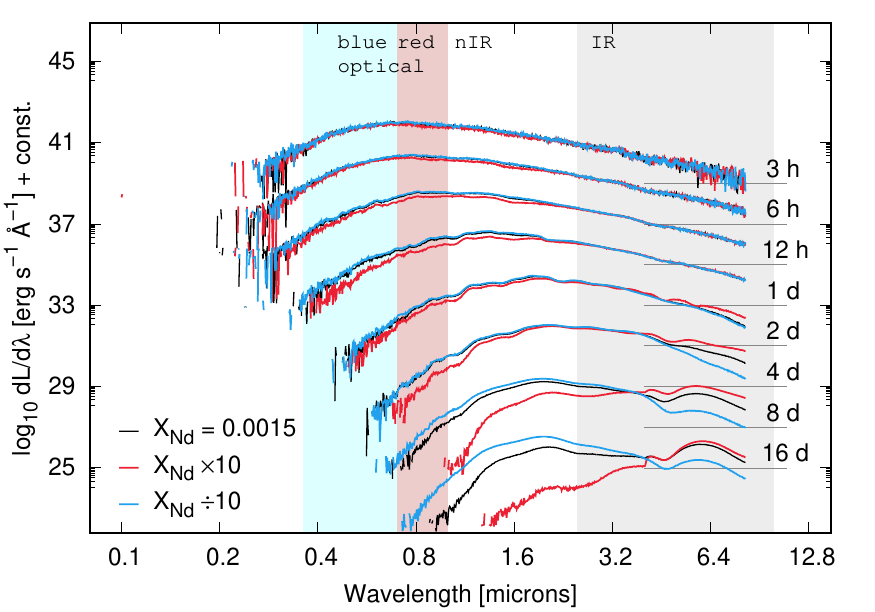} &
    \includegraphics[width=0.48\textwidth]{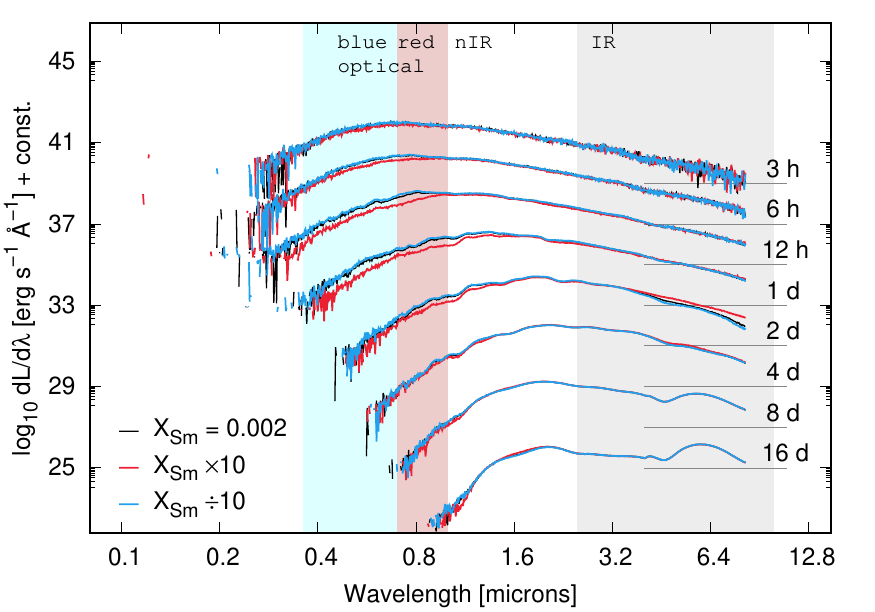}
  \end{tabular}
  \caption{Dependence of kilonova spectra on the mass fractions of Nd (left) and Sm (right) at a range of times.  The mass fractions of these individual species vary up and down by an order of magnitude; the total lanthanide fraction is kept constant. The top panels show variation from solar r-process residuals, and the bottom panels show variation from the main r-process (second to third peak) abundance pattern. As with most lanthanides, varying Sm does not dramatically alter the kilonova spectra.  However, Nd can dramatically shift the peak of the emission at late times (from $\sim 1\,\micron$ to $>5\,\micron$)}.
  \label{fig:nrspec}
\end{figure*}

To better understand this dependence, we study the effects of the Nd abundance in more detail.  Figure~\ref{fig:nrspec} shows the spectra from our neutron rich simulations when we vary Nd and Sm up and down by an order of magnitude while keeping the total lanthanide fraction constant.  Most lanthanides behave like Sm in that the spectra between $1-10\,\micron$ do not depend on an individual element's abundance; the thick forest of each species's lines contributes nearly equally across this band.  The exception to this basic trend is Nd, which pushes the late-time peak in the emission from $\sim 1\,\micron$ to $> 5\,\micron$.

\begin{figure*}
    \centering
    \includegraphics[width=0.48\textwidth]{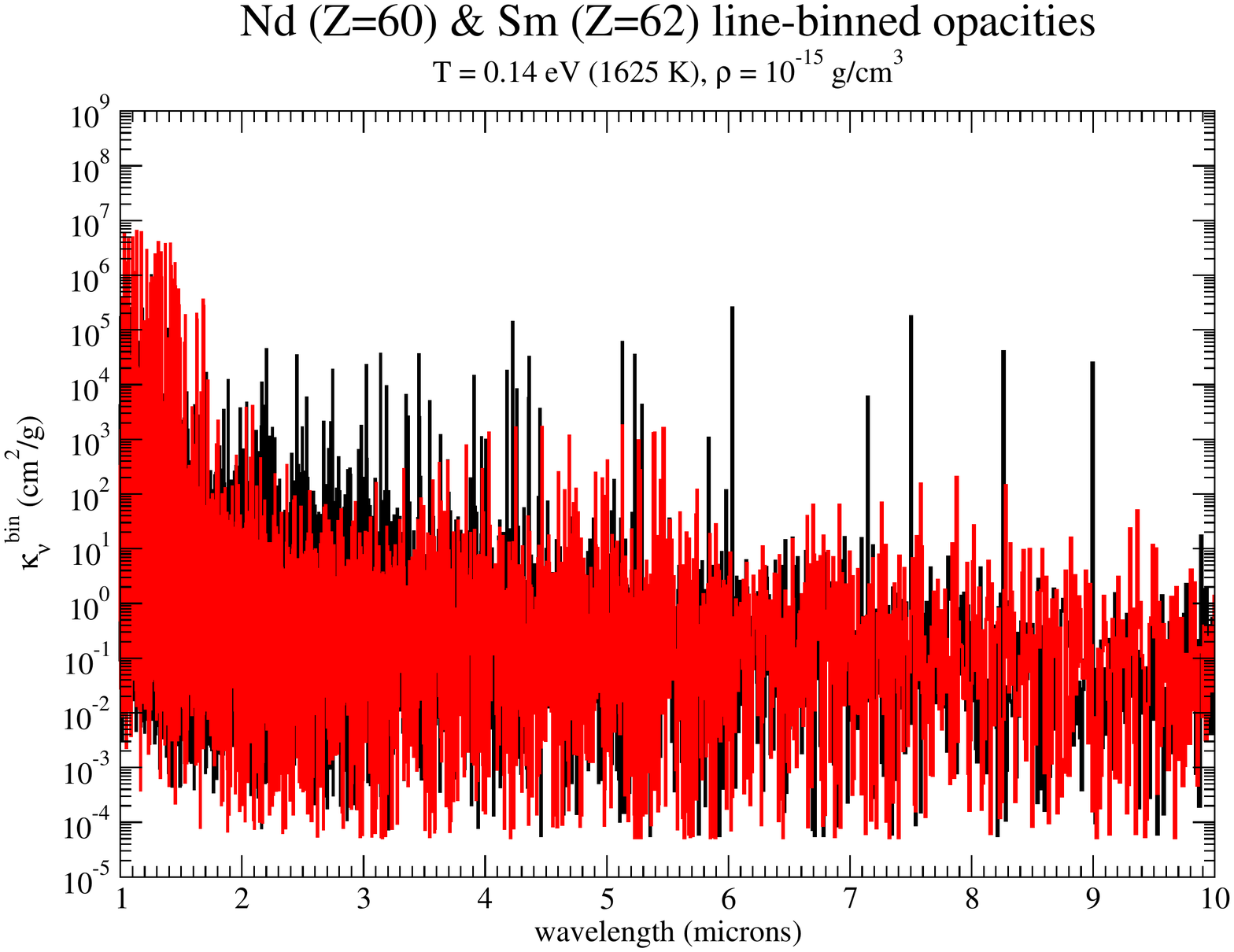}
    \includegraphics[width=0.48\textwidth]{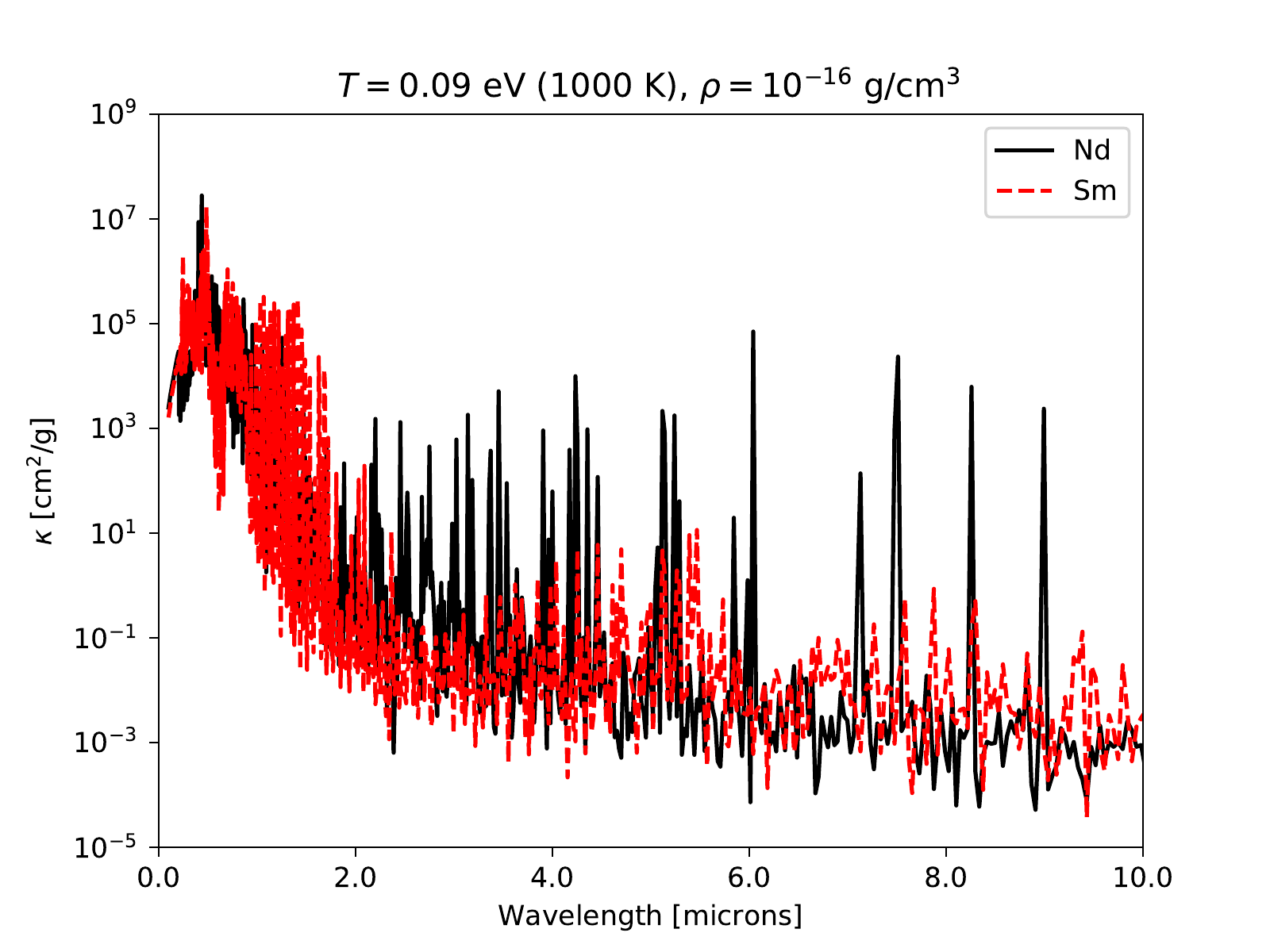}
    \caption{Opacity of Nd (black) versus Sm (red) in the wavelength range from $1-10\,\micron$.  Between $1.7-4\,\micron$, the Nd opacity is has roughly an order of magnitude greater opacity.  The left panel shows the full opacity from ~\cite{fontes19}, and the right panel shows this opacity once it is binned at the energy resolution of the simulations.}
    \label{fig:ndvssm}
\end{figure*}

The leverage of Nd on the spectra arises from its high opacity in the near infrared.  Figure~\ref{fig:ndvssm} shows the opacities of Nd and Sm between 1 and 10$\,\micron$.  Between 1.7 and 4$\,\micron$, the Nd opacity is nearly an order of magnitude greater.
This outsize opacity is a consequence of the energy level diagram for electrically neutral Nd, which is the dominant ionization stage under the relevant conditions.  The partially filled $4f^4$ electron ground configuration and the corresponding excited states result in atomic transitions that produce absorption lines in the 1.7 to 4.5$\,\micron$ range; these lines are strong compared to other lanthanides, as illustrated in Figures~6 and 8 of \citet{fontes19}.  In fact, those figures display significantly fewer absorption features from elements in the middle of the lanthanide series, such as Eu and Gd, despite the more complex energy level diagrams associated with their half-filled $4f^7$ ground state configurations.  The distinct absorption features of Nd---coupled with its relatively high abundance (Nd is an even element)---cause it to dominate the opacity in this region, and adjusting its abundance up/down results in a commensurate increase/decrease in the total opacity.  In contrast, increasing the fraction of Sm (an odd element) by a factor of 10 would not even double the opacity.  We note that the high Nd opacity may also explain the redward shift in the calculations by~\cite{fontes17}.  That work used Nd as a surrogate for many of the lanthanides, effectively increasing the opacity in this wavelength band by a factor of a few beyond what we would expect from a full lanthanide distribution.

\begin{figure}
  \centering
  \includegraphics[width=\columnwidth]{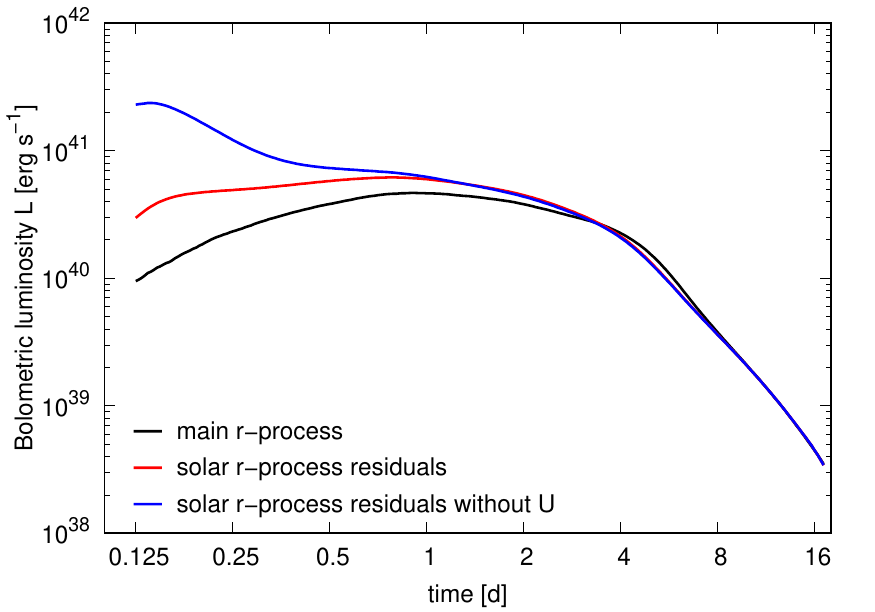}\\
  \includegraphics[width=\columnwidth]{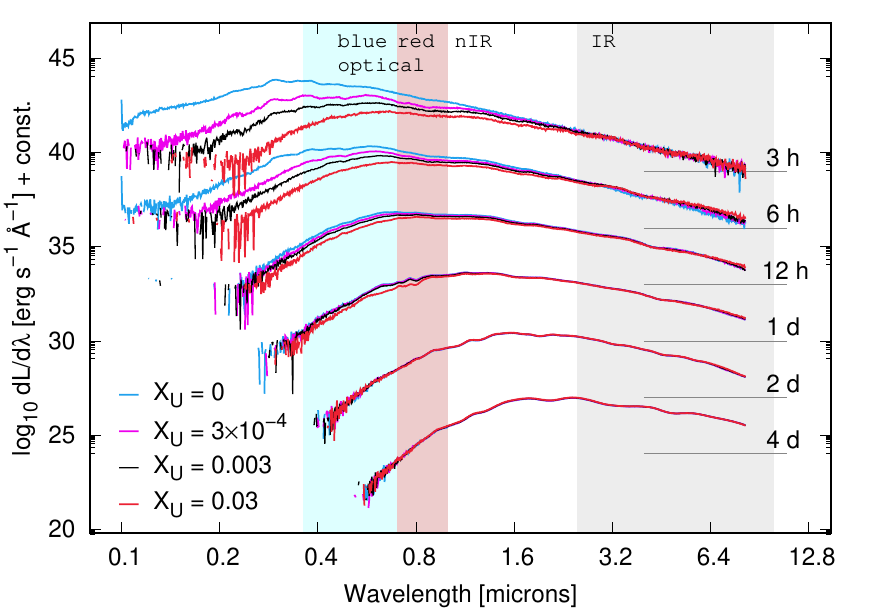}
  \caption{Impact of uranium on bolometric luminosity and spectra.  Black: synthetic main r-process composition.  Red: solar r-process residuals.  Blue: solar r-process composition with the uranium removed.}
    \label{fig:uraniumspec}
\end{figure}

In addition to the strong dependence on the Nd mass fraction, the early-time light curves ($<1\,{\rm d}$) are heavily influenced by the mass fraction of uranium (seen previously in Figure~\ref{fig:solarmag}).  We examined this effect by simulating the light curves with and without uranium.  We found that the presence of uranium dramatically lowers the early-time luminosity and suppresses the spectra in the optical and ultraviolet wavelengths.  Figure~\ref{fig:uraniumspec} displays these findings.

\begin{figure}
  \centering
  \includegraphics[width=\columnwidth]{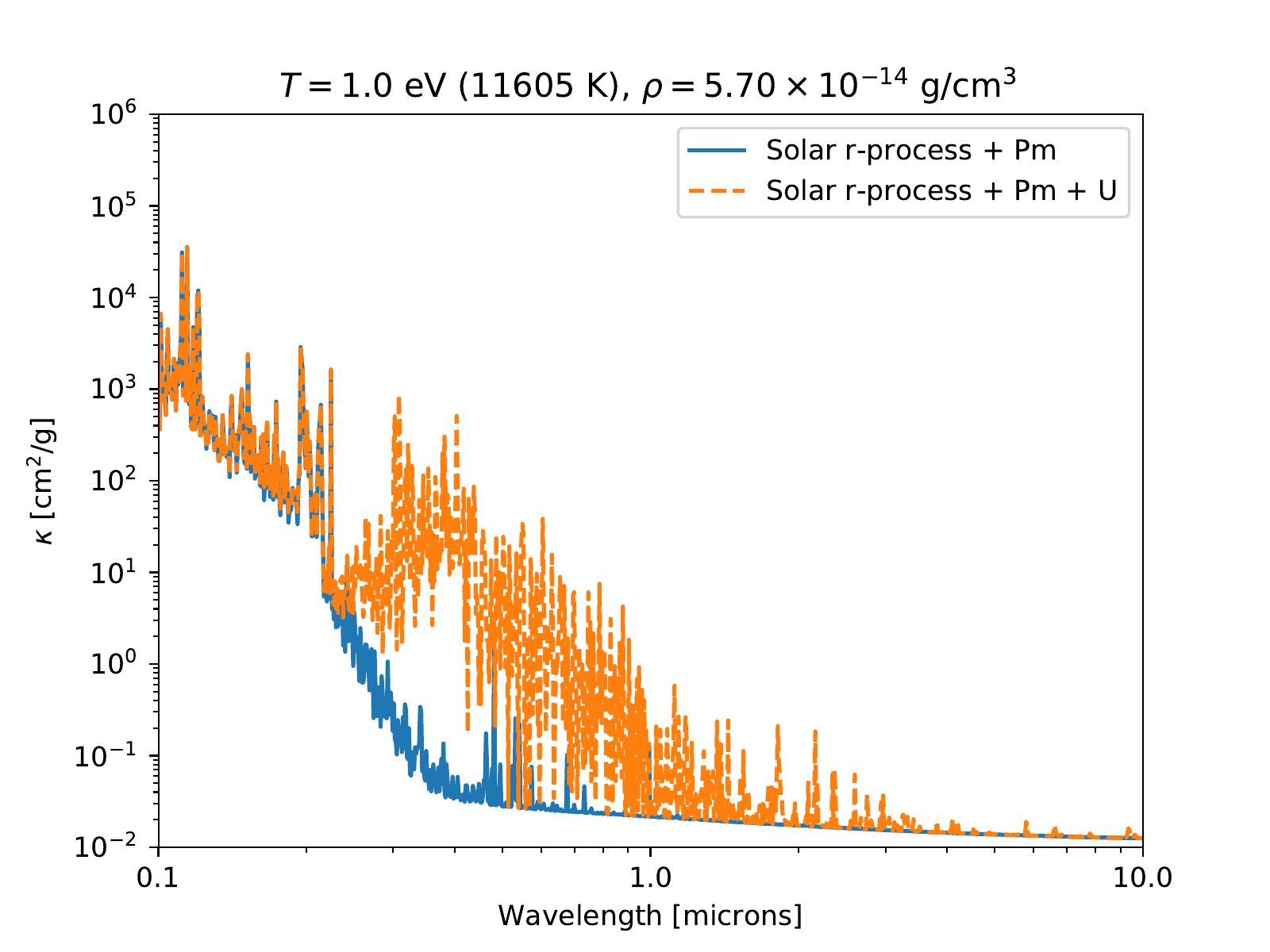}\\
  \includegraphics[width=\columnwidth]{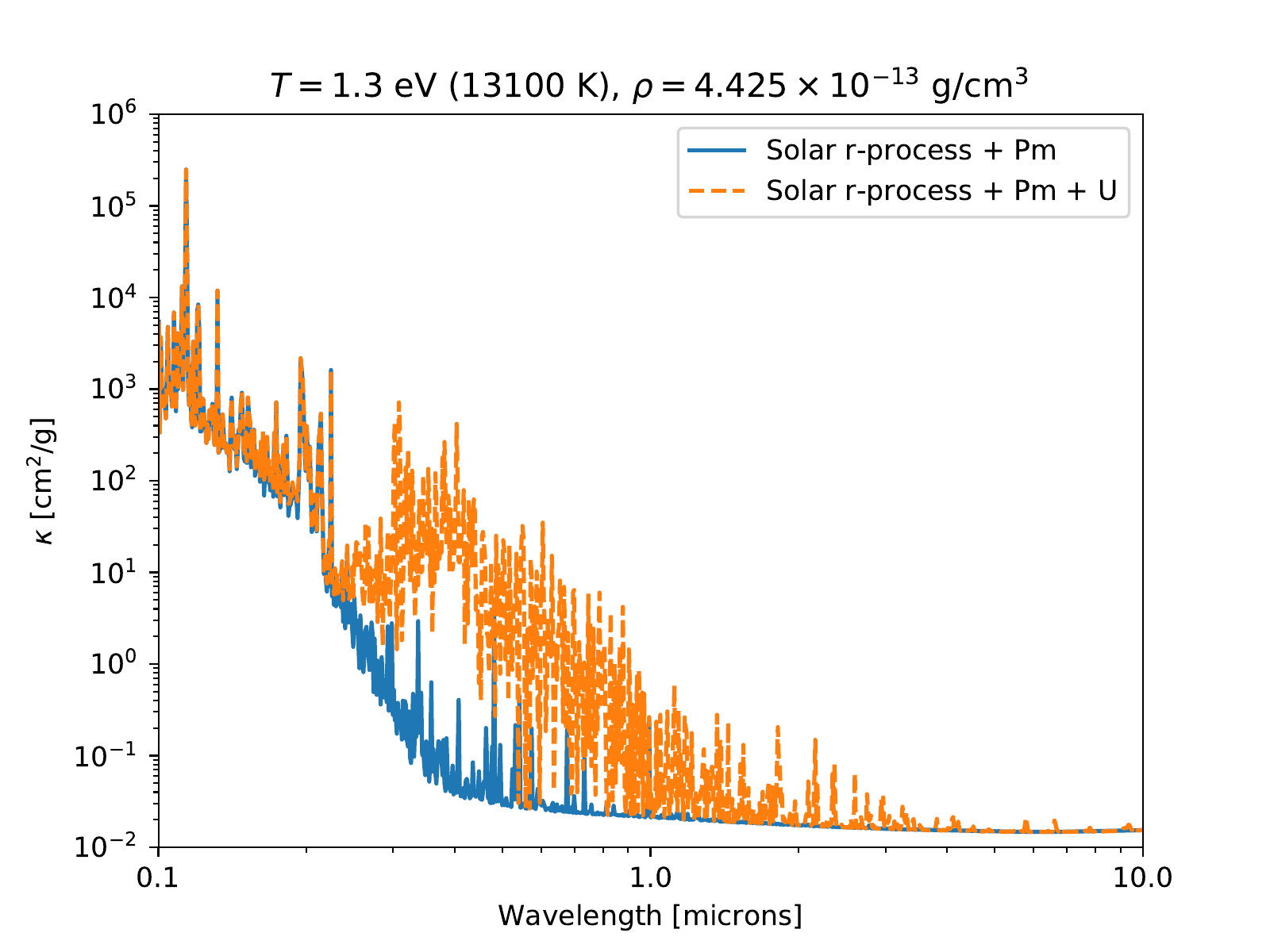}
  \caption{Opacity with and without uranium for two different temperature-density pairs.  Top: $T=1.0 {\rm \, eV}$, $\rho=5.70\times10^{-14} {\rm \, g \, cm^{-3}}$.  Bottom: $T=1.3 {\rm \, eV}$, $\rho=4.425\times10^{-13} {\rm \, g \, cm^{-3}}$.  Above $T\sim 1.0 {\rm \, eV}$, the lines in uranium dramatically dominate the opacity from $0.2-1\, \micron$, obscuring emission in this wavelength range.}
  \label{fig:uraniumopac}
\end{figure}

In order to illuminate these effects, we plot the ejecta opacities with and without uranium at temperatures and densities relevant for early times (Figure~\ref{fig:uraniumopac}).  Above a temperature of $T\sim 1.0$\, eV, the absorption lines in uranium dominate the opacity from $0.2-1\,\micron$, obscuring emission in this wavelength range.  Similar to Nd, this increase in opacity is related to the energy level diagram of the uranium ion stage that is dominant in the conditions of interest.  Note that U is a homologue of Nd, which means they reside in the same column of the periodic table, have similar electron structure, and hence are expected to display similar---though not identical---patterns in their absorption features.  Because U is a Nd analogue, we need to be careful about using it as a surrogate for all actinides, since Nd behaves differently from other lanthanides.

\subsection{Calculated Abundances}

\begin{figure*}[!htp]
  \begin{center}
  \begin{tabular}{cccc}
    \hspace{-9mm}
    \vspace{-2mm}\includegraphics[width=0.3\textwidth, trim=.2cm 3.5cm .05cm 4.5cm,clip]{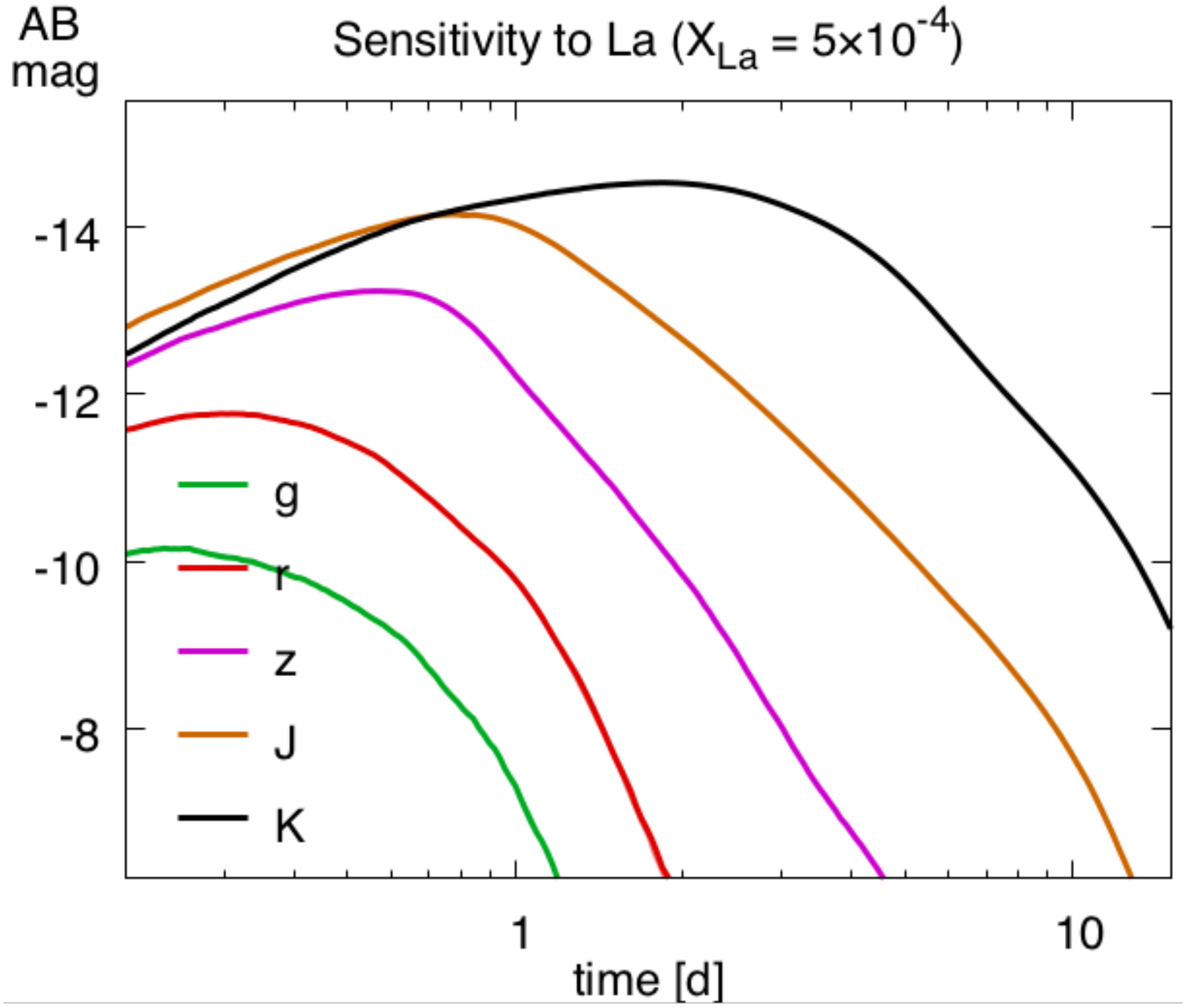} &
    \vspace{-2mm}\includegraphics[width=0.3\textwidth, trim=.2cm 3.5cm .05cm 4.5cm,clip]{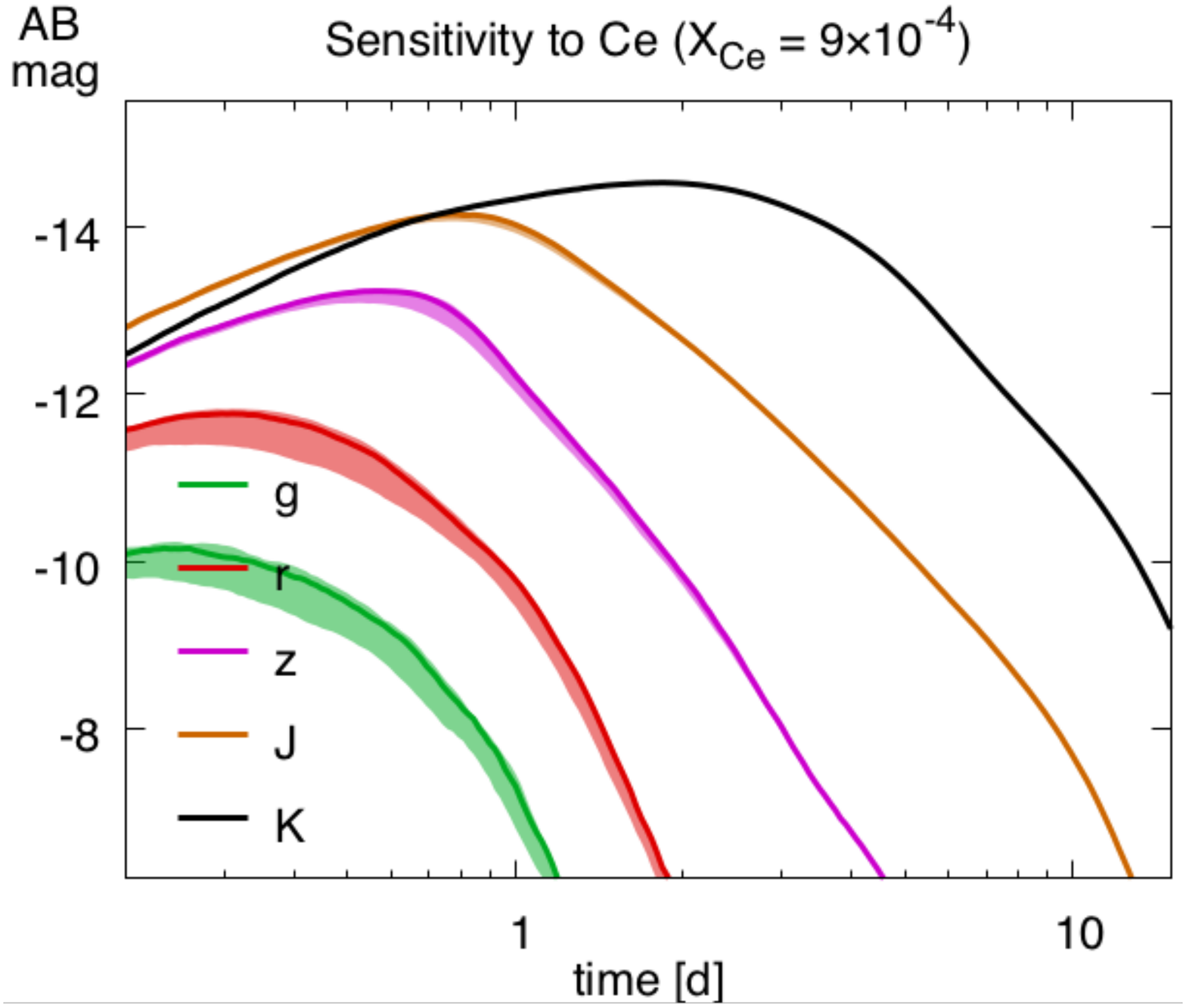} &
    \vspace{-2mm}\includegraphics[width=0.3\textwidth, trim=.2cm 3.5cm .05cm 4.5cm,clip]{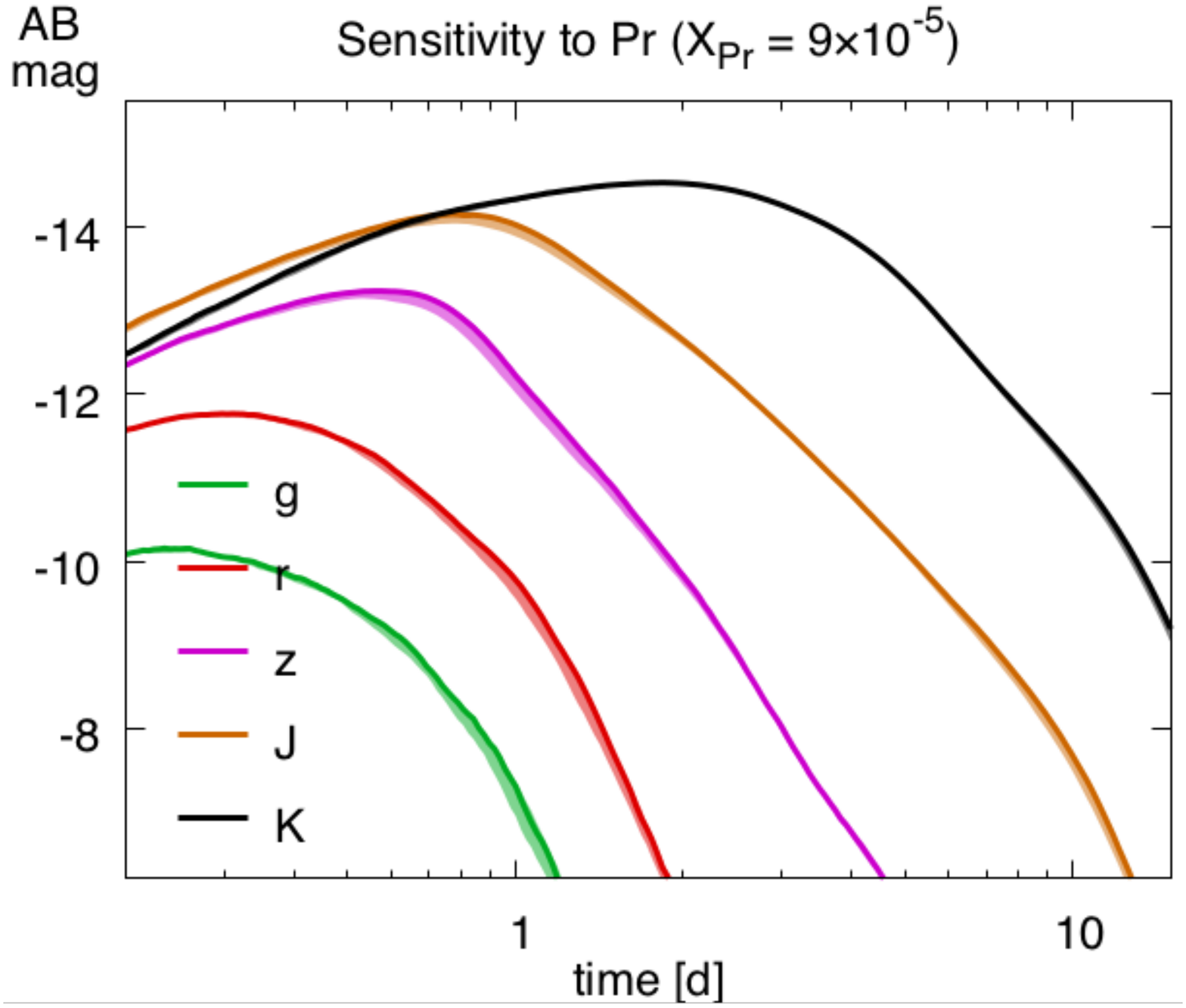}
    \\
    \hspace{-9mm}
    \vspace{-2mm}\includegraphics[width=0.3\textwidth, trim=.2cm 3.5cm .05cm 4.0cm,clip]{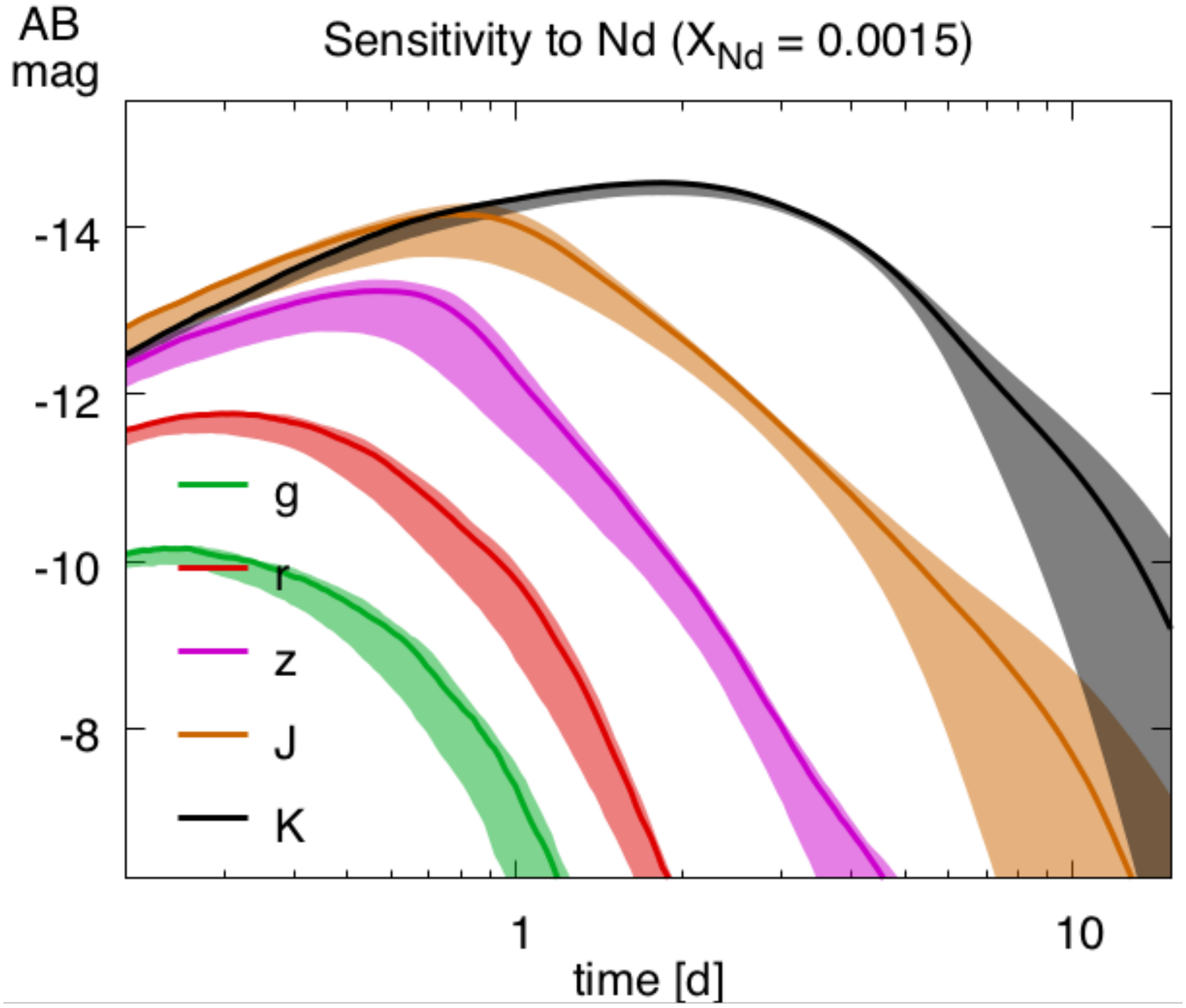} &
    \vspace{-2mm}\includegraphics[width=0.3\textwidth, trim=.2cm 3.5cm .05cm 4.0cm,clip]{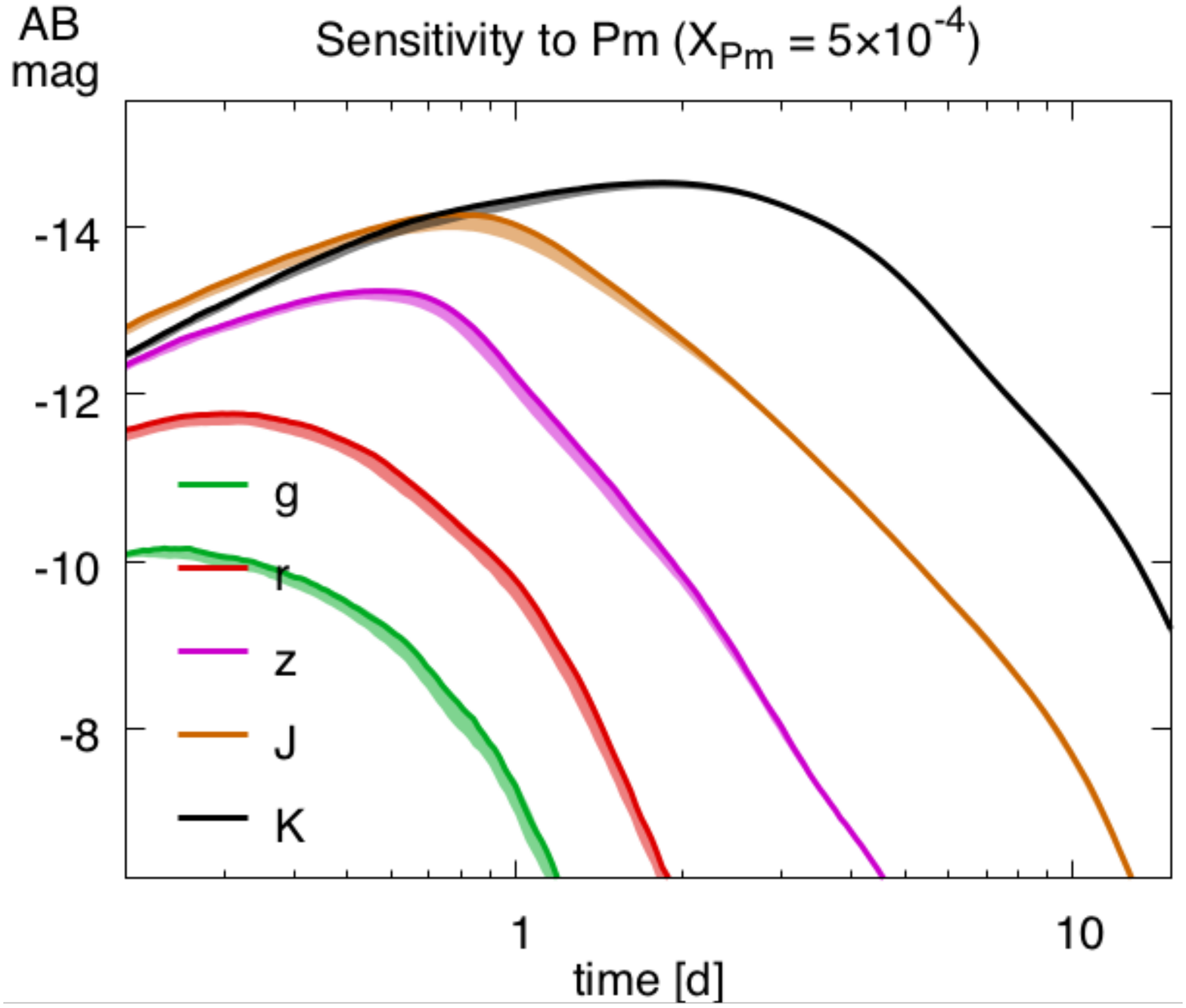} &
    \vspace{-2mm}\includegraphics[width=0.3\textwidth, trim=.2cm 3.5cm .05cm 4.0cm,clip]{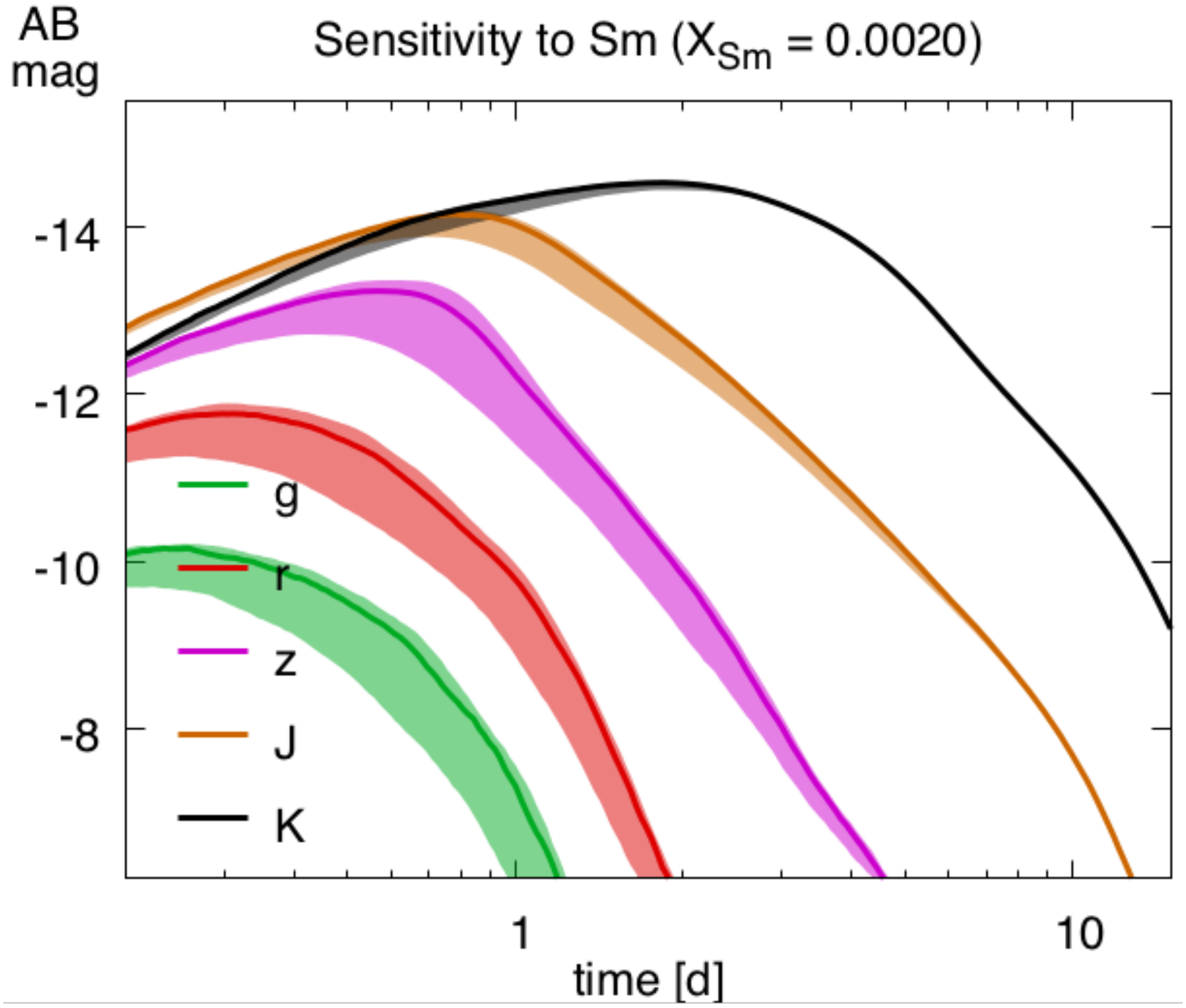}
    \\
    \hspace{-9mm}
    \vspace{-2mm}\includegraphics[width=0.3\textwidth, trim=.2cm 3.5cm .05cm 4.0cm,clip]{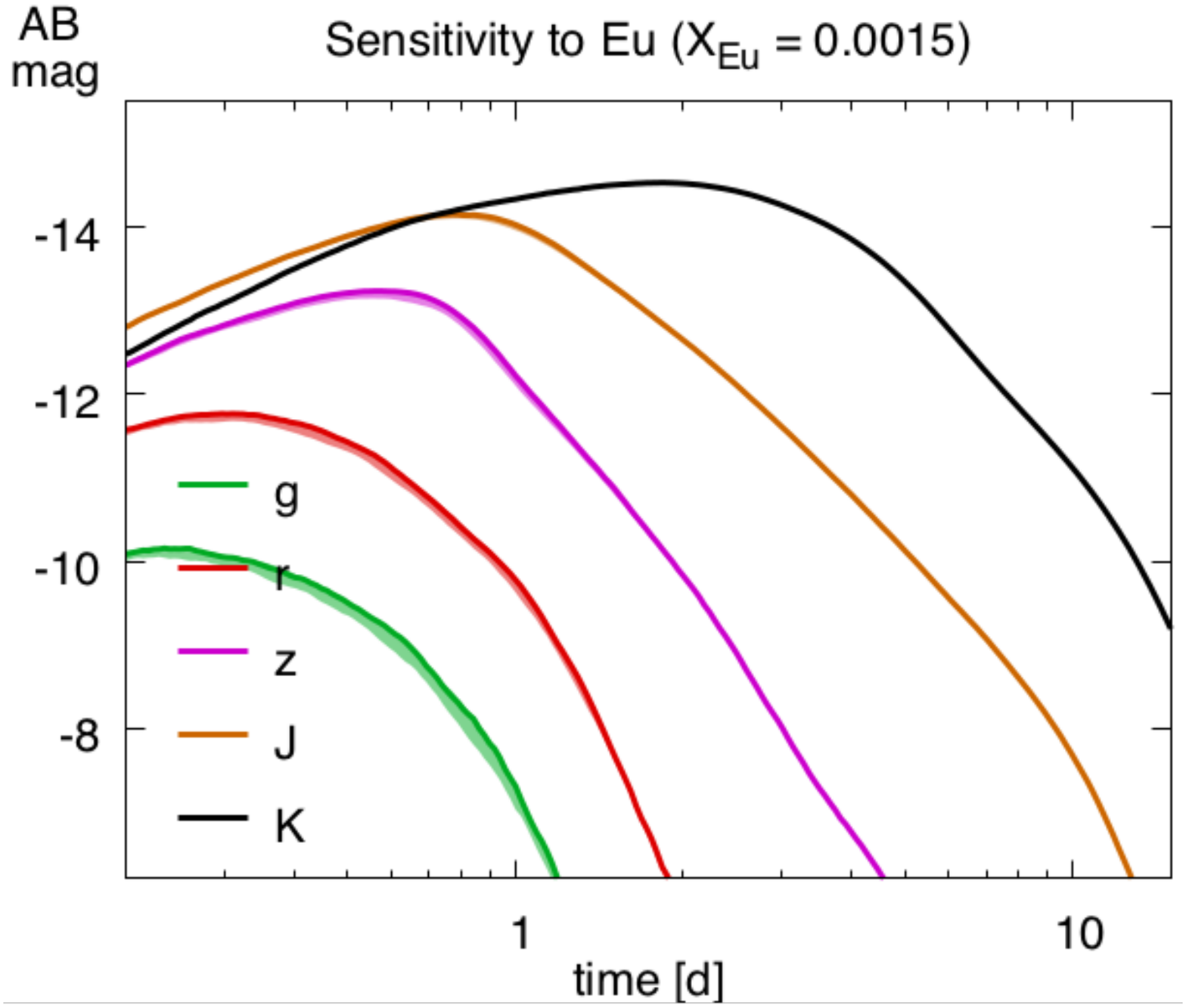} &
    \vspace{-2mm}\includegraphics[width=0.3\textwidth, trim=.2cm 3.5cm .05cm 4.0cm,clip]{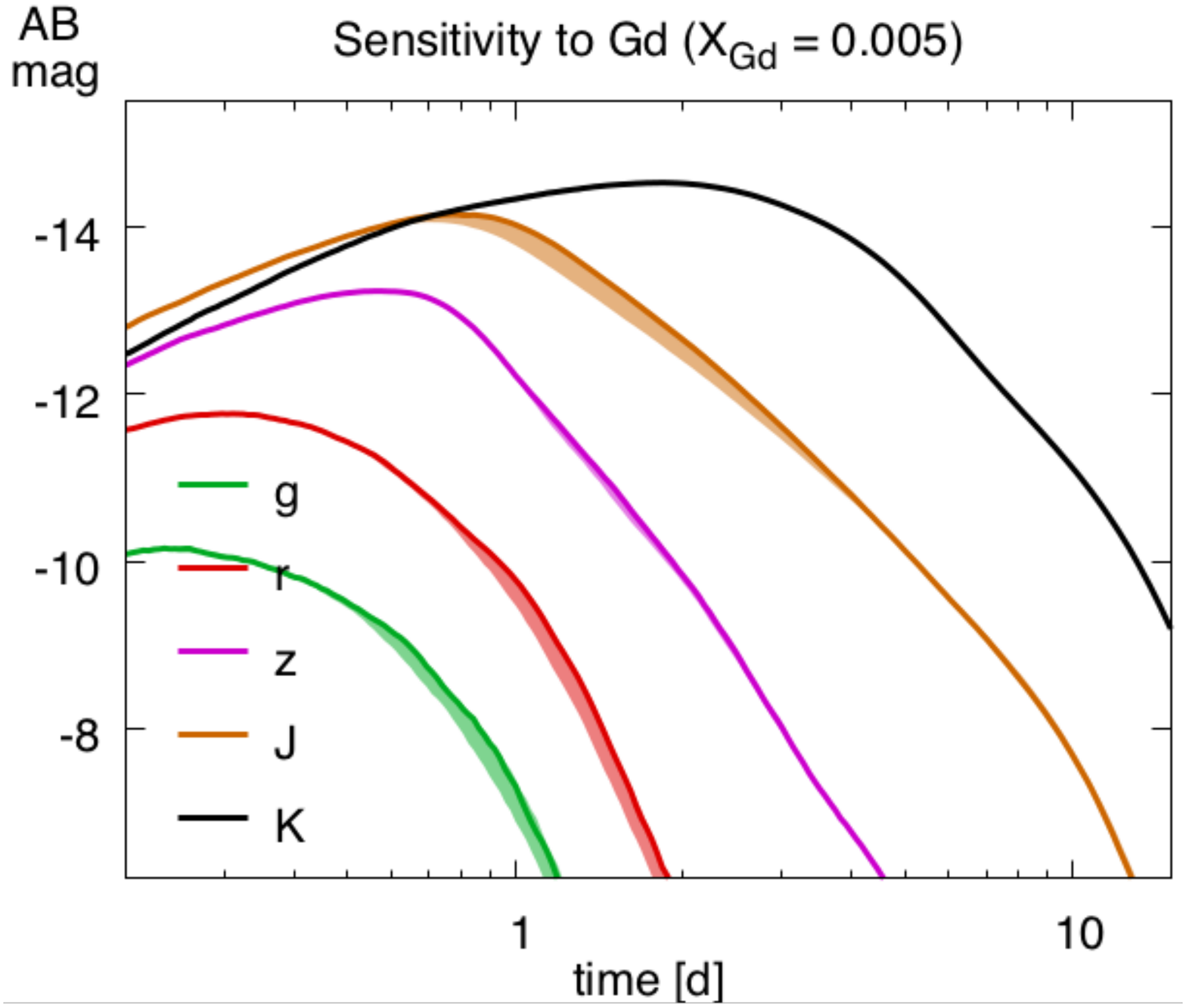} &
    \vspace{-2mm}\includegraphics[width=0.3\textwidth, trim=.2cm 3.5cm .05cm 4.0cm,clip]{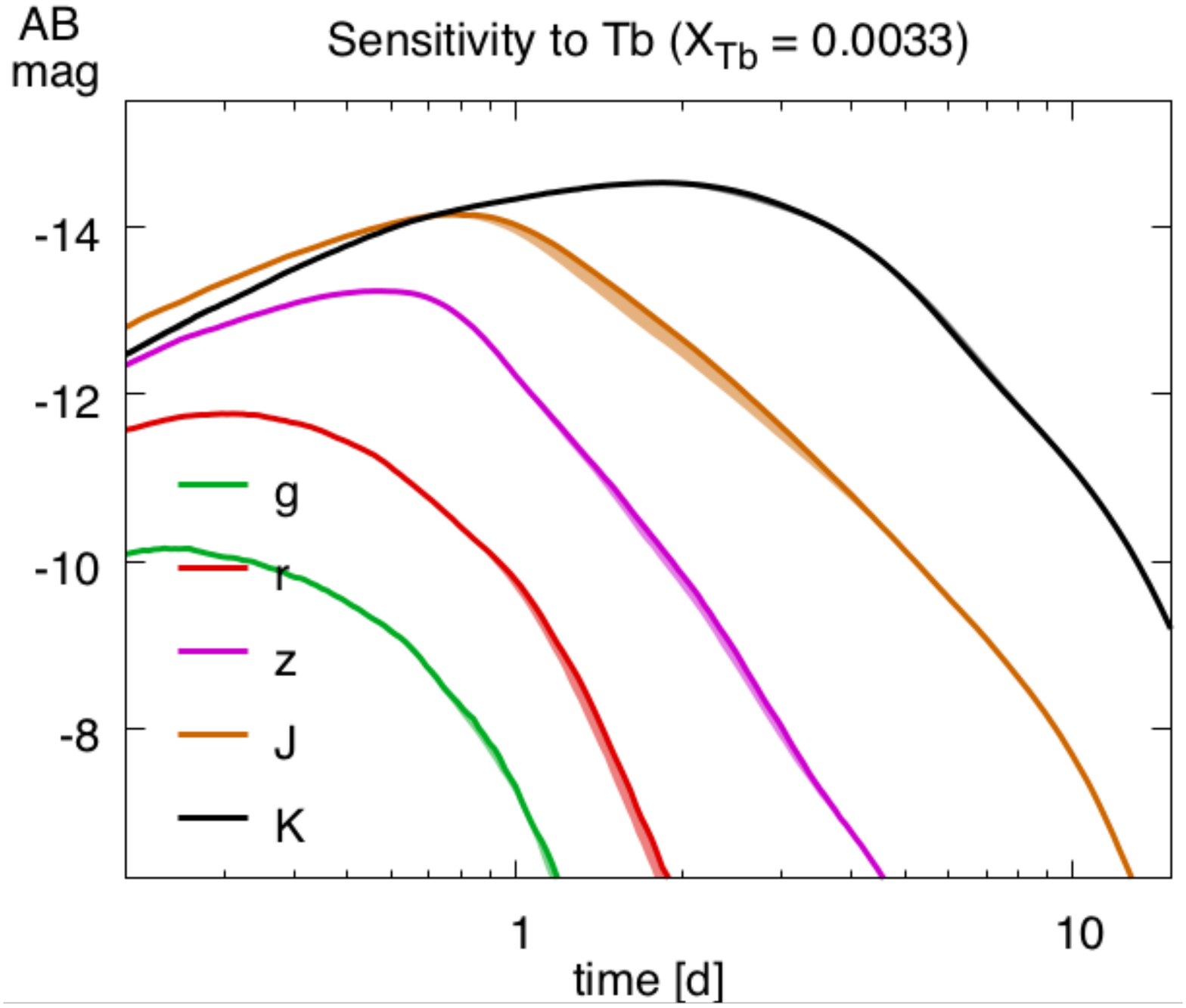}
    \\
    \hspace{-9mm}
    \vspace{-2mm}\includegraphics[width=0.3\textwidth, trim=.2cm 3.5cm .05cm 4.0cm,clip]{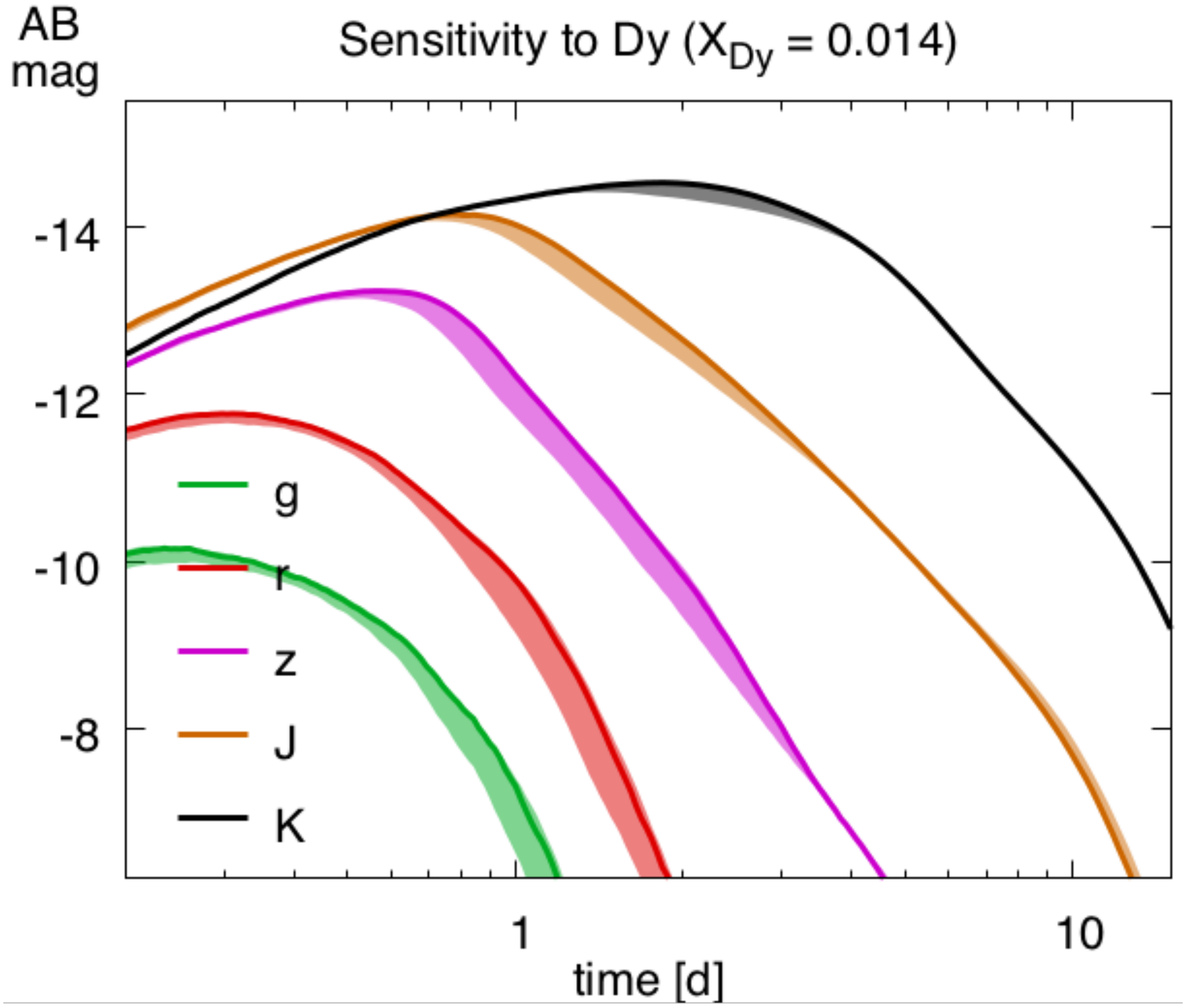} &
    \vspace{-2mm}\includegraphics[width=0.3\textwidth, trim=.2cm 3.5cm .05cm 4.0cm,clip]{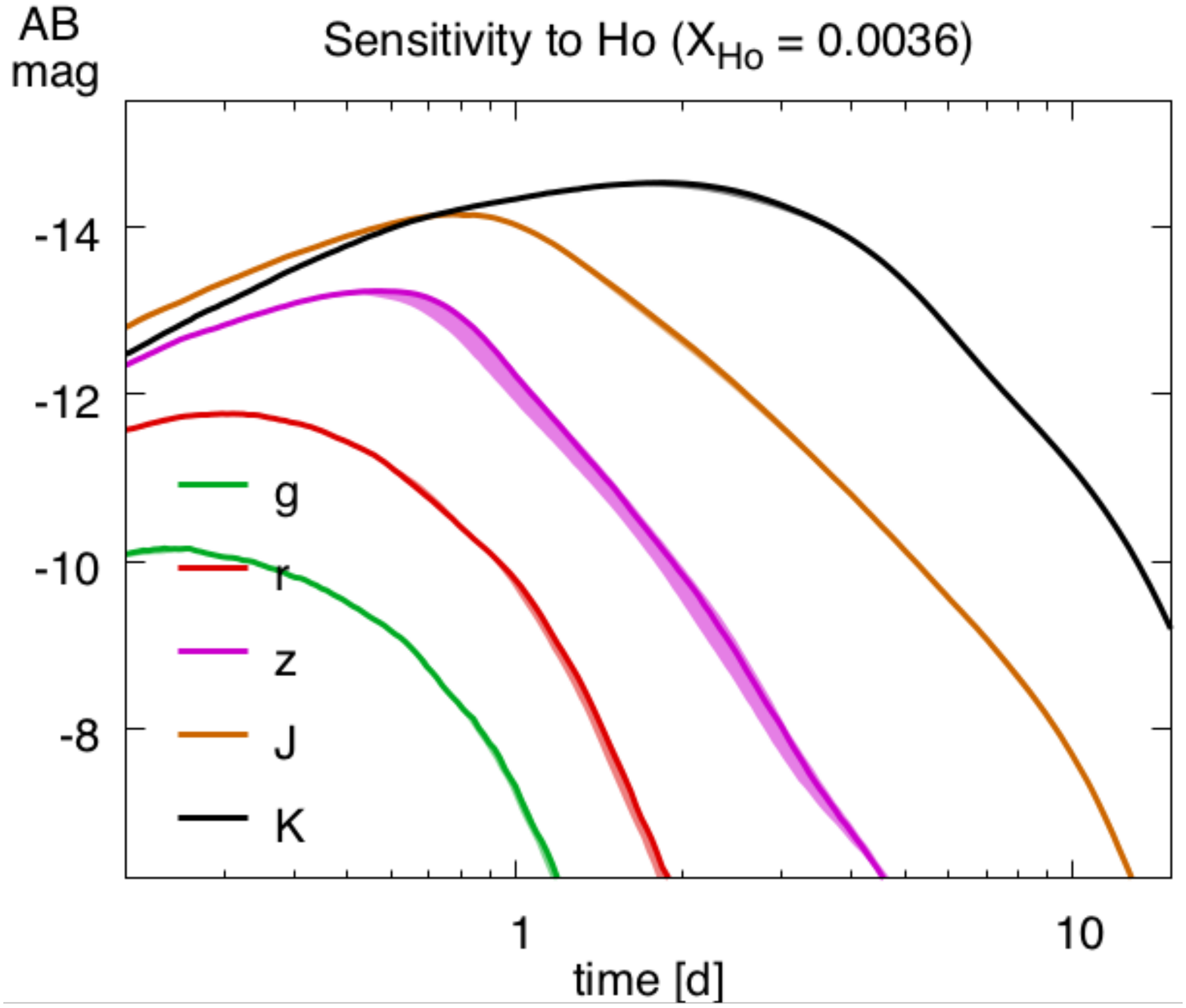} &
    \vspace{-2mm}\includegraphics[width=0.3\textwidth, trim=.2cm 3.5cm .05cm 4.0cm,clip]{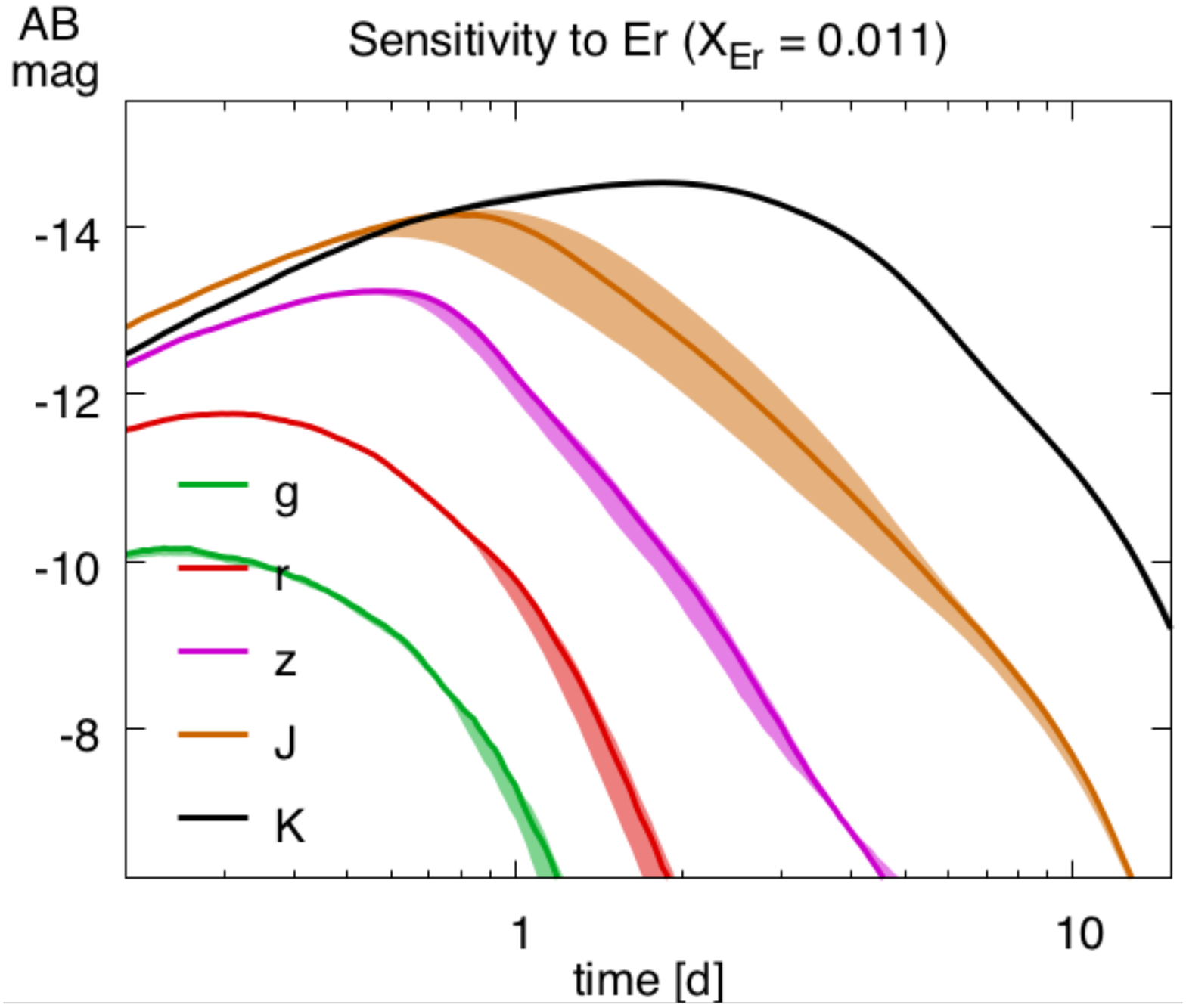}
    \\
    \hspace{-9mm}
    \vspace{-2mm}\includegraphics[width=0.3\textwidth, trim=.2cm 3.5cm .05cm 4.0cm,clip]{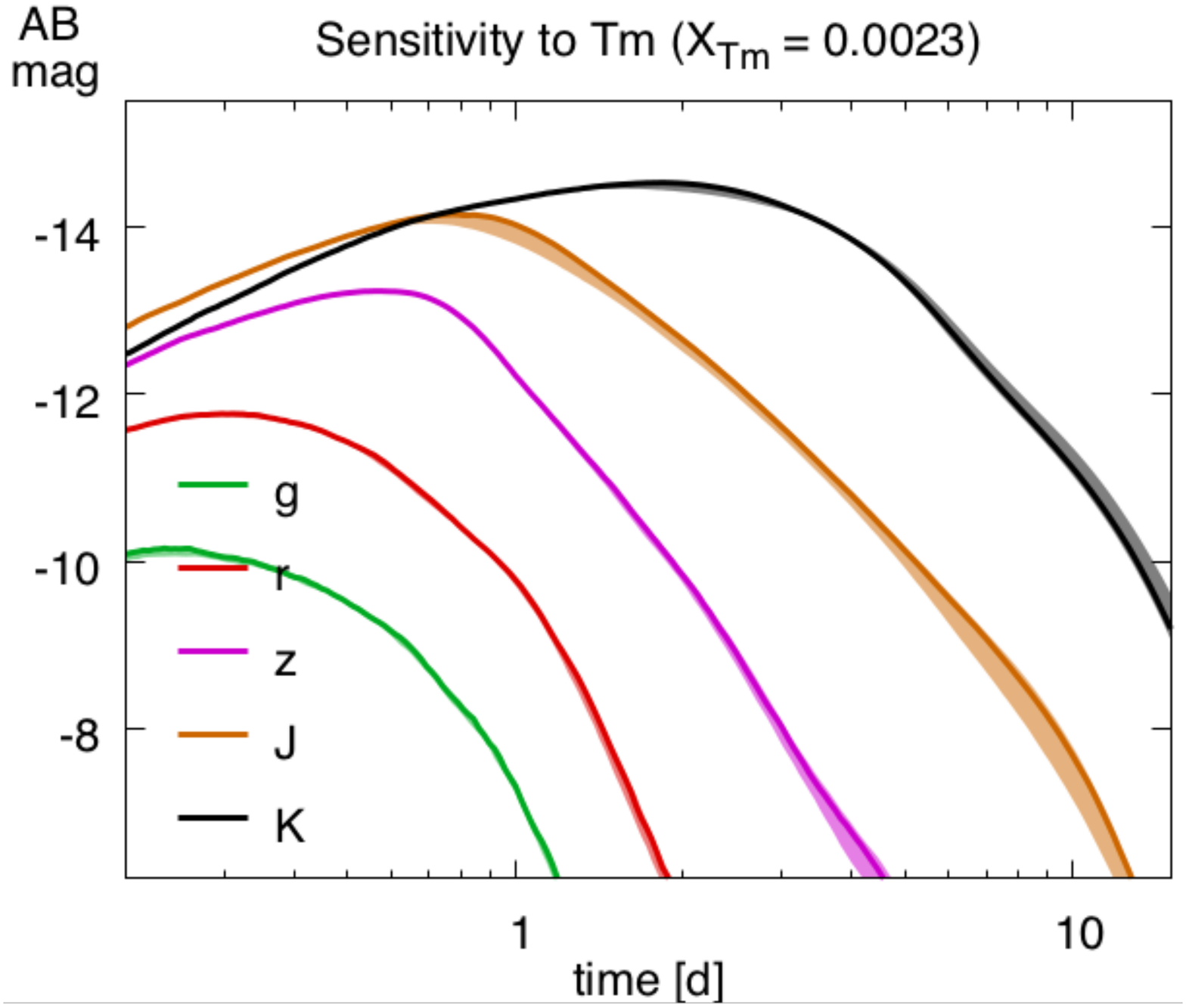} &
    \vspace{-2mm}\includegraphics[width=0.3\textwidth, trim=.2cm 3.5cm .05cm 4.0cm,clip]{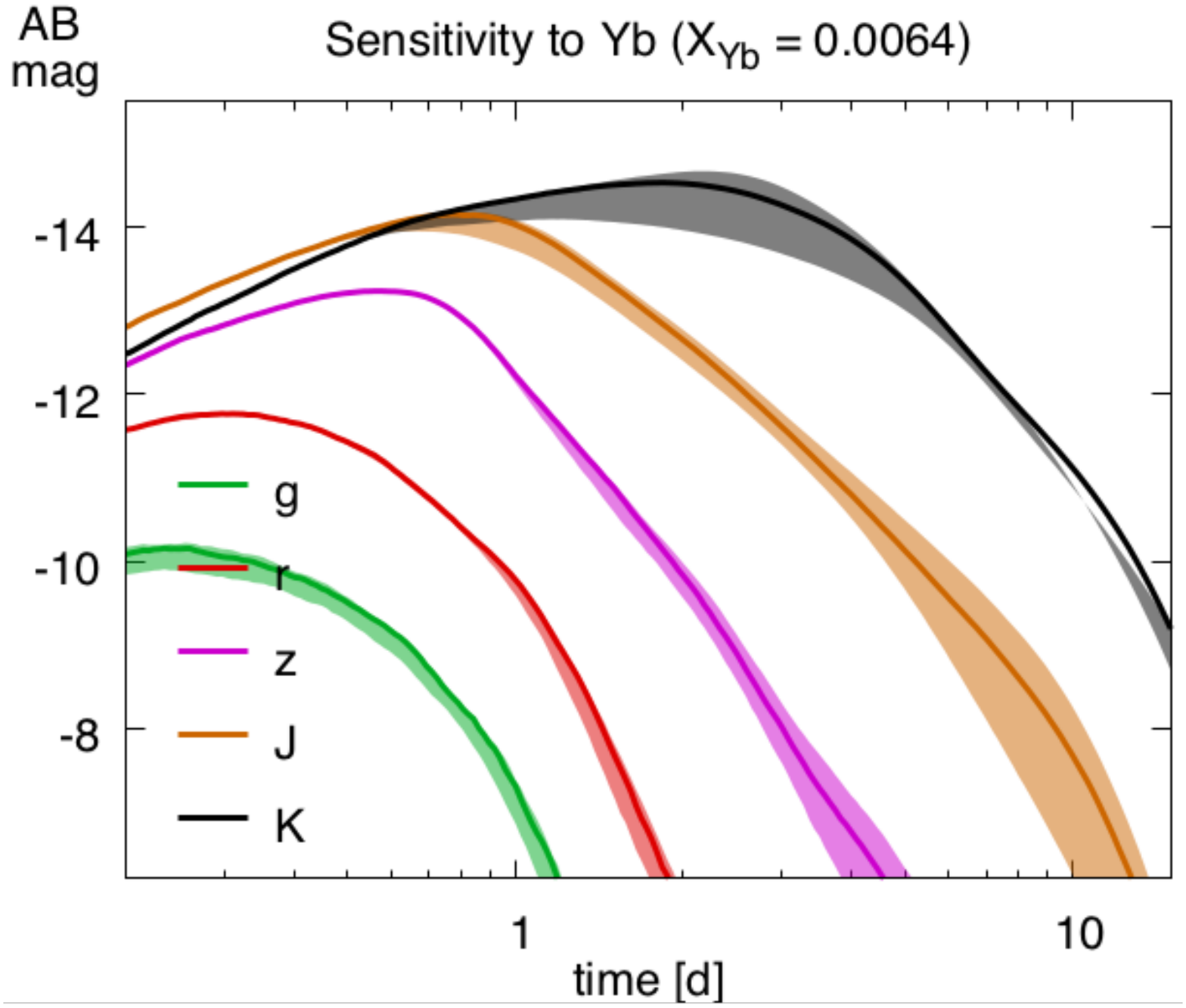} &
    \vspace{-2mm}\includegraphics[width=0.3\textwidth, trim=.2cm 3.5cm .05cm 4.0cm,clip]{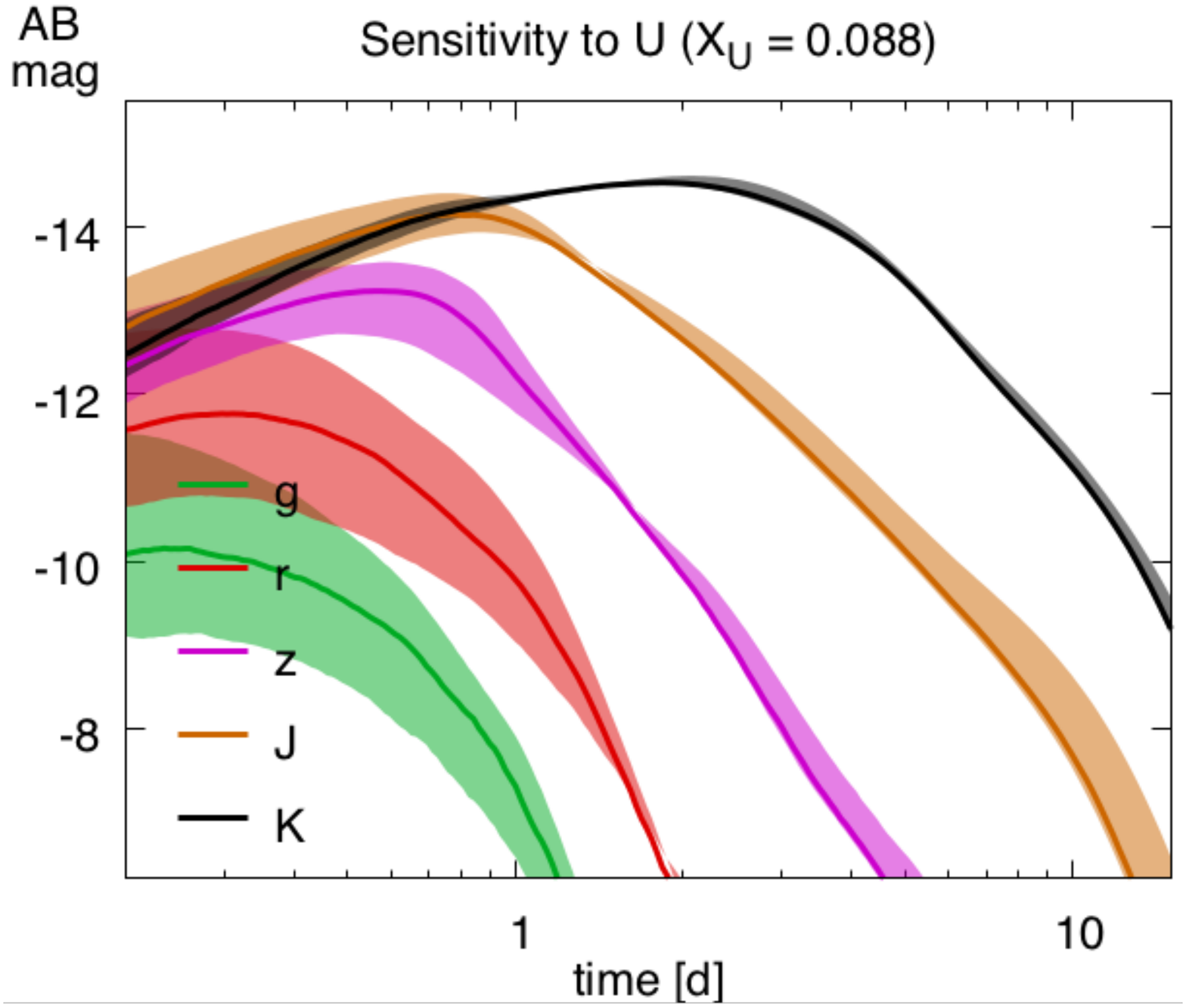}
  \end{tabular}
  \end{center}
  \caption{Sensitivity to the presence of individual lanthanides on the color band  magnitudes of a kilonova with yields derived from dynamical ejecta calculations.}
  \label{fig:dynmag}
\end{figure*}

We repeated our sensitivity study of the color band light curves using abundances from our kilonova dynamical ejecta models.  Figure~\ref{fig:dynmag} shows the band light curves for these models with each lanthanide (and uranium) abundance varied by a factor of ten up and down from the model abundance.  The individual abundance of each lanthanide is different in this composition model from its solar r-process value, and this can alter the effect of varying each by a factor of ten.  Nevertheless, the immense influence of the Nd and U abundances remains.

\section{Summary}

We presented a complete study of the dependence of the spectra and light curves from a kilonova on the individual lanthanide abundances.  For most species, the broad and dense forest of lanthanide lines means that varying its abundance does not affect the light curves or the spectra demonstrably.  However, the opacity of Nd is singularly dominant in the $\sim 1\,\micron$ to $\sim 5\,\micron$ range, and variations in its abundance can dramatically alter the color band light curves.  The abundance of Nd also moves the peak wavelength of the infrared emission, causing it to vary from $\sim 1.6 \, \micron$ for low Nd abundances to $\sim 6 \, \micron$ for high Nd abundances.  If we can sufficiently reduce the uncertainties in the ejecta properties, we can infer the abundance fraction of Nd by measuring this peak.

The spectra and light-curve data from these calculations is available at the Center for Theoretical Astrophysics website:  \url{https://ccsweb.lanl.gov/astro/transient/transients_astro.html}.

\section*{Acknowledgments}

This work was supported by the US Department of Energy through the Los Alamos
National Laboratory. Los Alamos National Laboratory is operated by Triad
National Security, LLC, for the National Nuclear Security Administration
of U.S.\ Department of Energy (Contract No.\ 89233218CNA000001).
Research presented in this article was supported by the Laboratory Directed
Research and Development program of Los Alamos National Laboratory under
project number 20190021DR.

\appendix

\section{Model tables: peak magnitudes, epochs and durations}
\label{tables:app}

In this section, we tabulate properties of our models.  These tables are divided into left and right sections, where each column of each section contains a grizyJHK filter.  The left section contains peak magnitudes, and the right section contains [time of peak magnitude] / [duration of peak magnitude].  Following~\cite{wollaeger18}, the duration of the peak is defined as the time it takes for that particular broadband light curve to drop in brightness by one magnitude.

\begin{table*}
\begin{adjustwidth}{-.25in}{-.25in}
\scriptsize
\caption{Properties of light curves for 2-3 peak dynamical ejecta abundance.
  The naming convention is ``SALanth\_xx\_[xo]10'', which implies element ``xx'' is
  multiplied (``x10'') or divided (``o10'') by 10 in the baseline 2-3 peak composition.}
\begin{tabular}{l|cccccccc|cccccccc}
\hline\hline
      & \multicolumn{8}{|c}{Peak magnitude, $m$}
      & \multicolumn{8}{|c}{Peak epoch $t_p$ [d] / duration $\Delta t_{\rm 1mag}$ [d]  }
\\
Model & g & r & i & z & y & J & H & K & g & r & i & z & y & J & H & K \\
\hline
SALanth\_br\_o10 & -10.2 & -11.8 & -12.8 & -13.2 & -13.5 & -14.1 & -14.4 & -14.5 & 0.23/0.60 & 0.31/0.70 & 0.42/0.78 & 0.57/1.00 & 0.63/1.10 & 0.82/1.60 & 1.18/2.82 & 1.82/4.66  \\
SALanth\_br\_x10 & -10.2 & -11.8 & -12.8 & -13.2 & -13.5 & -14.2 & -14.4 & -14.5 & 0.23/0.59 & 0.28/0.69 & 0.43/0.79 & 0.56/1.00 & 0.64/1.10 & 0.78/1.59 & 1.17/2.82 & 1.75/4.65  \\
SALanth\_ce\_o10 & -10.2 & -11.8 & -12.8 & -13.3 & -13.6 & -14.2 & -14.5 & -14.5 & 0.24/0.61 & 0.31/0.71 & 0.43/0.79 & 0.55/1.01 & 0.64/1.10 & 0.80/1.58 & 1.15/2.80 & 1.76/4.65  \\
SALanth\_ce\_x10 & -9.8 & -11.4 & -12.6 & -13.1 & -13.4 & -14.1 & -14.4 & -14.5 & 0.22/0.58 & 0.27/0.71 & 0.38/0.77 & 0.52/0.98 & 0.60/1.08 & 0.78/1.61 & 1.14/2.86 & 1.80/4.70  \\
SALanth\_dy\_o10 & -10.2 & -11.8 & -12.8 & -13.2 & -13.5 & -14.2 & -14.5 & -14.5 & 0.22/0.60 & 0.31/0.70 & 0.44/0.78 & 0.56/1.01 & 0.63/1.10 & 0.79/1.61 & 1.18/2.84 & 1.80/4.65  \\
SALanth\_dy\_x10 & -10.0 & -11.7 & -12.7 & -13.1 & -13.5 & -14.1 & -14.3 & -14.4 & 0.26/0.58 & 0.30/0.66 & 0.41/0.71 & 0.52/0.88 & 0.60/1.01 & 0.72/1.43 & 0.94/2.83 & 1.36/4.91  \\
SALanth\_er\_o10 & -10.2 & -11.8 & -12.8 & -13.2 & -13.6 & -14.2 & -14.4 & -14.5 & 0.25/0.60 & 0.31/0.70 & 0.44/0.79 & 0.53/1.01 & 0.68/1.21 & 0.88/1.90 & 1.09/2.83 & 1.78/4.64  \\
SALanth\_er\_x10 & -10.1 & -11.8 & -12.8 & -13.2 & -13.3 & -13.9 & -14.5 & -14.5 & 0.23/0.61 & 0.30/0.71 & 0.44/0.77 & 0.53/0.93 & 0.46/0.99 & 0.60/1.33 & 1.12/2.70 & 1.69/4.69  \\
SALanth\_eu\_o10 & -10.2 & -11.8 & -12.8 & -13.2 & -13.6 & -14.2 & -14.4 & -14.5 & 0.22/0.60 & 0.30/0.71 & 0.42/0.78 & 0.57/1.00 & 0.64/1.09 & 0.80/1.59 & 1.17/2.81 & 1.83/4.65  \\
SALanth\_eu\_x10 & -10.1 & -11.7 & -12.7 & -13.1 & -13.4 & -14.1 & -14.4 & -14.5 & 0.23/0.57 & 0.31/0.69 & 0.40/0.77 & 0.56/0.99 & 0.63/1.13 & 0.79/1.61 & 1.15/2.85 & 1.81/4.69  \\
SALanth\_fe\_o10 & -10.1 & -11.8 & -12.8 & -13.2 & -13.5 & -14.2 & -14.4 & -14.5 & 0.21/0.61 & 0.30/0.70 & 0.43/0.78 & 0.54/1.00 & 0.65/1.10 & 0.79/1.59 & 1.19/2.82 & 1.79/4.66  \\
SALanth\_fe\_x10 & -10.1 & -11.8 & -12.8 & -13.2 & -13.5 & -14.2 & -14.4 & -14.5 & 0.22/0.60 & 0.31/0.70 & 0.42/0.78 & 0.55/1.00 & 0.64/1.10 & 0.82/1.60 & 1.17/2.81 & 1.81/4.66  \\
SALanth\_gd\_o10 & -10.2 & -11.8 & -12.8 & -13.2 & -13.5 & -14.2 & -14.4 & -14.5 & 0.22/0.60 & 0.32/0.70 & 0.42/0.78 & 0.55/1.00 & 0.65/1.10 & 0.80/1.62 & 1.18/2.83 & 1.84/4.66  \\
SALanth\_gd\_x10 & -10.1 & -11.8 & -12.8 & -13.2 & -13.5 & -14.1 & -14.5 & -14.5 & 0.23/0.57 & 0.31/0.68 & 0.43/0.78 & 0.59/0.99 & 0.59/1.05 & 0.70/1.45 & 1.15/2.76 & 1.67/4.72  \\
SALanth\_ho\_o10 & -10.1 & -11.8 & -12.8 & -13.2 & -13.5 & -14.2 & -14.4 & -14.5 & 0.27/0.60 & 0.30/0.70 & 0.44/0.78 & 0.57/1.01 & 0.64/1.12 & 0.77/1.60 & 1.18/2.82 & 1.80/4.65  \\
SALanth\_ho\_x10 & -10.1 & -11.8 & -12.8 & -13.2 & -13.5 & -14.2 & -14.4 & -14.5 & 0.26/0.62 & 0.31/0.72 & 0.43/0.77 & 0.52/0.91 & 0.59/1.01 & 0.79/1.54 & 1.13/2.78 & 1.67/4.71  \\
SALanth\_la\_o10 & -10.2 & -11.8 & -12.8 & -13.2 & -13.6 & -14.2 & -14.4 & -14.5 & 0.24/0.60 & 0.31/0.70 & 0.43/0.78 & 0.55/1.00 & 0.64/1.09 & 0.79/1.59 & 1.15/2.82 & 1.82/4.65  \\
SALanth\_la\_x10 & -10.1 & -11.7 & -12.8 & -13.2 & -13.5 & -14.2 & -14.4 & -14.5 & 0.24/0.60 & 0.31/0.70 & 0.44/0.78 & 0.59/1.01 & 0.65/1.10 & 0.80/1.59 & 1.17/2.81 & 1.80/4.66  \\
SALanth\_nd\_o10 & -10.2 & -11.8 & -12.9 & -13.4 & -13.7 & -14.3 & -14.6 & -14.5 & 0.26/0.63 & 0.31/0.74 & 0.43/0.81 & 0.56/1.03 & 0.69/1.12 & 0.82/1.58 & 1.18/2.58 & 1.80/4.79  \\
SALanth\_nd\_x10 & -10.0 & -11.5 & -12.4 & -12.7 & -13.0 & -13.6 & -14.0 & -14.4 & 0.23/0.51 & 0.28/0.60 & 0.38/0.69 & 0.45/0.88 & 0.53/1.01 & 0.69/1.70 & 1.17/3.19 & 1.87/4.65  \\
SALanth\_pd\_o10 & -10.1 & -11.8 & -12.8 & -13.2 & -13.5 & -14.2 & -14.4 & -14.5 & 0.28/0.61 & 0.31/0.71 & 0.44/0.78 & 0.56/1.00 & 0.65/1.10 & 0.79/1.59 & 1.16/2.81 & 1.83/4.66  \\
SALanth\_pd\_x10 & -10.2 & -11.8 & -12.8 & -13.2 & -13.5 & -14.2 & -14.4 & -14.5 & 0.24/0.60 & 0.31/0.70 & 0.43/0.78 & 0.54/1.00 & 0.64/1.10 & 0.79/1.60 & 1.15/2.82 & 1.82/4.66  \\
SALanth\_pm\_o10 & -10.2 & -11.8 & -12.8 & -13.3 & -13.6 & -14.2 & -14.5 & -14.5 & 0.25/0.61 & 0.31/0.70 & 0.42/0.78 & 0.57/1.00 & 0.64/1.10 & 0.80/1.59 & 1.17/2.79 & 1.76/4.64  \\
SALanth\_pm\_x10 & -10.1 & -11.6 & -12.7 & -13.1 & -13.4 & -14.0 & -14.3 & -14.5 & 0.21/0.57 & 0.32/0.67 & 0.39/0.75 & 0.53/0.96 & 0.61/1.05 & 0.77/1.63 & 1.16/2.98 & 1.91/4.78  \\
SALanth\_pr\_o10 & -10.2 & -11.8 & -12.8 & -13.3 & -13.6 & -14.2 & -14.5 & -14.5 & 0.23/0.61 & 0.30/0.70 & 0.44/0.78 & 0.56/1.00 & 0.65/1.10 & 0.79/1.59 & 1.16/2.80 & 1.79/4.65  \\
SALanth\_pr\_x10 & -10.1 & -11.7 & -12.7 & -13.1 & -13.4 & -14.0 & -14.3 & -14.5 & 0.26/0.58 & 0.29/0.67 & 0.41/0.76 & 0.53/0.96 & 0.61/1.09 & 0.77/1.60 & 1.16/2.94 & 1.82/4.70  \\
SALanth\_se\_o10 & -10.0 & -11.7 & -12.7 & -13.2 & -13.5 & -14.1 & -14.4 & -14.5 & 0.25/0.60 & 0.30/0.69 & 0.41/0.77 & 0.54/0.97 & 0.61/1.07 & 0.77/1.57 & 1.15/2.84 & 1.75/4.70  \\
SALanth\_se\_x10 & -10.9 & -12.4 & -13.3 & -13.7 & -13.9 & -14.5 & -14.7 & -14.6 & 0.24/0.65 & 0.37/0.81 & 0.52/0.90 & 0.67/1.17 & 0.80/1.30 & 0.94/1.76 & 1.28/2.75 & 1.97/4.51  \\
SALanth\_sm\_o10 & -10.2 & -11.9 & -13.0 & -13.4 & -13.6 & -14.2 & -14.5 & -14.5 & 0.26/0.64 & 0.30/0.72 & 0.46/0.81 & 0.60/1.03 & 0.68/1.13 & 0.82/1.62 & 1.21/2.79 & 1.81/4.61  \\
SALanth\_sm\_x10 & -9.7 & -11.3 & -12.1 & -12.7 & -13.2 & -13.9 & -14.3 & -14.4 & 0.19/0.49 & 0.26/0.61 & 0.35/0.74 & 0.42/0.89 & 0.52/0.90 & 0.70/1.50 & 1.09/3.03 & 1.84/4.89  \\
SALanth\_tb\_o10 & -10.1 & -11.8 & -12.8 & -13.2 & -13.5 & -14.2 & -14.4 & -14.5 & 0.26/0.61 & 0.30/0.70 & 0.43/0.78 & 0.56/1.00 & 0.63/1.10 & 0.79/1.61 & 1.18/2.84 & 1.82/4.65  \\
SALanth\_tb\_x10 & -10.1 & -11.8 & -12.8 & -13.2 & -13.5 & -14.1 & -14.4 & -14.5 & 0.24/0.59 & 0.31/0.69 & 0.42/0.77 & 0.57/0.99 & 0.64/1.07 & 0.74/1.46 & 1.05/2.66 & 1.68/4.73  \\
SALanth\_te\_o10 & -9.7 & -11.3 & -12.4 & -12.9 & -13.2 & -13.9 & -14.2 & -14.4 & 0.22/0.57 & 0.28/0.65 & 0.38/0.73 & 0.49/0.91 & 0.54/0.99 & 0.70/1.50 & 1.09/2.89 & 1.63/4.85  \\
SALanth\_te\_x10 & -11.7 & -13.0 & -13.8 & -14.1 & -14.3 & -14.8 & -14.8 & -14.7 & 0.21/0.71 & 0.43/0.99 & 0.60/1.11 & 0.79/1.40 & 0.96/1.59 & 1.08/2.03 & 1.35/2.80 & 1.17/4.18  \\
SALanth\_tm\_o10 & -10.1 & -11.8 & -12.8 & -13.2 & -13.5 & -14.2 & -14.5 & -14.5 & 0.27/0.61 & 0.30/0.70 & 0.44/0.78 & 0.56/1.00 & 0.64/1.09 & 0.80/1.61 & 1.20/2.96 & 1.83/4.60  \\
SALanth\_tm\_x10 & -10.1 & -11.7 & -12.8 & -13.2 & -13.5 & -14.0 & -14.3 & -14.5 & 0.24/0.61 & 0.29/0.69 & 0.42/0.78 & 0.58/0.99 & 0.61/1.08 & 0.71/1.57 & 0.92/2.46 & 1.57/4.97  \\
SALanth\_u\_o10 & -11.6 & -12.8 & -13.3 & -13.6 & -13.8 & -14.4 & -14.5 & -14.5 & 0.18/0.45 & 0.26/0.63 & 0.41/0.85 & 0.59/1.01 & 0.65/1.09 & 0.76/1.46 & 1.13/2.66 & 1.69/4.58  \\
SALanth\_u\_x10 & -9.1 & -10.8 & -12.0 & -12.7 & -13.1 & -13.9 & -14.5 & -14.6 & 0.27/0.59 & 0.27/0.73 & 0.36/0.83 & 0.48/1.03 & 0.62/1.20 & 0.87/1.97 & 1.27/3.13 & 2.06/4.67  \\
SALanth\_yb\_o10 & -10.2 & -11.8 & -12.8 & -13.2 & -13.6 & -14.2 & -14.6 & -14.7 & 0.26/0.59 & 0.32/0.70 & 0.44/0.79 & 0.54/1.00 & 0.64/1.11 & 0.82/1.65 & 1.31/3.01 & 2.14/4.56  \\
SALanth\_yb\_x10 & -9.9 & -11.7 & -12.8 & -13.3 & -13.5 & -13.9 & -14.2 & -14.1 & 0.23/0.60 & 0.30/0.69 & 0.44/0.77 & 0.58/0.97 & 0.61/1.06 & 0.66/1.55 & 0.95/2.85 & 1.17/4.77  \\
SALanth\_zr\_o10 & -9.8 & -11.5 & -12.5 & -13.1 & -13.3 & -13.9 & -14.3 & -14.5 & 0.29/0.61 & 0.33/0.69 & 0.43/0.80 & 0.56/1.02 & 0.58/1.07 & 0.72/1.52 & 1.09/2.86 & 1.45/4.66  \\
SALanth\_zr\_x10 & -11.6 & -12.9 & -13.6 & -13.9 & -14.1 & -14.7 & -14.8 & -14.5 & 0.22/0.56 & 0.35/0.75 & 0.52/0.94 & 0.68/1.11 & 0.78/1.37 & 0.95/1.98 & 1.39/2.81 & 1.03/4.60  \\
\hline
\end{tabular}
\label{tb1:app}
\end{adjustwidth}
\label{tab:lanth}
\end{table*}

\begin{table*}
\begin{adjustwidth}{-.25in}{-.25in}
\scriptsize
\caption{Properties of light curves for solar r-process abundance, plus Pm and U.
  The naming convention is ``SolarUPm\_xx\_[xo]10'', which implies element ``xx'' is
  multiplied (``x10'') or divided (``o10'') by 10 in the baseline solar composition.}
\begin{tabular}{l|cccccccc|cccccccc}
\hline\hline
      & \multicolumn{8}{|c}{Peak magnitude, $m$}
      & \multicolumn{8}{|c}{Peak epoch $t_p$ [d] / duration $\Delta t_{\rm 1mag}$ [d]  }
\\
Model & g & r & i & z & y & J & H & K & g & r & i & z & y & J & H & K \\
\hline
SolarUPm\_baseline & -12.4 & -13.2 & -13.3 & -13.5 & -13.7 & -14.2 & -14.4 & -14.7 & 0.17/0.37 & 0.23/0.62 & 0.39/0.91 & 0.56/1.12 & 0.55/1.28 & 0.80/1.77 & 1.50/3.53 & 1.64/3.83  \\
SolarUPm\_br\_o10 & -12.0 & -13.0 & -13.2 & -13.4 & -13.5 & -14.1 & -14.3 & -14.6 & 0.17/0.37 & 0.22/0.59 & 0.36/0.86 & 0.52/1.05 & 0.52/1.20 & 0.75/1.70 & 1.43/3.55 & 1.64/3.90  \\
SolarUPm\_br\_x10 & -13.5 & -14.1 & -14.0 & -14.1 & -14.2 & -14.7 & -14.6 & -14.8 & 0.16/0.41 & 0.19/0.73 & 0.30/1.18 & 0.74/1.42 & 0.72/1.69 & 0.83/2.09 & 1.25/3.43 & 1.51/3.59  \\
SolarUPm\_ce\_o10 & -12.4 & -13.2 & -13.4 & -13.6 & -13.7 & -14.2 & -14.4 & -14.7 & 0.17/0.43 & 0.23/0.71 & 0.44/0.95 & 0.57/1.15 & 0.53/1.30 & 0.82/1.77 & 1.49/3.50 & 1.63/3.80  \\
SolarUPm\_ce\_x10 & -12.2 & -13.0 & -13.1 & -13.4 & -13.6 & -14.1 & -14.3 & -14.7 & 0.16/0.28 & 0.19/0.41 & 0.26/0.77 & 0.48/1.01 & 0.51/1.19 & 0.77/1.76 & 1.45/3.58 & 1.65/3.88  \\
SolarUPm\_dy\_o10 & -12.4 & -13.2 & -13.4 & -13.6 & -13.7 & -14.2 & -14.4 & -14.7 & 0.17/0.37 & 0.22/0.61 & 0.37/0.92 & 0.56/1.13 & 0.55/1.28 & 0.79/1.77 & 1.50/3.54 & 1.64/3.82  \\
SolarUPm\_dy\_x10 & -11.9 & -12.9 & -13.2 & -13.5 & -13.6 & -14.2 & -14.3 & -14.6 & 0.18/0.40 & 0.23/0.64 & 0.42/0.86 & 0.56/1.03 & 0.57/1.21 & 0.79/1.68 & 1.25/3.53 & 1.55/3.91  \\
SolarUPm\_er\_o10 & -12.4 & -13.2 & -13.3 & -13.5 & -13.7 & -14.2 & -14.3 & -14.7 & 0.17/0.37 & 0.22/0.62 & 0.40/0.92 & 0.56/1.13 & 0.55/1.36 & 0.81/1.88 & 1.48/3.56 & 1.63/3.81  \\
SolarUPm\_er\_x10 & -12.2 & -13.1 & -13.3 & -13.5 & -13.6 & -14.1 & -14.4 & -14.7 & 0.17/0.37 & 0.22/0.61 & 0.40/0.89 & 0.55/1.05 & 0.52/1.06 & 0.63/1.46 & 1.32/3.36 & 1.66/3.89  \\
SolarUPm\_eu\_o10 & -12.4 & -13.2 & -13.3 & -13.5 & -13.7 & -14.2 & -14.4 & -14.7 & 0.17/0.37 & 0.22/0.62 & 0.38/0.92 & 0.56/1.12 & 0.56/1.28 & 0.80/1.77 & 1.48/3.53 & 1.64/3.82  \\
SolarUPm\_eu\_x10 & -12.3 & -13.2 & -13.3 & -13.5 & -13.6 & -14.2 & -14.4 & -14.7 & 0.17/0.35 & 0.22/0.60 & 0.37/0.91 & 0.56/1.12 & 0.57/1.31 & 0.78/1.79 & 1.48/3.55 & 1.64/3.84  \\
SolarUPm\_gd\_o10 & -12.4 & -13.2 & -13.3 & -13.5 & -13.7 & -14.2 & -14.4 & -14.7 & 0.17/0.37 & 0.22/0.62 & 0.39/0.92 & 0.56/1.12 & 0.55/1.29 & 0.80/1.81 & 1.49/3.53 & 1.64/3.82  \\
SolarUPm\_gd\_x10 & -12.3 & -13.2 & -13.3 & -13.5 & -13.6 & -14.2 & -14.4 & -14.7 & 0.17/0.36 & 0.21/0.60 & 0.40/0.89 & 0.56/1.11 & 0.52/1.16 & 0.66/1.55 & 1.42/3.51 & 1.60/3.85  \\
SolarUPm\_ho\_o10 & -12.4 & -13.2 & -13.3 & -13.5 & -13.7 & -14.2 & -14.4 & -14.7 & 0.17/0.37 & 0.22/0.61 & 0.39/0.91 & 0.56/1.13 & 0.56/1.30 & 0.80/1.77 & 1.50/3.53 & 1.64/3.82  \\
SolarUPm\_ho\_x10 & -12.3 & -13.2 & -13.3 & -13.5 & -13.7 & -14.2 & -14.4 & -14.7 & 0.17/0.37 & 0.22/0.62 & 0.40/0.91 & 0.55/1.03 & 0.55/1.18 & 0.81/1.75 & 1.48/3.51 & 1.62/3.84  \\
SolarUPm\_la\_o10 & -12.4 & -13.2 & -13.3 & -13.5 & -13.7 & -14.2 & -14.4 & -14.7 & 0.18/0.37 & 0.22/0.61 & 0.40/0.92 & 0.57/1.12 & 0.55/1.28 & 0.79/1.77 & 1.49/3.53 & 1.64/3.82  \\
SolarUPm\_la\_x10 & -12.3 & -13.2 & -13.3 & -13.5 & -13.7 & -14.2 & -14.4 & -14.7 & 0.17/0.36 & 0.22/0.62 & 0.37/0.87 & 0.56/1.12 & 0.54/1.29 & 0.81/1.76 & 1.47/3.52 & 1.64/3.83  \\
SolarUPm\_nd\_o10 & -12.5 & -13.4 & -13.5 & -13.8 & -13.9 & -14.4 & -14.6 & -14.7 & 0.17/0.37 & 0.22/0.63 & 0.33/0.94 & 0.58/1.19 & 0.60/1.35 & 0.83/1.74 & 1.34/3.33 & 1.59/3.91  \\
SolarUPm\_nd\_x10 & -11.5 & -12.5 & -12.7 & -12.8 & -13.0 & -13.5 & -13.9 & -14.4 & 0.18/0.39 & 0.22/0.59 & 0.34/0.77 & 0.37/0.91 & 0.40/1.09 & 0.78/1.90 & 1.49/4.03 & 1.77/4.19  \\
SolarUPm\_pd\_o10 & -12.3 & -13.2 & -13.3 & -13.5 & -13.7 & -14.2 & -14.3 & -14.7 & 0.17/0.37 & 0.22/0.61 & 0.38/0.90 & 0.55/1.11 & 0.55/1.27 & 0.78/1.76 & 1.48/3.53 & 1.64/3.84  \\
SolarUPm\_pm\_o10 & -12.4 & -13.3 & -13.4 & -13.6 & -13.7 & -14.2 & -14.4 & -14.7 & 0.17/0.37 & 0.22/0.61 & 0.37/0.91 & 0.56/1.13 & 0.53/1.29 & 0.80/1.76 & 1.49/3.51 & 1.62/3.79  \\
SolarUPm\_pm\_x10 & -12.0 & -12.8 & -13.1 & -13.4 & -13.5 & -14.0 & -14.3 & -14.5 & 0.18/0.38 & 0.23/0.65 & 0.41/0.88 & 0.55/1.08 & 0.58/1.23 & 0.78/1.84 & 1.51/3.65 & 1.74/4.04  \\
SolarUPm\_pr\_o10 & -12.4 & -13.2 & -13.4 & -13.7 & -13.8 & -14.3 & -14.4 & -14.7 & 0.17/0.38 & 0.22/0.63 & 0.40/0.94 & 0.60/1.16 & 0.65/1.32 & 0.82/1.75 & 1.55/3.42 & 1.61/3.79  \\
SolarUPm\_pr\_x10 & -12.2 & -13.0 & -13.0 & -13.1 & -13.2 & -13.7 & -14.1 & -14.6 & 0.17/0.35 & 0.22/0.58 & 0.31/0.75 & 0.39/0.93 & 0.36/1.11 & 0.72/1.86 & 1.30/3.83 & 1.65/3.98  \\
SolarUPm\_se\_o10 & -11.7 & -12.7 & -13.0 & -13.2 & -13.4 & -13.9 & -14.2 & -14.5 & 0.18/0.37 & 0.22/0.56 & 0.35/0.80 & 0.48/0.98 & 0.48/1.12 & 0.70/1.65 & 1.36/3.59 & 1.65/3.99  \\
SolarUPm\_se\_x10 & -13.9 & -14.5 & -14.3 & -14.3 & -14.3 & -14.8 & -14.7 & -14.8 & 0.17/0.43 & 0.20/0.74 & 0.28/1.26 & 0.77/1.51 & 0.76/1.82 & 0.77/2.17 & 1.19/3.32 & 1.41/3.44  \\
SolarUPm\_sm\_o10 & -12.6 & -13.2 & -13.4 & -13.6 & -13.7 & -14.2 & -14.4 & -14.7 & 0.17/0.36 & 0.24/0.69 & 0.45/0.97 & 0.61/1.16 & 0.56/1.35 & 0.81/1.80 & 1.51/3.50 & 1.63/3.78  \\
SolarUPm\_sm\_x10 & -11.5 & -12.8 & -13.0 & -13.2 & -13.5 & -14.0 & -14.3 & -14.6 & 0.18/0.36 & 0.19/0.45 & 0.24/0.69 & 0.37/0.90 & 0.48/0.97 & 0.71/1.63 & 1.45/3.68 & 1.76/4.05  \\
SolarUPm\_tb\_o10 & -12.4 & -13.2 & -13.3 & -13.5 & -13.7 & -14.2 & -14.4 & -14.7 & 0.17/0.37 & 0.22/0.61 & 0.39/0.91 & 0.56/1.12 & 0.55/1.28 & 0.80/1.77 & 1.50/3.54 & 1.63/3.82  \\
SolarUPm\_tb\_x10 & -12.3 & -13.2 & -13.3 & -13.5 & -13.7 & -14.2 & -14.3 & -14.7 & 0.17/0.37 & 0.22/0.61 & 0.39/0.91 & 0.56/1.12 & 0.56/1.27 & 0.79/1.71 & 1.37/3.46 & 1.65/3.87  \\
SolarUPm\_te\_o10 & -12.2 & -13.1 & -13.3 & -13.5 & -13.6 & -14.2 & -14.3 & -14.6 & 0.17/0.37 & 0.22/0.60 & 0.39/0.89 & 0.54/1.09 & 0.54/1.25 & 0.78/1.74 & 1.47/3.54 & 1.64/3.85  \\
SolarUPm\_te\_x10 & -13.1 & -13.8 & -13.8 & -13.9 & -14.0 & -14.5 & -14.5 & -14.8 & 0.17/0.39 & 0.21/0.69 & 0.37/1.07 & 0.67/1.30 & 0.64/1.52 & 0.85/1.96 & 1.50/3.50 & 1.61/3.70  \\
SolarUPm\_tm\_o10 & -12.4 & -13.2 & -13.3 & -13.5 & -13.7 & -14.2 & -14.4 & -14.7 & 0.17/0.37 & 0.22/0.61 & 0.38/0.92 & 0.56/1.12 & 0.53/1.29 & 0.79/1.78 & 1.54/3.62 & 1.63/3.79  \\
SolarUPm\_tm\_x10 & -12.3 & -13.2 & -13.3 & -13.5 & -13.7 & -14.2 & -14.3 & -14.7 & 0.17/0.37 & 0.22/0.62 & 0.39/0.91 & 0.56/1.12 & 0.56/1.26 & 0.76/1.69 & 1.20/3.22 & 1.66/3.96  \\
SolarUPm\_u\_o10 & -13.3 & -13.8 & -13.6 & -13.7 & -13.8 & -14.3 & -14.4 & -14.7 & 0.16/0.32 & 0.18/0.49 & 0.22/0.87 & 0.53/1.10 & 0.51/1.25 & 0.75/1.73 & 1.45/3.50 & 1.61/3.81  \\
SolarUPm\_u\_x10 & -11.0 & -12.3 & -13.0 & -13.3 & -13.4 & -14.1 & -14.4 & -14.7 & 0.19/0.52 & 0.27/0.73 & 0.46/0.87 & 0.59/1.14 & 0.66/1.34 & 0.86/1.86 & 1.53/3.59 & 1.69/3.86  \\
SolarUPm\_yb\_o10 & -12.4 & -13.2 & -13.3 & -13.5 & -13.7 & -14.2 & -14.4 & -14.7 & 0.17/0.38 & 0.22/0.61 & 0.39/0.92 & 0.56/1.13 & 0.55/1.30 & 0.85/1.96 & 1.58/3.64 & 1.87/3.97  \\
SolarUPm\_yb\_x10 & -12.1 & -13.2 & -13.3 & -13.5 & -13.7 & -14.1 & -14.1 & -14.5 & 0.16/0.35 & 0.21/0.60 & 0.36/0.87 & 0.55/1.07 & 0.56/1.23 & 0.59/1.55 & 1.01/3.24 & 1.40/3.94  \\
SolarUPm\_zr\_o10 & -12.3 & -13.2 & -13.3 & -13.5 & -13.7 & -14.2 & -14.3 & -14.7 & 0.17/0.37 & 0.22/0.64 & 0.39/0.99 & 0.56/1.19 & 0.54/1.31 & 0.78/1.77 & 1.50/3.53 & 1.63/3.72  \\
SolarUPm\_zr\_x10 & -12.6 & -13.4 & -13.5 & -13.6 & -13.8 & -14.3 & -14.5 & -14.6 & 0.17/0.37 & 0.21/0.57 & 0.38/0.85 & 0.57/1.04 & 0.58/1.27 & 0.83/1.82 & 1.49/3.52 & 1.83/4.03  \\
\hline
\end{tabular}
\label{tab:solaru}
\end{adjustwidth}
\end{table*}

\begin{table*}
\begin{adjustwidth}{-.25in}{-.25in}
\scriptsize
\caption{Properties of light curves for solar r-process abundance, varying the overall lanthanide fraction.
  The naming convention is ``SolarRP\_lanth\_x1em[1-9]'', which implies the total lanthanide abundance is
  divided by $10^{1}-10^{9}$ in the baseline solar composition.
  The models with ``\_1ep1'' and ``\_1ep2'' correspond to $X_{\rm lanth}=0.25$ and 0.75 in Figure~\ref{fig:lanthspec}.}
\begin{tabular}{l|cccccccc|cccccccc}
\hline\hline
      & \multicolumn{8}{|c}{Peak magnitude, $m$}
      & \multicolumn{8}{|c}{Peak epoch $t_p$ [d] / duration $\Delta t_{\rm 1mag}$ [d]  }
\\
Model & g & r & i & z & y & J & H & K & g & r & i & z & y & J & H & K \\
\hline
SolarRP\_lanth\_x1ep2 & -7.6 & -9.5 & -10.4 & -11.0 & -11.2 & -12.0 & -12.9 & -13.8 & 0.14/0.29 & 0.16/0.34 & 0.22/0.41 & 0.26/0.55 & 0.27/0.63 & 0.43/1.71 & 1.11/4.20 & 2.28/5.21  \\
SolarRP\_lanth\_x1ep1 & -9.3 & -10.9 & -11.8 & -12.1 & -12.3 & -12.9 & -13.5 & -14.2 & 0.19/0.40 & 0.24/0.49 & 0.30/0.57 & 0.36/0.71 & 0.36/0.78 & 0.51/1.57 & 1.07/3.91 & 1.85/4.55  \\
SolarRP\_lanth\_x1em1 & -14.3 & -14.9 & -14.7 & -14.4 & -14.5 & -15.0 & -14.8 & -14.7 & 0.17/0.44 & 0.24/0.74 & 0.30/1.14 & 0.73/1.47 & 0.37/1.78 & 0.72/2.21 & 1.08/2.94 & 1.16/3.06  \\
SolarRP\_lanth\_x1em2 & -15.5 & -15.6 & -15.2 & -15.0 & -15.0 & -15.2 & -14.4 & -14.0 & 0.26/0.52 & 0.37/0.83 & 0.58/1.22 & 0.62/1.37 & 0.62/1.51 & 0.82/2.23 & 0.90/2.17 & 0.97/2.22  \\
SolarRP\_lanth\_x1em3 & -15.6 & -15.4 & -15.0 & -14.8 & -14.7 & -14.8 & -13.6 & -12.9 & 0.30/0.69 & 0.33/1.15 & 0.33/1.49 & 0.33/1.56 & 0.36/1.53 & 0.97/2.94 & 1.05/1.53 & 0.34/2.63  \\
SolarRP\_lanth\_x1em4 & -15.6 & -15.3 & -15.0 & -14.8 & -14.7 & -14.6 & -13.5 & -12.9 & 0.30/0.74 & 0.33/1.26 & 0.33/1.60 & 0.34/0.72 & 0.36/0.72 & 1.00/1.43 & 0.35/1.43 & 0.34/0.75  \\
SolarRP\_lanth\_x1em5 & -15.6 & -15.3 & -15.0 & -14.8 & -14.7 & -14.6 & -13.5 & -12.9 & 0.30/0.74 & 0.33/1.28 & 0.33/1.61 & 0.34/0.70 & 0.36/0.70 & 1.00/1.40 & 0.35/1.42 & 0.34/0.73  \\
SolarRP\_lanth\_x1em6 & -15.6 & -15.3 & -15.0 & -14.8 & -14.7 & -14.6 & -13.5 & -12.9 & 0.30/0.74 & 0.32/1.28 & 0.33/1.61 & 0.34/0.70 & 0.36/0.70 & 1.00/1.40 & 0.34/1.42 & 0.34/0.73  \\
SolarRP\_lanth\_x1em7 & -15.6 & -15.3 & -15.0 & -14.8 & -14.7 & -14.6 & -13.5 & -12.9 & 0.30/0.74 & 0.32/1.28 & 0.33/1.61 & 0.33/0.70 & 0.36/0.70 & 1.00/1.41 & 0.34/1.42 & 0.36/0.73  \\
SolarRP\_lanth\_x1em8 & -15.6 & -15.3 & -15.0 & -14.8 & -14.7 & -14.6 & -13.5 & -12.9 & 0.30/0.74 & 0.32/1.28 & 0.33/1.61 & 0.34/0.70 & 0.36/0.70 & 1.00/1.40 & 0.35/1.42 & 0.35/0.73  \\
\hline
\end{tabular}
\label{tab:solarla}
\end{adjustwidth}
\end{table*}

\bibliography{refs}

\begin{thebibliography}{}
\expandafter\ifx\csname natexlab\endcsname\relax\def\natexlab#1{#1}\fi
\providecommand{\url}[1]{\href{#1}{#1}}

\bibitem[{{Abbott} {et~al.}(2017{\natexlab{a}}){Abbott}, {Abbott}, {Abbott},
  {Acernese}, {Ackley}, {Adams}, {Adams}, {Addesso}, {Adhikari}, {Adya}, \&
  et~al.}]{abbott17a}
{Abbott}, B.~P., {Abbott}, R., {Abbott}, T.~D., {et~al.} 2017{\natexlab{a}},
  \apjl, 848, L12

\bibitem[{{Abbott} {et~al.}(2017{\natexlab{b}}){Abbott}, {Abbott}, {Abbott},
  {Acernese}, {Ackley}, {Adams}, {Adams}, {Addesso}, {Adhikari}, {Adya}, \&
  et~al.}]{abbott17b}
---. 2017{\natexlab{b}}, \apjl, 848, L13

\bibitem[{{Abbott} {et~al.}(2017{\natexlab{c}}){Abbott}, {Abbott}, {Abbott},
  {Acernese}, {Ackley}, {Adams}, {Adams}, {Addesso}, {Adhikari}, {Adya}, \&
  et~al.}]{abbott17c}
---. 2017{\natexlab{c}}, \apjl, 850, L39

\bibitem[{{Arcavi} {et~al.}(2017){Arcavi}, {Hosseinzadeh}, {Howell}, {McCully},
  {Poznanski}, {Kasen}, {Barnes}, {Zaltzman}, {Vasylyev}, {Maoz}, \&
  {Valenti}}]{2017Natur.551...64A}
{Arcavi}, I., {Hosseinzadeh}, G., {Howell}, D.~A., {et~al.} 2017, \nat, 551, 64

\bibitem[{{Bauswein} {et~al.}(2013){Bauswein}, {Goriely}, \&
  {Janka}}]{bauswein13a}
{Bauswein}, A., {Goriely}, S., \& {Janka}, H.-T. 2013, \apj, 773, 78

\bibitem[{{Bloom} {et~al.}(1999){Bloom}, {Sigurdsson}, \& {Pols}}]{bloom99}
{Bloom}, J.~S., {Sigurdsson}, S., \& {Pols}, O.~R. 1999, \mnras, 305, 763

\bibitem[{{Bovard} {et~al.}(2017){Bovard}, {Martin}, {Guercilena}, {Arcones},
  {Rezzolla}, \& {Korobkin}}]{bovard17}
{Bovard}, L., {Martin}, D., {Guercilena}, F., {et~al.} 2017, \prd, 96, 124005

\bibitem[{{Chornock} {et~al.}(2017){Chornock}, {Berger}, {Kasen},
  {Cowperthwaite}, {Nicholl}, {Villar}, {Alexander}, {Blanchard}, {Eftekhari},
  {Fong}, {Margutti}, {Williams}, {Annis}, {Brout}, {Brown}, {Chen}, {Drout},
  {Farr}, {Foley}, {Frieman}, {Fryer}, {Herner}, {Holz}, {Kessler}, {Matheson},
  {Metzger}, {Quataert}, {Rest}, {Sako}, {Scolnic}, {Smith}, \&
  {Soares-Santos}}]{2017ApJ...848L..19C}
{Chornock}, R., {Berger}, E., {Kasen}, D., {et~al.} 2017, \apjl, 848, L19

\bibitem[{{C{\^o}t{\'e}} {et~al.}(2018){C{\^o}t{\'e}}, {Fryer}, {Belczynski},
  {Korobkin}, {Chru{\'s}li{\'n}ska}, {Vassh}, {Mumpower}, {Lippuner},
  {Sprouse}, {Surman}, \& {Wollaeger}}]{cote18}
{C{\^o}t{\'e}}, B., {Fryer}, C.~L., {Belczynski}, K., {et~al.} 2018, \apj, 855,
  99

\bibitem[{{Cowperthwaite} {et~al.}(2017){Cowperthwaite}, {Berger}, {Villar},
  {Metzger}, {Nicholl}, {Chornock}, {Blanchard}, {Fong}, {Margutti},
  {Soares-Santos}, {Alexander}, {Allam}, {Annis}, {Brout}, {Brown}, {Butler},
  {Chen}, {Diehl}, {Doctor}, {Drout}, {Eftekhari}, {Farr}, {Finley}, {Foley},
  {Frieman}, {Fryer}, {Garc{\'{\i}}a-Bellido}, {Gill}, {Guillochon}, {Herner},
  {Holz}, {Kasen}, {Kessler}, {Marriner}, {Matheson}, {Neilsen}, {Quataert},
  {Palmese}, {Rest}, {Sako}, {Scolnic}, {Smith}, {Tucker}, {Williams},
  {Balbinot}, {Carlin}, {Cook}, {Durret}, {Li}, {Lopes}, {Louren{\c c}o},
  {Marshall}, {Medina}, {Muir}, {Mu{\~n}oz}, {Sauseda}, {Schlegel}, {Secco},
  {Vivas}, {Wester}, {Zenteno}, {Zhang}, {Abbott}, {Banerji}, {Bechtol},
  {Benoit-L{\'e}vy}, {Bertin}, {Buckley-Geer}, {Burke}, {Capozzi}, {Carnero
  Rosell}, {Carrasco Kind}, {Castander}, {Crocce}, {Cunha}, {D'Andrea}, {da
  Costa}, {Davis}, {DePoy}, {Desai}, {Dietrich}, {Drlica-Wagner}, {Eifler},
  {Evrard}, {Fernandez}, {Flaugher}, {Fosalba}, {Gaztanaga}, {Gerdes},
  {Giannantonio}, {Goldstein}, {Gruen}, {Gruendl}, {Gutierrez}, {Honscheid},
  {Jain}, {James}, {Jeltema}, {Johnson}, {Johnson}, {Kent}, {Krause}, {Kron},
  {Kuehn}, {Nuropatkin}, {Lahav}, {Lima}, {Lin}, {Maia}, {March}, {Martini},
  {McMahon}, {Menanteau}, {Miller}, {Miquel}, {Mohr}, {Neilsen}, {Nichol},
  {Ogando}, {Plazas}, {Roe}, {Romer}, {Roodman}, {Rykoff}, {Sanchez},
  {Scarpine}, {Schindler}, {Schubnell}, {Sevilla-Noarbe}, {Smith}, {Smith},
  {Sobreira}, {Suchyta}, {Swanson}, {Tarle}, {Thomas}, {Thomas}, {Troxel},
  {Vikram}, {Walker}, {Wechsler}, {Weller}, {Yanny}, \&
  {Zuntz}}]{2017ApJ...848L..17C}
{Cowperthwaite}, P.~S., {Berger}, E., {Villar}, V.~A., {et~al.} 2017, \apjl,
  848, L17

\bibitem[{{Evans} {et~al.}(2017){Evans}, {Cenko}, {Kennea}, {Emery}, {Kuin},
  {Korobkin}, {Wollaeger}, {Fryer}, {Madsen}, {Harrison}, {Xu}, {Nakar},
  {Hotokezaka}, {Lien}, {Campana}, {Oates}, {Troja}, {Breeveld}, {Marshall},
  {Barthelmy}, {Beardmore}, {Burrows}, {Cusumano}, {D'A{\`i}}, {D'Avanzo},
  {D'Elia}, {de Pasquale}, {Even}, {Fontes}, {Forster}, {Garcia}, {Giommi},
  {Grefenstette}, {Gronwall}, {Hartmann}, {Heida}, {Hungerford}, {Kasliwal},
  {Krimm}, {Levan}, {Malesani}, {Melandri}, {Miyasaka}, {Nousek}, {O'Brien},
  {Osborne}, {Pagani}, {Page}, {Palmer}, {Perri}, {Pike}, {Racusin}, {Rosswog},
  {Siegel}, {Sakamoto}, {Sbarufatti}, {Tagliaferri}, {Tanvir}, \&
  {Tohuvavohu}}]{2017Sci...358.1565E}
{Evans}, P.~A., {Cenko}, S.~B., {Kennea}, J.~A., {et~al.} 2017, Science, 358,
  1565

\bibitem[{{Fern{\'a}ndez} \& {Metzger}(2013)}]{2013MNRAS.435..502F}
{Fern{\'a}ndez}, R., \& {Metzger}, B.~D. 2013, \mnras, 435, 502

\bibitem[{{Fong} \& {Berger}(2013)}]{fong13}
{Fong}, W., \& {Berger}, E. 2013, \apj, 776, 18

\bibitem[{{Fontes} {et~al.}(2019){Fontes}, {Fryer}, {Hungerford}, {Wollaeger},
  \& {Korobkin}}]{fontes19}
{Fontes}, C.~J., {Fryer}, C.~L., {Hungerford}, A.~L., {Wollaeger}, R.~T., \&
  {Korobkin}, O. 2019, arXiv e-prints, arXiv:1904.08781

\bibitem[{{Fontes} {et~al.}(2017){Fontes}, {Fryer}, {Hungerford}, {Wollaeger},
  {Rosswog}, \& {Berger}}]{fontes17}
{Fontes}, C.~J., {Fryer}, C.~L., {Hungerford}, A.~L., {et~al.} 2017, arXiv
  e-prints, arXiv:1702.02990

\bibitem[{{Fontes} {et~al.}(2015){Fontes}, {Zhang}, {Abdallah}, {Clark},
  {Kilcrease}, {Colgan}, {Cunningham}, {Hakel}, {Magee}, \&
  {Sherrill}}]{fontes2015b}
{Fontes}, C.~J., {Zhang}, H.~L., {Abdallah}, J., {et~al.} 2015, {Journal of
  Physics B: Atomic, Molecular and Optical Physics}, 48, 144014

\bibitem[{{Foucart} {et~al.}(2018){Foucart}, {Duez}, {Kidder}, {Nguyen},
  {Pfeiffer}, \& {Scheel}}]{2018PhRvD..98f3007F}
{Foucart}, F., {Duez}, M.~D., {Kidder}, L.~E., {et~al.} 2018, \prd, 98, 063007

\bibitem[{{Freiburghaus} {et~al.}(1999){Freiburghaus}, {Rosswog}, \&
  {Thielemann}}]{freiburghaus99}
{Freiburghaus}, C., {Rosswog}, S., \& {Thielemann}, F.-K. 1999, \apjl, 525,
  L121.
\newblock \url{http://adsabs.harvard.edu/abs/1999ApJ...525L.121F}

\bibitem[{{Fryer}(2009)}]{fryer09}
{Fryer}, C.~L. 2009, \apj, 699, 409

\bibitem[{{Fryer} {et~al.}(1999){Fryer}, {Woosley}, \& {Hartmann}}]{fryer99}
{Fryer}, C.~L., {Woosley}, S.~E., \& {Hartmann}, D.~H. 1999, \apj, 526, 152

\bibitem[{{Fryer} {et~al.}(2006){Fryer}, {Young}, \& {Hungerford}}]{fryer06}
{Fryer}, C.~L., {Young}, P.~A., \& {Hungerford}, A.~L. 2006, \apj, 650, 1028

\bibitem[{{Just} {et~al.}(2015){Just}, {Bauswein}, {Ardevol Pulpillo},
  {Goriely}, \& {Janka}}]{2015MNRAS.448..541J}
{Just}, O., {Bauswein}, A., {Ardevol Pulpillo}, R., {Goriely}, S., \& {Janka},
  H.-T. 2015, \mnras, 448, 541

\bibitem[{{Kasen} {et~al.}(2017){Kasen}, {Metzger}, {Barnes}, {Quataert}, \&
  {Ramirez-Ruiz}}]{2017Natur.551...80K}
{Kasen}, D., {Metzger}, B., {Barnes}, J., {Quataert}, E., \& {Ramirez-Ruiz}, E.
  2017, \nat, 551, 80

\bibitem[{{Kasliwal} {et~al.}(2017){Kasliwal}, {Korobkin}, {Lau}, {Wollaeger},
  \& {Fryer}}]{kasliwal17}
{Kasliwal}, M.~M., {Korobkin}, O., {Lau}, R.~M., {Wollaeger}, R., \& {Fryer},
  C.~L. 2017, \apjl, 843, L34

\bibitem[{{Korobkin} {et~al.}(2012){Korobkin}, {Rosswog}, {Arcones}, \&
  {Winteler}}]{korobkin12}
{Korobkin}, O., {Rosswog}, S., {Arcones}, A., \& {Winteler}, C. 2012, \mnras,
  426, 1940

\bibitem[{{Lattimer} \& {Schramm}(1974)}]{lattimer74}
{Lattimer}, J.~M., \& {Schramm}, D.~N. 1974, \apjl, 192, L145.
\newblock \url{http://adsabs.harvard.edu/abs/1974ApJ...192L.145L}

\bibitem[{{Metzger} \& {Fern{\'a}ndez}(2014)}]{metzger14}
{Metzger}, B.~D., \& {Fern{\'a}ndez}, R. 2014, \mnras, 441, 3444

\bibitem[{{Miller} {et~al.}(2019){Miller}, {Ryan}, \& {Dolence}}]{miller19}
{Miller}, J.~M., {Ryan}, B.~R., \& {Dolence}, J.~C. 2019, \apjs, 241, 30

\bibitem[{{Mooley} {et~al.}(2018{\natexlab{a}}){Mooley}, {Deller}, {Gottlieb},
  {Nakar}, {Hallinan}, {Bourke}, {Frail}, {Horesh}, {Corsi}, \&
  {Hotokezaka}}]{mooley18a}
{Mooley}, K.~P., {Deller}, A.~T., {Gottlieb}, O., {et~al.} 2018{\natexlab{a}},
  \nat, 561, 355

\bibitem[{{Mooley} {et~al.}(2018{\natexlab{b}}){Mooley}, {Frail}, {Dobie},
  {Lenc}, {Corsi}, {De}, {Nayana}, {Makhathini}, {Heywood}, {Murphy}, {Kaplan},
  {Chandra}, {Smirnov}, {Nakar}, {Hallinan}, {Camilo}, {Fender}, {Goedhart},
  {Groot}, {Kasliwal}, {Kulkarni}, \& {Woudt}}]{mooley18b}
{Mooley}, K.~P., {Frail}, D.~A., {Dobie}, D., {et~al.} 2018{\natexlab{b}},
  \apjl, 868, L11

\bibitem[{{Mumpower} {et~al.}(2018){Mumpower}, {Kawano}, {Sprouse}, {Vassh},
  {Holmbeck}, {Surman}, \& {M{\"o}ller}}]{mumpower18}
{Mumpower}, M.~R., {Kawano}, T., {Sprouse}, T.~M., {et~al.} 2018, \apj, 869, 14

\bibitem[{Mumpower {et~al.}(2017)Mumpower, McLaughlin, Surman, \&
  Steiner}]{mumpower17}
Mumpower, M.~R., McLaughlin, G., Surman, R., \& Steiner, A.~W. 2017, Journal of
  Physics G: Nuclear and Particle Physics, 44, 034003

\bibitem[{{Mumpower} {et~al.}(2016){Mumpower}, {Surman}, {McLaughlin}, \&
  {Aprahamian}}]{mumpower16}
{Mumpower}, M.~R., {Surman}, R., {McLaughlin}, G.~C., \& {Aprahamian}, A. 2016,
  Progress in Particle and Nuclear Physics, 86, 86

\bibitem[{{Narayan} {et~al.}(1992){Narayan}, {Paczynski}, \&
  {Piran}}]{narayan92}
{Narayan}, R., {Paczynski}, B., \& {Piran}, T. 1992, \apjl, 395, L83

\bibitem[{{Nicholl} {et~al.}(2017){Nicholl}, {Berger}, {Kasen}, {Metzger},
  {Elias}, {Brice{\~n}o}, {Alexander}, {Blanchard}, {Chornock},
  {Cowperthwaite}, {Eftekhari}, {Fong}, {Margutti}, {Villar}, {Williams},
  {Brown}, {Annis}, {Bahramian}, {Brout}, {Brown}, {Chen}, {Clemens},
  {Dennihy}, {Dunlap}, {Holz}, {Marchesini}, {Massaro}, {Moskowitz},
  {Pelisoli}, {Rest}, {Ricci}, {Sako}, {Soares-Santos}, \&
  {Strader}}]{2017ApJ...848L..18N}
{Nicholl}, M., {Berger}, E., {Kasen}, D., {et~al.} 2017, \apjl, 848, L18

\bibitem[{Orford {et~al.}(2018)Orford, Vassh, Clark, McLaughlin, Mumpower,
  Savard, Surman, Aprahamian, Buchinger, Burkey, {et~al.}}]{orford18}
Orford, R., Vassh, N., Clark, J., {et~al.} 2018, Physical review letters, 120,
  262702

\bibitem[{{Perego} {et~al.}(2017){Perego}, {Radice}, \&
  {Bernuzzi}}]{2017ApJ...850L..37P}
{Perego}, A., {Radice}, D., \& {Bernuzzi}, S. 2017, \apjl, 850, L37

\bibitem[{{Popham} {et~al.}(1999){Popham}, {Woosley}, \& {Fryer}}]{popham99}
{Popham}, R., {Woosley}, S.~E., \& {Fryer}, C. 1999, \apj, 518, 356

\bibitem[{{Radice} {et~al.}(2016){Radice}, {Galeazzi}, {Lippuner}, {Roberts},
  {Ott}, \& {Rezzolla}}]{2016MNRAS.460.3255R}
{Radice}, D., {Galeazzi}, F., {Lippuner}, J., {et~al.} 2016, \mnras, 460, 3255

\bibitem[{{Smartt} {et~al.}(2017){Smartt}, {Chen}, {Jerkstrand}, {Coughlin},
  {Kankare}, {Sim}, {Fraser}, {Inserra}, {Maguire}, {Chambers}, {Huber},
  {Kr{\"u}hler}, {Leloudas}, {Magee}, {Shingles}, {Smith}, {Young}, {Tonry},
  {Kotak}, {Gal-Yam}, {Lyman}, {Homan}, {Agliozzo}, {Anderson}, {Angus},
  {Ashall}, {Barbarino}, {Bauer}, {Berton}, {Botticella}, {Bulla}, {Bulger},
  {Cannizzaro}, {Cano}, {Cartier}, {Cikota}, {Clark}, {De Cia}, {Della Valle},
  {Denneau}, {Dennefeld}, {Dessart}, {Dimitriadis}, {Elias-Rosa}, {Firth},
  {Flewelling}, {Fl{\"o}rs}, {Franckowiak}, {Frohmaier}, {Galbany},
  {Gonz{\'a}lez-Gait{\'a}n}, {Greiner}, {Gromadzki}, {Guelbenzu},
  {Guti{\'e}rrez}, {Hamanowicz}, {Hanlon}, {Harmanen}, {Heintz}, {Heinze},
  {Hernandez}, {Hodgkin}, {Hook}, {Izzo}, {James}, {Jonker}, {Kerzendorf},
  {Klose}, {Kostrzewa-Rutkowska}, {Kowalski}, {Kromer}, {Kuncarayakti},
  {Lawrence}, {Lowe}, {Magnier}, {Manulis}, {Martin-Carrillo}, {Mattila},
  {McBrien}, {M{\"u}ller}, {Nordin}, {O'Neill}, {Onori}, {Palmerio},
  {Pastorello}, {Patat}, {Pignata}, {Podsiadlowski}, {Pumo}, {Prentice}, {Rau},
  {Razza}, {Rest}, {Reynolds}, {Roy}, {Ruiter}, {Rybicki}, {Salmon}, {Schady},
  {Schultz}, {Schweyer}, {Seitenzahl}, {Smith}, {Sollerman}, {Stalder},
  {Stubbs}, {Sullivan}, {Szegedi}, {Taddia}, {Taubenberger}, {Terreran}, {van
  Soelen}, {Vos}, {Wainscoat}, {Walton}, {Waters}, {Weiland}, {Willman},
  {Wiseman}, {Wright}, {Wyrzykowski}, \& {Yaron}}]{2017Natur.551...75S}
{Smartt}, S.~J., {Chen}, T.-W., {Jerkstrand}, A., {et~al.} 2017, \nat, 551, 75

\bibitem[{{Surman} \& {McLaughlin}(2004)}]{surman04}
{Surman}, R., \& {McLaughlin}, G.~C. 2004, \apj, 603, 611

\bibitem[{{Surman} {et~al.}(2006){Surman}, {McLaughlin}, \& {Hix}}]{surman06}
{Surman}, R., {McLaughlin}, G.~C., \& {Hix}, W.~R. 2006, \apj, 643, 1057

\bibitem[{{Surman} {et~al.}(2008){Surman}, {McLaughlin}, {Ruffert}, {Janka}, \&
  {Hix}}]{surman08}
{Surman}, R., {McLaughlin}, G.~C., {Ruffert}, M., {Janka}, H.-T., \& {Hix},
  W.~R. 2008, \apjl, 679, L117

\bibitem[{{Tanaka} {et~al.}(2017){Tanaka}, {Utsumi}, {Mazzali}, {Tominaga},
  {Yoshida}, {Sekiguchi}, {Morokuma}, {Motohara}, {Ohta}, {Kawabata}, {Abe},
  {Aoki}, {Asakura}, {Baar}, {Barway}, {Bond}, {Doi}, {Fujiyoshi}, {Furusawa},
  {Honda}, {Itoh}, {Kawabata}, {Kawai}, {Kim}, {Lee}, {Miyazaki}, {Morihana},
  {Nagashima}, {Nagayama}, {Nakaoka}, {Nakata}, {Ohsawa}, {Ohshima}, {Okita},
  {Saito}, {Sumi}, {Tajitsu}, {Takahashi}, {Takayama}, {Tamura}, {Tanaka},
  {Terai}, {Tristram}, {Yasuda}, \& {Zenko}}]{2017PASJ...69..102T}
{Tanaka}, M., {Utsumi}, Y., {Mazzali}, P.~A., {et~al.} 2017, \pasj, 69, 102

\bibitem[{{Tanvir} {et~al.}(2017){Tanvir}, {Levan},
  {Gonz{\'a}lez-Fern{\'a}ndez}, {Korobkin}, {Mandel}, {Rosswog}, {Hjorth},
  {D'Avanzo}, {Fruchter}, {Fryer}, {Kangas}, {Milvang-Jensen}, {Rosetti},
  {Steeghs}, {Wollaeger}, {Cano}, {Copperwheat}, {Covino}, {D'Elia}, {de Ugarte
  Postigo}, {Evans}, {Even}, {Fairhurst}, {Figuera Jaimes}, {Fontes}, {Fujii},
  {Fynbo}, {Gompertz}, {Greiner}, {Hodosan}, {Irwin}, {Jakobsson},
  {J{\o}rgensen}, {Kann}, {Lyman}, {Malesani}, {McMahon}, {Melandri},
  {O'Brien}, {Osborne}, {Palazzi}, {Perley}, {Pian}, {Piranomonte}, {Rabus},
  {Rol}, {Rowlinson}, {Schulze}, {Sutton}, {Th{\"o}ne}, {Ulaczyk}, {Watson},
  {Wiersema}, \& {Wijers}}]{tanvir17}
{Tanvir}, N.~R., {Levan}, A.~J., {Gonz{\'a}lez-Fern{\'a}ndez}, C., {et~al.}
  2017, \apjl, 848, L27

\bibitem[{{Thielemann} {et~al.}(2017){Thielemann}, {Eichler}, {Panov}, \&
  {Wehmeyer}}]{2017ARNPS..6701916T}
{Thielemann}, K., F., {Eichler}, M., {Panov}, I.~V., \& {Wehmeyer}, B. 2017,
  Annual Review of Nuclear and Particle Science, 67, annurev

\bibitem[{{Troja} {et~al.}(2017){Troja}, {Piro}, {van Eerten}, {Wollaeger},
  {Im}, {Fox}, {Butler}, {Cenko}, {Sakamoto}, {Fryer}, {Ricci}, {Lien}, {Ryan},
  {Korobkin}, {Lee}, {Burgess}, {Lee}, {Watson}, {Choi}, {Covino}, {D'Avanzo},
  {Fontes}, {Gonz{\'a}lez}, {Khandrika}, {Kim}, {Kim}, {Lee}, {Lee}, {Kutyrev},
  {Lim}, {S{\'a}nchez-Ram{\'{\i}}rez}, {Veilleux}, {Wieringa}, \&
  {Yoon}}]{troja17}
{Troja}, E., {Piro}, L., {van Eerten}, H., {et~al.} 2017, \nat, 551, 71

\bibitem[{{Vassh} {et~al.}(2018){Vassh}, {Vogt}, {Surman}, {Randrup},
  {Sprouse}, {Mumpower}, {Jaffke}, {Shaw}, {Holmbeck}, {Zhu}, \&
  {McLaughlin}}]{vassh19}
{Vassh}, N., {Vogt}, R., {Surman}, R., {et~al.} 2018, arXiv e-prints,
  arXiv:1810.08133

\bibitem[{Vil{\'e}n {et~al.}(2018)Vil{\'e}n, Kelly, Kankainen, Brodeur,
  Aprahamian, Canete, Eronen, Jokinen, Kuta, Moore, {et~al.}}]{vilen18}
Vil{\'e}n, M., Kelly, J., Kankainen, A., {et~al.} 2018, Physical review
  letters, 120, 262701

\bibitem[{{Wollaeger} \& {van Rossum}(2014)}]{wollaeger2014}
{Wollaeger}, R.~T., \& {van Rossum}, D.~R. 2014, \apjs, 214, 28

\bibitem[{{Wollaeger} {et~al.}(2013){Wollaeger}, {van Rossum}, {Graziani},
  {Couch}, {Jordan}, {Lamb}, \& {Moses}}]{wollaeger2013}
{Wollaeger}, R.~T., {van Rossum}, D.~R., {Graziani}, C., {et~al.} 2013, \apjs,
  209, 36

\bibitem[{{Wollaeger} {et~al.}(2018){Wollaeger}, {Korobkin}, {Fontes},
  {Rosswog}, {Even}, {Fryer}, {Sollerman}, {Hungerford}, {van Rossum}, \&
  {Wollaber}}]{wollaeger18}
{Wollaeger}, R.~T., {Korobkin}, O., {Fontes}, C.~J., {et~al.} 2018, \mnras,
  478, 3298

\bibitem[{{Zhu} {et~al.}(2018){Zhu}, {Wollaeger}, {Vassh}, {Surman}, {Sprouse},
  {Mumpower}, {M{\"o}ller}, {McLaughlin}, {Korobkin}, {Kawano}, {Jaffke},
  {Holmbeck}, {Fryer}, {Even}, {Couture}, \& {Barnes}}]{zhu18}
{Zhu}, Y., {Wollaeger}, R.~T., {Vassh}, N., {et~al.} 2018, \apjl, 863, L23

\end{thebibliography}

\end{document}